\documentclass[useAMS,usenatbib,usegraphicx]{mn2e}

\newcommand{\lsun}{L$_\odot$}
\newcommand{\rsun}{R$_\odot$}
\newcommand{\msun}{M$_\odot$}
\newcommand{\kmps}{km\,s$^{-1}$}
\newcommand{\teff}{$T_{\rm eff}$}

\usepackage{txfonts}
\usepackage{lscape}

\title[Hot subdwarf binaries]{New binaries among UV-selected, hot subdwarf stars and population properties \thanks{Based on observations made with ESO telescopes at the La Silla Paranal Observatory
under programmes 076.D-0355, 077.D-0515, 078.D-0098, 086.D-0714, 089.D-0864, 090.D-0012 and 093.D-0273.}}

\author[Kawka et al.]{A. Kawka$^{1}$\thanks{E-mail:
kawka@asu.cas.cz (AK); vennes@asu.cas.cz (SV); otoole@aao. gov.au (SO); pnemeth1981@gmail.com (PN); donna.burton@usq.edu.au (DB);
ejk@saao.ac.za (EK); dibnob@saao.ac.za (DAHB)} 
\thanks{Visiting Astronomer, Kitt Peak National 
Observatory, National Optical Astronomy Observatory, which is operated by the 
Association of Universities for Research in Astronomy (AURA) under cooperative 
agreement with the National Science Foundation.}, 
S. Vennes$^{1}$\footnotemark[2]\footnotemark[3],
S. O'Toole$^{2}$\footnotemark[2], 
P. N\'emeth$^{3}$\footnotemark[2]\footnotemark[3], 
D. Burton$^{4}$\footnotemark[2],
\newauthor E. Kotze$^{5,6}$\footnotemark[2] and 
D.A.H. Buckley$^{5,7}$\footnotemark[2]\\
$^{1}$Astronomick\'y \'ustav AV \v{C}R, Fri\v{c}ova 298,CZ-251 65 Ond\v{r}ejov,
Czech Republic\\
$^{2}$Australian Astronomical Observatory, P.O. Box 915, 1670 North Ryde NSW,
Australia\\
$^{3}$Dr. Remeis--Sternwarte, Institute for Astronomy, University
Erlangen-N\"{u}rnberg, Sternwartstr. 7, 96049 Bamberg, Germany\\
$^{4}$Faculty of Sciences, University of Southern Queensland, Toowoomba, 
QLD 4350, Australia\\
$^{5}$South African Astronomical Observatory, Observatory Road, Observatory 7935, South Africa\\
$^{6}$Department of Astronomy, University of Cape Town, Rondebosch 7770, Cape Town, South Africa\\
$^{7}$South African Large Telescope, PO Box 9, Observatory 7935, South Africa}
\begin{document}

\date{}

\pagerange{\pageref{firstpage}--\pageref{lastpage}} \pubyear{2015}

\maketitle

\label{firstpage}

\begin{abstract}
We have measured the orbital parameters of seven close binaries, including six new objects, in a 
radial velocity survey of 38 objects comprising a hot subdwarf 
star with orbital periods ranging from $\sim0.17$ to 3~d. 
One new system, GALEX~J2205$-$3141, shows reflection on a M dwarf companion. Three 
other objects show significant short-period variations, but their orbital 
parameters could not be constrained. Two systems comprising a hot 
subdwarf paired with a bright main-sequence/giant companion display 
short-period photometric variations possibly due to irradiation or 
stellar activity and are also short-period candidates. All except two candidates 
were drawn from a selection of subluminous stars in the {\it Galaxy 
Evolution Explorer} ultraviolet sky survey. Our new 
identifications also include a low-mass subdwarf B star and likely progenitor 
of a low mass white dwarf (GALEX~J0805$-$1058) paired with an unseen, 
possibly substellar, companion. The mass functions of the newly identified 
binaries imply minimum secondary masses ranging from 0.03 to 0.39\,\msun.
Photometric time series suggest that, apart from GALEX~J0805$-$1058 and J2205$-$3141,
the companions are most likely white dwarfs.
We update the binary population statistics: 
Close to 40 per cent of hot subdwarfs have a companion. Also, we found that the secondary mass distribution shows a low-mass peak 
attributed to late-type dwarfs, and a higher-mass peak and tail 
distribution attributed to white dwarfs and a few spectroscopic 
composites. Also, we found that the population kinematics imply an 
old age and include a few likely halo population members. 
\end{abstract}


\section{Introduction}

Hot subdwarf stars \citep[see a review by ][]{heb2009} are core helium burning stars with very thin hydrogen envelopes
and belong to the extreme horizontal branch (EHB). The mass of most hot subdwarfs is about
0.5\,\msun.
The origin of EHB stars, i.e, the
hot, hydrogen-rich (sdB) and helium-rich subdwarf (sdO) stars, is
closely linked to binarity. \citet{men1976} first proposed
that sdB stars are formed in close binary systems and \citet{dor1993} inferred
the presence of an extremely thin hydrogen envelope ($<0.001$\,\msun). \citet{han2002,han2003}
proposed three formation channels for sdB stars
through binary interaction, i.e., common envelope (CE), Roche lobe overflow (RLOF), and binary
merger. \citet{han2003} predict a binary fraction
of 76 - 89 per cent with orbital periods ranging from 0.5~hr
to 500~d. However, they caution that the observed frequency
could be much lower due to selection effects. The proposed formation
channels also predict single sdB stars that form via the
merger of two helium white dwarfs. Approximately 11 - 26 per cent
of subdwarfs are expected to form via this merger channel \citep{han2003}.

Formation channels of helium-rich (He-sdO) stars are not as well
defined. \citet{jus2011} proposed that these objects may form in a
close double degenerate binary with the massive component accreting from a helium
white dwarf companion and initiating helium-shell burning. A small number of sdO
stars are known to exist as companions to Be stars \citep{gie1998,pet2008,pet2013}.
These sdO stars are formed through close binary interaction where the more massive
primary star begins mass transfer onto its less massive companion during its 
shell-hydrogen burning phase. The result of this mass transfer leaves a spun up Be 
star with an sdO companion \citep{pol1991}.

Cool companions to hot subdwarf stars can be revealed as infrared
excess in the spectral energy distribution (SED). \citet{the1995} and \citet{ull1998}
detected infrared excess in over
20 per cent of the hot subdwarf stars studied in their sample. \citet{gir2012}
explored photometric surveys that cover a wide wavelength
range, from the Galaxy Evolution Explorer ($GALEX$) ultraviolet
survey through to the infrared, the Two Micron All Sky Survey (2MASS) and the UKIRT Infrared Deep Sky Survey (UKIDSS), 
and searched for main-sequence companions to hot subdwarf stars. They found that
the most common companions to hot subdwarfs have a spectral type
between F0 and K0, while M-type companions were found to be
much rarer.

Radial velocity surveys \citep[e.g.,][]{max2001,mor2003,cop2011,gei2011a}
of sdB stars have shown that approximately half of all sdB stars reside in
close binary systems with either a cool main-sequence star or a
white dwarf companion. These surveys target binary systems with periods of a few hours to $\approx 30$ days. \citet{nap2004}
reported a binary fraction of 39 per cent of sdB stars
from the ESO Supernovae type Ia Progenitor surveY (SPY). \citet{cop2011}
estimated a higher binary fraction of 46 - 56
per cent from their survey of sdB stars selected from the Palomar-Green
and Edinburgh-Cape surveys.

A few rare sdB stars are found in close orbit with a massive
white dwarf ($M_{\rm WD} \ga 0.9$\,\msun), making them Type Ia
supernova progenitors. These systems would first evolve to AM
CVn systems before detonating either as a Type Ia 
or the less energetic Type~.Ia (Iax) supernova \citep{bil2007,fin2010,sol2010}. The
first such candidate is KPD 1930+2752 \citep{max2000a,gei2007},
with a second candidate, GALEX~J1411$-$3053 (CD$-$30~11223),
discovered as part of our radial velocity survey of $GALEX$
selected hot subdwarf stars \citep{ven2012}.

Some sdB stars in close binary systems have stellar parameters that fall below
the zero-age horizontal branch and probably did not initiate helium burning. Such
objects have very low masses 
($\approx 0.2$\,\msun) and are the progenitors of 
extremely low mass (ELM) white dwarfs, which will in time evolve into AM CVn systems. 
If the companion to these low mass stars is a massive enough white dwarf, then 
the system may become a Type Ia supernova. The first known low mass sdB star,  
HD~188112, was discovered by \citet{heb2003}. 

\citet{ahm2004} discovered the first double subdwarf binary, PG~1544$+$488. 
This helium rich sdB (He-sdB) binary remains, at the present time, unique.
The mass ratio determined from the velocity semi-amplitude of
the components show that they 
have a similar mass which suggests that the system emerged from a CE comprised of two nearly identical red giant cores \citep{sen2014}.
Alternatively, \citet{lan2004} interpreted the peculiar atmospheric composition of He-sdB stars, such as PG~1544$+$488, 
with evolutionary models involving a delayed 
helium-core flash and convective mixing while descending on the white dwarf cooling track. 
Similarly, HE~0301$-$3039 is a close binary consisting of two sdO stars \citep{lis2004,str2007} that
may be the outcome of double-core CE
evolution \citep{jus2011}.

Surveys of hot subdwarfs involving photometric time series
have uncovered several more low 
mass sdB stars. Kepler observations revealed that KIC 6614501 is another low mass 
sdB plus white dwarf system \citep{sil2012}. Also, \citet{max2014} presented 17 
eclipsing systems from Wide Angle Search for Planets (WASP) survey that are 
likely to contain a pre-helium white dwarf, similar to the system 
1SWASP~J024743.37$-$251549.2 \citep{max2011}. Follow-up spectroscopy for six
of these systems confirmed them to be main-sequence A stars with very low
mass ($\approx 0.2$\,\msun) pre-He white dwarfs currently experiencing hydrogen-shell burning.

Wider binaries (orbital periods $\sim$ years) containing a sdB star with a cool 
main-sequence companion were reported by \citet{bar2012,bar2013a} and \citet{vos2013}. 
The predicted period distribution by \citet{han2003} is bimodal with some B to F type companions in
the longer-period range: The relative frequency of
short- to long-period binaries depends on the actual value of the critical
mass ratio for stable mass transfer; this ratio may be set with
a study of potential subdwarf plus A-star binaries.
\citet{che2013} showed that these long period binaries are the result of
stable RLOF.

\citet{ven2011} and \citet{nem2012} presented a new sample of sdB stars selected from the
$GALEX$ all-sky survey and we conducted a radial velocity
survey of a subsample of stars from this selection. The first two systems 
(GALEX~J0321$+$4727 and GALEX~J2349$+$3844) discovered as part of this survey
were presented by \citet{kaw2010a}, followed by the aforementioned short-period system
GALEX~J1411$-$3053 \citep{ven2012,gei2013a}. Additional spectroscopic and photometric
observations of the first two systems were presented in \citet{kaw2012a} along with
a progress report on the other systems that were observed as part of this 
programme. The photometric observations confirmed the reflection effect in 
GALEX~J0321$+$4727 originally reported by \citet{kaw2010a} and based on 
Northern Sky Variability Survey (NSVS)
photometry. The observations also showed that both GALEX~J0321$+$4727 and 
GALEX~J2349$+$3844 are V2093 Her type pulsating subdwarfs \citep{gre2003}. 

In this paper, we present spectroscopic and photometric observations of a sample 
of $GALEX$-selected hot subdwarf stars with the aim of determining their binary 
properties. 
Sections 2.1 and 2.2 present details of
our spectroscopic observations, while Section 2.3 present archival photometric time series. 
In Section 3 we 
present an analysis of stellar properties (3.1), and of binary properties
supplemented by our analysis of photometric time series (3.2).
Finally, we present a review of the properties of known binaries comprising a hot subdwarf star, including
the properties of the components (Section 4.1), the population kinematics (4.2), and the properties of
some outstanding individual cases (4.3), followed by a summary of the present work (4.4). 

\begin{table*}
\centering
\begin{minipage}{\textwidth}
\caption{Target summary. \label{tbl_sum}}
\renewcommand{\footnoterule}{\vspace*{-15pt}}
\renewcommand{\thefootnote}{\alph{footnote}}
\begin{tabular}{llcccl}
\hline
        & Other names &  \teff         & $\log{g}$              & $\log{(\rm He/H)}$          & Notes \footnotemark[1]\footnotetext[1]{RV: confirmed radial velocity variable star; IR: SED of the stars shows significant IR excess.} \\
GALEX~J &             &    (K)                 & c.g.s.                 &                         &    \\
\hline
004759.6$+$033742 & BPS BS 17579-0012, PB 6168     & $38620^{+2250}_{-970}$ & $6.14^{+0.22}_{-0.18}$ & $-2.63^{+0.44}_{-1.17}$ & sdB+F6V; IR; nearby star\\
004729.4$+$095855 & HD~4539, HIP 3701              & $24650^{+590}_{-200}$  & $5.38^{+0.03}_{-0.05}$ & $-2.42^{+0.20}_{-0.07}$ & \\
004917.2$+$205640 & PG 0046+207                    & $27520^{+500}_{-450}$  & $5.55^{+0.07}_{-0.06}$ & $-2.48^{+0.16}_{-0.23}$ & \\
005956.7$+$154419 & HIP 4666, PG 0057+155, PHL 932 & $33530^{+190}_{-310}$  & $5.83^{+0.04}_{-0.05}$ & $-1.69^{+0.06}_{-0.04}$ &  \\
020656.1$+$143900 & CHSS~3497                      & $30310^{+660}_{-80}$   & $5.77^{+0.05}_{-0.06}$ & $-2.61^{+0.15}_{-0.24}$ &  \\
023251.9$+$441126 & FBS 0229+439                   & $33260^{+420}_{-380}$  & $5.73^{+0.09}_{-0.10}$ & $-1.70^{+0.08}_{-0.12}$ &  \\
040105.3$-$322348 & CD-32 1567, EC 03591-3232      & $30490^{+250}_{-220}$  & $5.71^{+0.06}_{-0.04}$ & $-1.92^{+0.06}_{-0.04}$ & \\
050018.9$+$091203 & HS 0457+0907                   & $36270^{+490}_{-1130}$ & $5.75^{+0.15}_{-0.13}$ & $-1.46^{+0.14}_{-0.15}$ & \\
050735.7$+$034814 &                                & $23990^{+630}_{-610}$  & $5.42^{+0.08}_{-0.11}$ & $-3.05^{+0.48}_{-0.78}$ & Ca\,H\&K, RV \\
061325.3$+$342053 &                                & $34250^{+330}_{-390}$  & $5.75^{+0.10}_{-0.06}$ & $-1.28^{+0.04}_{-0.08}$ & RV \\
065736.7$-$732447 & CPD-73 420                     & $29940^{+900}_{-160}$  & $5.45^{+0.07}_{-0.15}$ & $< -3.21$               & nearby star \\
070331.5$+$623626 & FBS 0658+627                   & $28750^{+370}_{-340}$  & $5.40^{+0.07}_{-0.04}$ & $-2.76^{+0.22}_{-0.26}$ & \\
071646.9$+$231930 & TYC 1909-865-1                 & $11140/9310$           & $4.39/3.67$            & ... &  close B+A\,V binary, RV \\
075147.1$+$092526 &                                & $30620^{+490}_{-460}$  & $5.74^{+0.11}_{-0.12}$ & $-2.49^{+0.27}_{-0.30}$ & nearby star (6~arcsec), RV \\
080510.9$-$105834 & TYC 5417-2552-1                & $22320^{+330}_{-280}$  & $5.68^{+0.03}_{-0.06}$ & $ < -3.44$              & ELM WD progenitor, RV \\
081233.6$+$160123 &                                & $31580^{+440}_{-490}$  & $5.56^{+0.10}_{-0.13}$ & $ < -2.90$              & RV \\
104148.6$-$073034 & TYC 5492-642-1                 & $27440^{+620}_{-450}$  & $5.63^{+0.09}_{-0.06}$ & $-2.44^{+0.16}_{-0.23}$ & \\
111422.0$-$242130 & EC 11119-2405, TYC 6649-111-1  & $23430^{+480}_{-450}$  & $5.29^{+0.08}_{-0.07}$ & $-2.46^{+0.19}_{-0.31}$ & \\
135629.2$-$493403 & CD-48 8608, TYC 8271-627-1     & $33070^{+230}_{-660}$  & $5.74^{+0.07}_{-0.16}$ & $-2.75^{+0.25}_{-0.43}$ & sdB+G8V; IR \\
140747.6$+$310318 & BPS BS 16082-0122              & $24900^{+50}_{-3050}$  & $4.25^{+0.03}_{-0.09}$ & $-1.18^{+0.08}_{-0.09}$ & high-$\varv$ early B \\
141133.3$+$703737 & TYC 4406-666-1                 & $21170^{+1500}_{-1110}$& $5.55^{+0.31}_{-0.23}$ & $<-2.36$                & sdB+F; IR; ELM WD progenitor?  \\
142126.5$+$712427 & TYC 4406-285-1                 & $25620^{+320}_{-220}$  & $5.67\pm0.04$          & $<-3.7$                 &  \\
142747.2$-$270108 & EC 14248-2647, TYC 6740-942-1  & $31880^{+360}_{-290}$  & $5.70^{+0.05}_{-0.08}$ & $-1.71^{+0.05}_{-0.11}$ &  \\
143519.8$+$001352 & TYC 325-452-1, PG 1432+004     & $23090^{+780}_{-250}$  & $5.28^{+0.08}_{-0.08}$ & $-2.39^{+0.18}_{-0.20}$ &  \\
163201.4$+$075940 & TYC 960-1373-1, PG 1629+081    & $38110^{+570}_{-680}$  & $5.38^{+0.06}_{-0.09}$ & $-2.71^{+0.27}_{-0.29}$ & nearby star, RV \\
173153.7$+$064706 &                                & $27780^{+1030}_{-470}$ & $5.35^{+0.18}_{-0.07}$ & $<-2.53$                & RV \\
173651.2$+$280635 & TYC 2084-448-1                 & $36160^{+6500}_{-4200}$& $5.24^{+0.84}_{-0.84}$ & $-1.09^{+0.69}_{-1.34}$ & sdB+F7V; IR; variable \\
175340.5$-$500741 &                                & $32430^{+880}_{-570}$  & $5.95^{+0.18}_{-0.18}$ & $-2.25^{+0.31}_{-1.04}$ & sdB+F7V; IR \\
184559.8$-$413826 &                                & $35930^{+840}_{-4770}$ & $5.23^{+0.27}_{-0.23}$ & $+2.10^{+1.10}_{-0.38}$ & sdO; He\,{\sc i} spectrum \\
190211.7$-$513005 & CD-51 11879, TYC 8386-1370-1, LSE 263 & $72300^{+5380}_{-3260}$& $5.49^{+0.11}_{-0.11}$ & $+0.02^{+2.10}_{-0.03}$ & sdO; He\,{\sc ii} spectrum \\
190302.4$-$352828 & BPS CS 22936-0293              & $32100^{+1760}_{-1260}$ & $5.26^{+0.31}_{-0.30}$ & $<-1.96$ & RV \\
191109.2$-$140651 & TYC 5720-292-1                 & $55970^{+4540}_{-1780}$ & $5.69^{+0.71}_{-0.09}$ & $+0.25^{+0.70}_{-0.60}$ & sdO; He\,{\sc ii} spectrum \\
203850.3$-$265750 & TYC 6916-251-1                 & $58450^{+4600}_{-7920}$ & $5.04^{+0.39}_{-0.17}$ & $-1.13^{+0.27}_{-0.29}$ & sdO+G3.5III; IR; variable \\
215340.4$-$700430 & EC 21494-7018, TYC 9327-1311-1 & $23720^{+260}_{-230}$   & $5.65^{+0.03}_{-0.02}$ & $-3.22^{+0.13}_{-1.15}$ & ELM WD progenitor?\\
220551.8$-$314105 & TYC 7489-686-1, BPS CS 30337-0074 & $28650^{+930}_{-80}$ & $5.68^{+0.01}_{-0.03}$ & $-2.09^{+0.12}_{-0.03}$ & reflection, RV \\
225444.1$-$551505 &                                & $31070^{+150}_{-190}$   & $5.80^{+0.04}_{-0.06}$ & $-2.47^{+0.15}_{-0.13}$ & RV \\
233451.7$+$534701 & TYC4000-216-1                  & $35680^{+340}_{-250}$   & $5.91^{+0.07}_{-0.06}$ & $-1.43\pm0.07$          &  \\
234421.6$-$342655 & CD-35 15910, HE 2341-3443      & $28390^{+410}_{-120}$   & $5.39^{+0.05}_{-0.03}$ & $-3.07^{+0.21}_{-0.26}$ & \\
\hline
  & Other names &  \teff                 & $\log{g}$              & $\log{(\rm He/H)}$          & Notes \\
J       &             &    (K)                 & c.g.s.                 &                         &    \\
\hline
123723.5$+$250400 & Feige 66                       & $34300^{+160}_{-180}$  & $5.82\pm0.04$          & $-1.51^{+0.05}_{-0.07}$ & \\
160011.8$-$643330 & TYC 9044-1653-1                & $34640^{+590}_{-580}$  & $6.02^{+0.08}_{-0.11}$ & $-0.30^{+0.05}_{-0.04}$ &  \\
\hline
\end{tabular}
\end{minipage}
\end{table*}

\section{Sample selection and observations}

Table~\ref{tbl_sum} lists the stars originally included in our radial velocity 
survey with notable properties described in Section 3.1. The sample includes 38
spectroscopically confirmed hot subdwarf stars, and two objects that were
respectively identified as an early B star and a A\,V+B\,V binary. The early
B star GALEX~J1407+3103 is notable for its high radial velocity, while the 
close A\,V+B\,V pair GALEX~J0716+2319 shows significant radial velocity 
variations on a short time scale. All except two objects were randomly 
selected from our catalogue of $GALEX$/Guide Star Catalogue ultraviolet-excess objects
\citep{ven2011,nem2012}. Briefly, the source catalogue includes bright objects ($N_{UV}<14$) with an ultraviolet excess
($N_{UV}-V<0.5$). The latter criterion still allows for the selection of hot subdwarf plus F/G dwarf pairs \citep[see ][]{ven2011}.
Two additional stars that were not observed by $GALEX$, including a blue-excess object \citep{jim2011},
are listed at the bottom of Table~\ref{tbl_sum} with J2000 coordinates.

The $GALEX$ name corresponds to the coordinates of the
ultraviolet source detected in the near ultraviolet (NUV) band (Section 2.3); for convenience, the 
names are abbreviated to 4 digits right ascension
and declination. The ultraviolet coordinates are
generally close to the Guide Star Catalog (GSC2.3.2) optical coordinates ($<$1~arcsec), but, in a few cases, offsets as large as
4 to 9~arcsec occurred (GALEX~J1421+7124, J1427$-$2701, J1902$-$5130, J2344$-$3426). 
Despite the offsets, the ultraviolet and optical sources must be one and the same.
These offsets cannot be attributed to a high proper-motion and are most likely
due to a distorted point spread function (PSF) in bright off-centred sources in the $GALEX$ images (Section 2.3).

Throughout this paper we will refer to the hot subdwarf as the primary and its companion as the secondary.
Table~\ref{tbl_sum} lists some notable particularities 
such as the presence of a nearby star, whether unrelated or physically associated to the hot subdwarf, a bright main-sequence companion, or photometric variability due to
reflection on a late-type companion or stellar activity (see Section 3.1). 
Most stars display H\,{\sc i}-dominated line spectra, but we also
noted the presence of He-rich subdwarfs characterized by He\,{\sc i} and 
He\,{\sc ii}-dominated line spectra. The stellar parameters of a handful of subdwarfs locate them below the
zero-age EHB (ZAEHB) and these objects are likely progenitors of ELM white dwarfs (Sections 3.1 and 3.2).

\subsection{Intermediate to high-dispersion spectroscopy for radial velocity measurements}

Our first extensive set of observations was obtained with the Wide Field Spectrograph 
\citep[WiFeS,][]{dop2007} attached to the 2.3~m telescope at the Siding Spring
Observatory (SSO). The observations were conducted on UT 2011 July 14 to 18,
UT 2011 December 2 to 3 and UT 2012 April 27 to 30. 
We used the B3000 and R7000 gratings with a slit width
of 1~arcsec that provided spectral ranges
of 3200-5900 \AA\ at a resolution of $R=\lambda/\Delta\lambda=3000$ and 5300-7000 \AA\ at $R=7000$,
respectively. The RT560 dichroic beam splitter separated the incoming light 
into its red and blue components. WiFeS is an image-slicing spectrograph with 25 slitlets ($38\times1$ arcsec)
and depending on the seeing, the target can cover a few slitlets. The signal-to-noise ratio (S/N) of
each observation was maximized by extracting the spectrum from the most significant ($\la 6$)
traces. Each trace was wavelength and flux calibrated prior to co-addition.
The spectra were wavelength calibrated using NeAr arc spectra that were 
obtained either prior to or following each observation.

Next, our second set of observations was obtained using the Ritchey-Chr\'etien Focus (R.-C.) Spectrograph attached 
to the 4~m telescope at Kitt Peak National Observatory (KPNO) on UT 4 - 6 
January 2012. We used the KPC24 grating in second order combined with the T2KA 
CCD to provide a spectral range of 6030 - 6720 \AA\ and a dispersion 
of 0.52 \AA\ pixel$^{-1}$. The slit width was set to 1.5~arcsec which provided a 
resolution of $\sim$0.9 \AA\ or $R=7000$. Contamination from third order was removed using
the GG495 filter. The spectra were wavelength calibrated using HeNeAr spectra
which were obtained following each observation.

We obtained a third set of observations using the ESO Faint Object Spectrograph
and Camera (EFOSC2) attached to the 3.6~m New Technology Telescope (NTT) at
La Silla Observatory in September 2012. We used grism number 20 centred on 
H$\alpha$ providing a spectral range from 6040 to 7140 \AA\ and a dispersion
of $0.55$ \AA\ pixel$^{-1}$. We set the slit width to 0.7~arcsec resulting
in a $2$ \AA\ resolution or $R=3500$. 
Next, we obtained additional spectra 
with EFOSC2 on the NTT on UT 31 July and 1 August 2014. We used grism number 
19 that provided a spectral range from 4435 to 5120 \AA\ and, after binning $2\times2$, a dispersion of 
0.67 \AA\ per binned pixel. The 
slit width was set to 1~arcsec resulting in a resolution of $\sim$2 \AA\ or $R\approx2000$.
Additional EFOSC2 spectra of GALEX~J1731+0647 were extracted from the ESO archive
(programme 090.D-0012, PI S. Geier). The data were also obtained with grism 19, but
binned $2\times1$ resulting in a dispersion of 0.34 \AA\ pixel$^{-1}$. The slit width
was set to 1~arcsec resulting in a resolution of $\sim$2 \AA.
All spectra were wavelength calibrated using
HeAr arc spectra which were obtained following each observation.

Also, we obtained a fourth set of spectra using the grating spectrograph attached
to the 1.9~m telescope at the South African Astronomical Observatory (SAAO) on
UT 2014 February 11. We used the 1200 lines/mm grating with a blaze
wavelength of 6800 \AA. This arrangement provided a range of 6023 to 6782 \AA\
with a dispersion of 0.439 \AA\ per pixel. The slit width was set to 1.05~arcsec
resulting in $R=7000$, or a resolution of $\approx$1\AA\ at H$\alpha$. 
A CuNe comparison arc was obtained following each target
observation.

\begin{figure*}
\includegraphics[viewport=40 45 530 545, clip,width=0.9\textwidth]{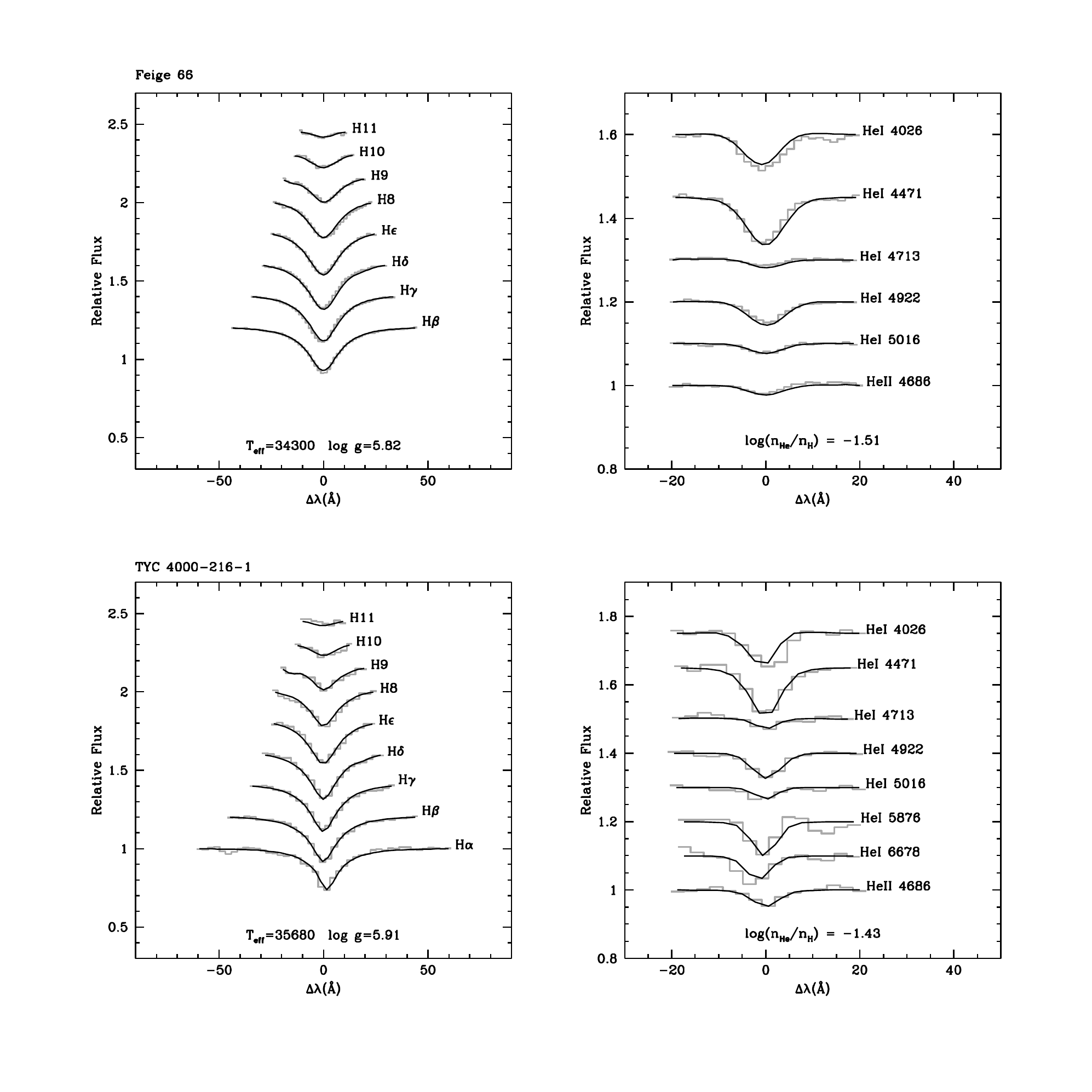}
\caption{Line profile analysis of the WHT spectrum of Feige~66 (top) and KPNO spectrum of TYC4000-216-1 (bottom), labelled
with best-fitting parameters.
\label{fig_Feige66}}
\end{figure*}

We assembled a fifth data set with observations of the bright objects 
HD~4539 (GALEX~J0047$+$0958), the spectro-photometric standard Feige~66, GALEX~J1421$+$7124, GALEX~J1736$+$2806,
and GALEX~J2334$+$5347 using the 2m telescope at Ond\v{r}ejov
Observatory. The observing configuration and procedure are described in
\citet{kaw2010a}. Briefly, for each star 
we obtained a series of spectra centred on H$\alpha$.
We used the 830.77 lines mm$^{-1}$ grating with a SITe $2030\times 800$ CCD, 
this resulted in a spectral resolution of $R = 13\,000$. Each target exposure
was immediately followed by a ThAr comparison arc. 

Finally, and introducing our sixth and most recent observation programme, 
we obtained three high-dispersion echelle spectra
of the short period binary GALEX~J2254$-$5515. From UT 24 November to 4 December 2014 we used the 
Fiber-fed Extended Range Optical Spectrograph (FEROS) attached to the
2.2~m telescope at La Silla. The spectra range from $\approx$3600 to $\approx$9200 \AA\ at a
resolution of $R\approx 48,000$.

We supplemented our data sets with archival spectra.
We extracted processed FEROS data from the ESO archive. 
The spectra were obtained under the programmes 076.D-0355, 077.D-0515, 078.D-0098 (PI: L. Morales-Rueda)
and 086.D-0714 (PI: S. Geier).

We also extracted spectra from the Isaac Newton Group (ING) Archive. The first set of
data (GALEX~J1632+0759 and GALEX~J1731+0647) was obtained with the Intermediate Dispersion Spectrograph (IDS) 
attached to the Isaac Newton Telescope (INT) on UT 17 May 2013 (run numbers 
984456 and 984458) and on UT 19 May 2013 (run numbers 984760, 984762 and 984763). 
The spectra were obtained with the R1200B grating which
resulted in a useful range of 3900 to 5200 \AA\ and a
dispersion of 0.48 \AA\ and delivering a resolution of 1.5 \AA\ assuming a 3-pixel full-width at half-maximum (FWHM).
The spectra were wavelength calibrated using CuAr and CuNe arcs
and adjacent exposures were co-added to obtain the final radial velocity.
A second set of data (GALEX~J1632+0759) was obtained with the William Herschel Telescope (WHT) and 
the Intermediate dispersion Spectrograph and Imaging System (ISIS) on
UT 26 August 2010 (run numbers 1483813 and 1483814). The spectra were obtained
with the R600B and R600R gratings and calibrated with CuAr and CuNe arcs resulting in 
useful ranges of 3500-5100 \AA\ and 5500-7030 \AA\ and dispersion of 0.88 \AA\ per binned pixel in the blue ($2\times2$)
and 0.49 \AA\ pixel$^{-1}$ in the red (binned $2\times1$), corresponding to
spectral resolutions of 1.7 \AA\ in the blue and 1.5 \AA\ in the red assuming
a 3-pixel FWHM.

On average, a high signal-to-noise ratio was achieved with EFOSC2 on the NTT ($\overline{\rm S/N}\approx 100$), 
the R.-C. Spectrograph on the KPNO 4~m telescope ($\overline{\rm S/N}\approx 60$),
and WiFeS on the SSO 2.3~m telescope ($\overline{\rm S/N}\approx 80$). A lower signal-to-noise ratio
was achieved with the coud\'e spectrograph on the Ond\v{r}ejov 2~m telescope, FEROS on the 
MPG~2.2~m telescope (La~Silla), and the grating spectrograph on the SAAO 1.9~m telescope ($\overline{\rm S/N}\approx 30$).
The lower S/N achieved at Ond\v{r}ejov and La~Silla is largely compensated by the higher dispersion resulting
in comparable or superior velocity accuracy (see next Section). More than 70 per cent of our spectra had a
S/N$\gtrsim 40$ and spectra with ill-defined hydrogen or helium lines (S/N$\lesssim 15$) were rejected.

\subsubsection{Tests of the wavelength and velocity scales}

We performed a series of tests of the wavelength scale of relevant spectra using the O\,{\sc i} 
sky emission lines and atmospheric molecular absorption bands. 

Diffuse O\,{\sc i}$\lambda$6300.304 emission helps set the accuracy of the wavelength 
scale, particularly in low- to intermediate-dispersion spectra. 
A strong emission line is detected in 93 per cent of all usable EFOSC2 spectra,
95 per cent of all KPNO and SSO spectra, and nearly all SAAO spectra. A short exposure time as well
as the appearance of scattered moonlight usually limit the usefulness of this template.
The O\,{\sc i} velocity averaged $\varv$(O\,{\sc i})$=0.0$\,\kmps\ at KPNO with a dispersion
$\sigma_\varv$(O\,{\sc i})$=2.1$\,\kmps, $\varv$(O\,{\sc i})$=1.9$\,\kmps\ with EFOSC2 and
a dispersion $\sigma_\varv$(O\,{\sc i})$=5.4$\,\kmps,
$\varv$(O\,{\sc i})$=3.7$\,\kmps\ at SSO and a dispersion $\sigma_\varv$(O\,{\sc i})$=7.4$\,\kmps,
and $\varv$(O\,{\sc i})$=4.6$\,\kmps\ at SAAO and a dispersion $\sigma_\varv$(O\,{\sc i})$=4.4$\,\kmps.
The emission line appeared blended in most spectra obtained during bright time at SSO.
Based on this analysis the expected accuracy should be of the order of 2-7\,\kmps.
The accuracy of the wavelength scale using coud\'e or echelle spectrographs is normally
of the order of 1\,\kmps\ or better.

Systematic velocity shifts are expected following an improper placement of the star on the
slit, particularly if the stellar image is much narrower than the slit width.
Excellent seeing conditions are often encountered at La Silla and KPNO.
We cross-correlated telluric absorption features in the KPNO and EFOSC2 spectra
with a telluric template of identical
spectral resolution. We measured an average velocity of $-0.8$\,\kmps\ with
a dispersion of 13.4\,\kmps\ in the EFOSC2 spectra and an average
velocity of 2.4\,\kmps\ with a dispersion of 10.0\,\kmps\ in the KPNO
spectra. Velocity deviations of up to 50\,\kmps\ were found in a few well-exposed EFOSC2 spectra:
We corrected the measured stellar velocities at La Silla using the telluric template velocities.

In summary, after applying telluric corrections, we estimate that errors in stellar velocity
measurements due to various systematic effects are better than $\sim$10\,\kmps\ provided that
the photospheric lines are well defined.  Ultimately, the accuracy of the wavelength is verifiable
using actual stellar data and by plotting the velocity dispersion distribution (Section 3.1.3).

\subsection{Low-dispersion spectroscopy for stellar parameter determinations}

For stars not listed in \citet{nem2012}, we obtained additional low dispersion
spectra with the R.-C. spectrograph attached to the 4m telescope at KPNO on 
UT 2013 July 12 (GALEX J1421$+$7124) and 2014 May 24 (GALEX J2334$+$5347). 
We used the KPC-10A grating and T2KA CCD with a dispersion of 2.77 \AA\ pixel$^{-1}$
in first order and centred on 5875 \AA. We used the order sorting filter
WG360 and set the slit width at 1.5~arcsec resulting in a spectral resolution of 
$\approx 5.5$ \AA. The spectra were wavelength calibrated using the HeNeAr arc.

We extracted a set of spectra of the spectro-photometric standard Feige~66 from the ING archive. These spectra were
obtained with ISIS
attached to the WHT (run numbers 133198, 133200, 133201, 
133223, 133226, 133227). The spectra were obtained using the R300B grating in 
the blue arm providing a dispersion of 1.54 \AA\ pixel$^{-1}$ and a spectral 
range from 3620 to 5190 \AA. The slit width was set to 2.4~arcsec for each
observation which corresponds to a resolution of $\approx 7.5$ \AA. The spectra
were wavelength calibrated using a CuAr arc.

Details of the low-dispersion spectroscopy obtained of other objects in the present sample are given by
\citet{ven2011} and \citet{nem2012}.

\subsection{Photometry and imaging}

We compiled available optical and infrared photometric measurements and combined them with 
the $GALEX$ NUV and FUV photometry from the all-sky imaging survey (AIS) to build a SED for each object in the sample.
The ultraviolet data were collected from the site {\tt galex.stsci.edu/GalexView/}. \citet{mor2007} present details of the
instrument calibration. Table~\ref{tbl_phot} in the Appendix lists the $GALEX$ magnitudes, along
with the available $V$ magnitudes as well as 2MASS \citep{skr2006} and Wide-field Infrared Survey Explorer \citep[$WISE$,][]{wri2010} 
infrared measurements.

The PSF of $WISE$ images ranges from 6 to 12~arcsec in the 3 to 24$\mu$m wavelength range, while the PSF in 2MASS images is close to
2.5~arcsec. Because of its relatively broad PSF, stars located within its range and identified in higher-resolution imaging
are certainly contaminating the SED in the mid-IR range.
 
Also, we extracted photometric time series from the SuperWASP \citep[SWASP,][]{pol2006} public archive, 
NSVS \citep{woz2004}, 
All Sky Automated Survey 
\citep[ASAS;][]{poj1997} and Catalina surveys \citep{dra2009}. 
The Catalina photometry is unfiltered. Bright targets ($<12$~mag) are often saturated, but the photometric measurements are
more precise with faint targets ($>14$~mag) than those obtained in the other three surveys consulted.
The SWASP images are filtered (4000-7000 \AA).
Light curve analysis of SWASP data is valuable because of the large number of measurements obtained for individual
targets. We obtained ASAS times series in the V band and the NSVS images are unfiltered.

The calibrated $GALEX$ magnitudes are obtained from the count rates extracted using elliptical apertures (fuv\_flux\_auto, nuv\_flux\_auto)
fitted to the actual stellar profiles and
converted into the AB system.
The average $GALEX$ PSF is matched approximately by Gaussian functions with
FWHM of 5.3 and 4.2~arcsec in NUV and FUV images, respectively, and a positional accuracy of $\approx0.5$~arcsec.
However, several factors affect the reliability of the $GALEX$ photometric magnitudes.
The $GALEX$ imaging quality varies with the detector position with a strong dependency
on the radial distance from the image centre. We recorded the target distance to the
centre of the field of view (fov\_radius), as well as the actual FWHM values in the FUV and NUV
images (fuv\_fwhm\_world, nuv\_fwhm\_world) for each target. Measurements with a radial distance outside of 
$0^\circ.4$ combined with a large PSF ($>0^\circ.01$) or measurements with an exceedingly large PSF ($>0^\circ.04$) are marked in Table~\ref{tbl_phot} as possibly unreliable. Finally, bright objects with unreliable non-linearity corrections outside the range
of validity are marked.
Non-linearity effects dominate the photometric error: A 10 per cent loss is observed at $N_{UV}=13.9$ and $F_{UV}=13.7$
so that most measurements in the present selection are affected. \citet{mor2007} and \cite{cam2014} propose correction
algorithms that are nearly identical. \citet{cam2014} presented a calibration sample sufficiently large
to allow us to evaluate the scatter in the synthetic versus measured magnitude relations. For example, this scatter
is of the order of 0.35 and 0.4 mag at $N_{UV}$ and $F_{UV}=13$, respectively. We adopted $GALEX$ magnitudes
adjusted using the correction algorithm of \citet{cam2014} with errors estimated using the scatter in these corrections for a
given magnitude.

The EFOSC2 acquisition images provide a deep, high-spatial resolution view of the
fields surrounding target stars. These images were obtained with the Loral/Lesser 2048$\times$2048
CCD. With a focal plane scale of 8.6~arcsec~mm$^{-1}$ and a pixel size of 15 $\mu$m,
the sky images are sampled with a pixel size of $0.129\times0.129$ arcsec$^2$, or, after binning $2\times2$,
a pixel size of $0.258\times0.258$ arcsec$^2$. The images allow for the identification of physical companions or
unrelated, nearby stars.

Fig.~\ref{fig_sed1}, Fig.~\ref{fig_sed2} and Fig.~\ref{fig_sed3} in the 
Appendix compare all available photometry to synthetic spectra computed using 
stellar parameters listed in Table~\ref{tbl_sum}.

\section{Analysis}

We present, in order, the properties of the sample including an overview of
the stellar parameters (\teff, $\log{g}$, $\log{\rm He/H}$) and evolutionary
history, the characteristics
of the SEDs and photometric time series, and the radial velocity data set.
We identify new binary candidates and present an analysis of individual binary properties 
from the combined data sets. 

\subsection{Sample properties}

\begin{figure}
\includegraphics[width=1.0\columnwidth]{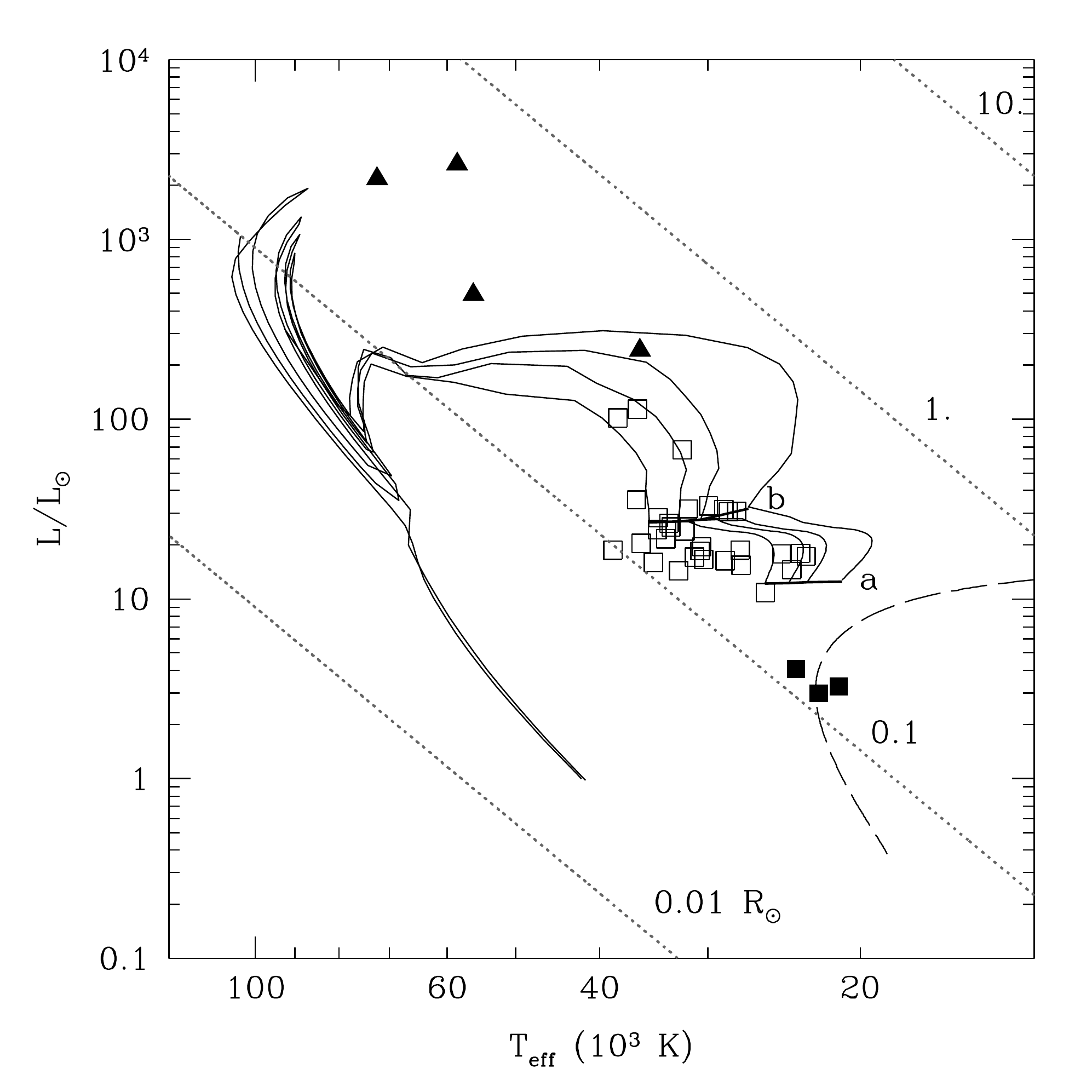}
\caption{Physical properties, luminosity versus effective temperature, of the sample: sdO stars are shown
with full triangles while sdB stars are shown with open squares (assuming 0.47\,\msun) or
full squares (assuming 0.234\,\msun). 
The zero-age EHB is labelled ``a'' while the terminal-age EHB
is labelled ``b''. Evolutionary tracks computed by \citet{dor1993} with a helium core mass of
0.469\,\msun\ and hydrogen envelopes of, left to right, 0.002, 0.004, 0.006, and 0.01\,\msun\ are shown
with full lines. The cooling track from \citet{dri1998} for progenitors of 
ELM white dwarfs of 0.234\,\msun\ is
shown prior to hydrogen shell flashes with a dashed line. Lines of constant radii at 
0.01, 0.1, 1, and 10\,\rsun\ are labelled accordingly.
\label{fig_sample}}
\end{figure}

Table~\ref{tbl_sum} lists the atmospheric parameters obtained from \citet{nem2012}. 
The Balmer line analysis for three additional objects (Feige 66, GALEX~J1421$+$7124 
and GALEX~J2334$+$5347) is based on the model grids of \citet{ven2011}.
Best fitting parameters (\teff, $\log{g}$, $\log{\rm He/H}$) are obtained using $\chi^2$ minimization techniques with the
observed line profiles (He\,{\sc i,ii} and H\,{\sc i}) being simultaneously adjusted to
interpolated spectra from the model grid.
Examples of Balmer and helium line analyses are shown in Fig.~\ref{fig_Feige66}.

Fig.~\ref{fig_sample} shows properties of the sample presently investigated. Using 
the effective temperature ($T_{\rm eff}$) and surface gravity ($g$) we
determined the total luminosity (in \lsun) by adopting for most objects a 
sample-average mass of 0.47\,\msun\ \citep{fon2012}.  
\begin{displaymath}
L = 4\pi R^2 \sigma T_{\rm eff}^4,
\end{displaymath}
where $\sigma$ is the Stefan-Boltzmann constant and the radius ($R$) is calculated using
\begin{displaymath}
R = \sqrt{\frac{GM}{g}},
\end{displaymath}
where $M$ is the subdwarf mass and $G$ is the gravitational constant.

The sdB stars form a sequence of approximately constant luminosity, $L=$10-30\,\lsun or $M_V=4.3$ ($\sigma=0.9$) mag, and located between
the ZAEHB and the TAEHB while a few ageing sdB stars and all He-rich sdO stars set out on a higher luminosity
excursion beyond the stable He-burning stage. The objects
lying below the ZAEHB ($L<10$\,\lsun) with a low-temperature and a high-gravity, GALEX~J0805$-$1058 and, tentatively,
J1411+7037 and J2153$-$7004,
were singled-out and were attributed a mass of 0.23\,\msun\ based on their likely evolutionary status \citep{dri1998}.
Most objects lie to the left of the EHB tracks suggesting that their hydrogen layer is thinner than 
0.002\,\msun, or, possibly, that their surface gravity is overestimated. 
To investigate the latter possibility, we compared the results of a model
atmosphere analysis using the hydrogen Stark broadening tables of
\citet{lem1997}, employed in the present work, to those of \citet{tre2009}, which include improved treatment of merging atomic energy levels.
We found that 
improvements in Stark broadening theory may account for a shift of
$\Delta\log{g}=+0.08$ dex near 30\,000~K \citep[see, e.g., ][]{ost2014,
tel2014b} in agreement with a shift of $\Delta\log{g}=+0.06$ dex
found at 40\,000~K by \citet{kle2011}.
These systematic shifts are notable, but
still cannot explain the model offsets apparent in Fig.~\ref{fig_sample}.
Metallicity has little effect on temperature and gravity below
\teff$=35\,000$~K as demonstrated by \citet{lat2014}.
It is worth noting that, while a sdB mass of 0.47\,\msun\ may be typical, it
does not necessarily apply to all objects (e.g., ELM progenitors).

On the other hand, \citet{sch2014} pointed out that current evolutionary models fail to
reproduce some observed properties of EHB stars, such as the core mass derived
from asteroseismology, and concluded that evolutionary models must be updated to
match observed seismic and spectroscopic stellar parameters.
\citet{sch2014} found that very high convective overshooting would be needed to reproduce the seismic
core mass but that it
would, quite improbably, double the EHB lifetime.
Therefore, they conclude that the general treatment of convection in evolutionary
models needs updating, and that new opacity tables and diffusion
calculations are required.

\subsubsection{Overview of the SEDs}

\begin{figure}
\includegraphics[width=1.0\columnwidth]{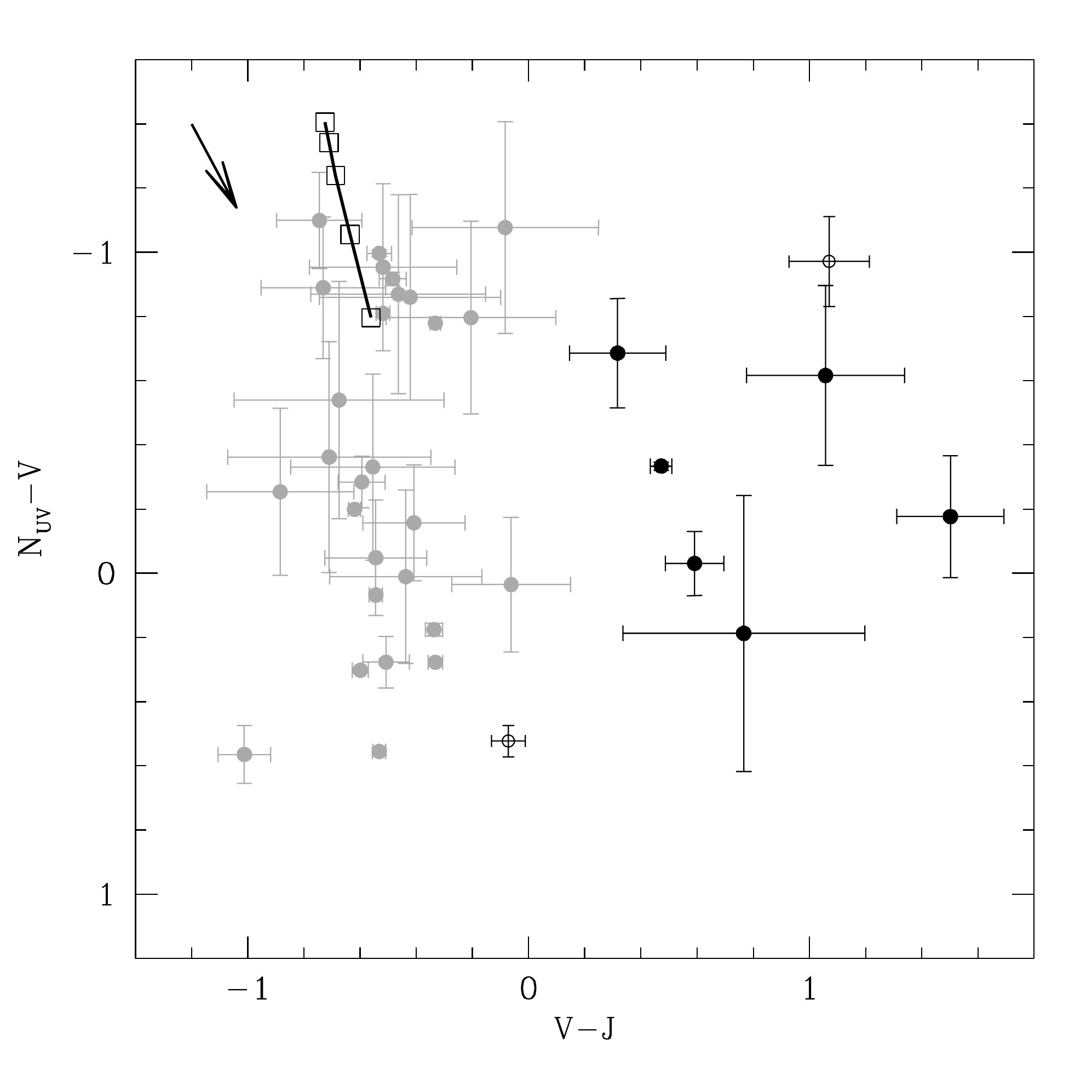}
\caption{$NUV-V$ versus $V-J$ colour-colour diagram. Stars with a composite IR 
excess are shown with full black circles and stars with an IR excess due to a 
nearby star are shown with open circles (see Section 3.1.1), 
while all others are shown in grey. Models at 24, 28, 32, 36, and
$40\times10^3$ K ($\log{g}=5.7$, $\log{\rm He/H}=-2.5$) are shown, from bottom to top, with open squares linked 
by a full line. The effect of interstellar extinction ($E_{B-V}=0.05$) on the colours is shown with an arrow in the
upper left corner.
\label{fig_nmv_vmj}}
\end{figure}

\begin{figure}
\includegraphics[width=1.0\columnwidth]{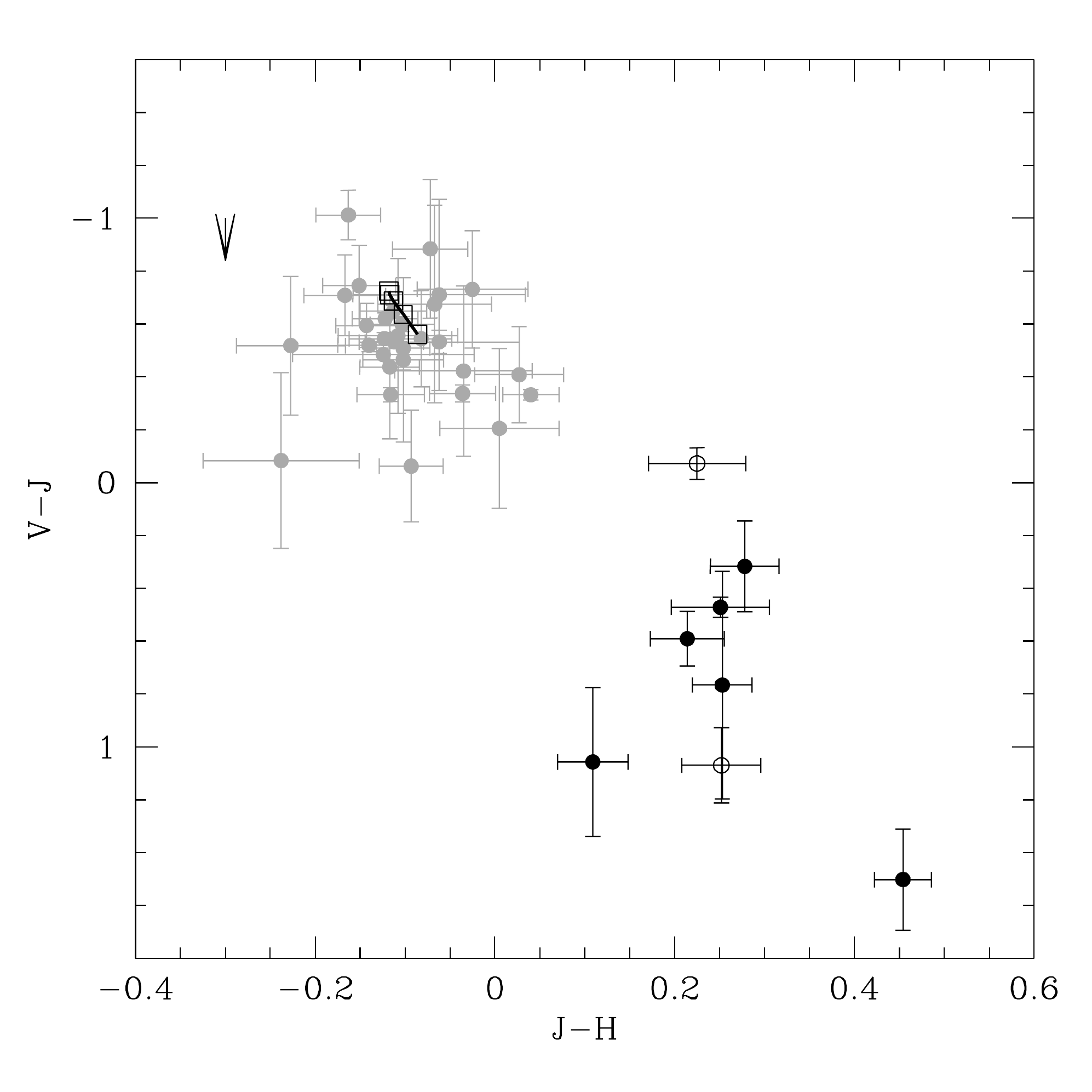}
\caption{Same as Fig.~\ref{fig_nmv_vmj} but showing $V-J$ versus $J-H$. 
\label{fig_vmj_jmh}}
\end{figure}

\begin{figure}
\includegraphics[width=1.0\columnwidth]{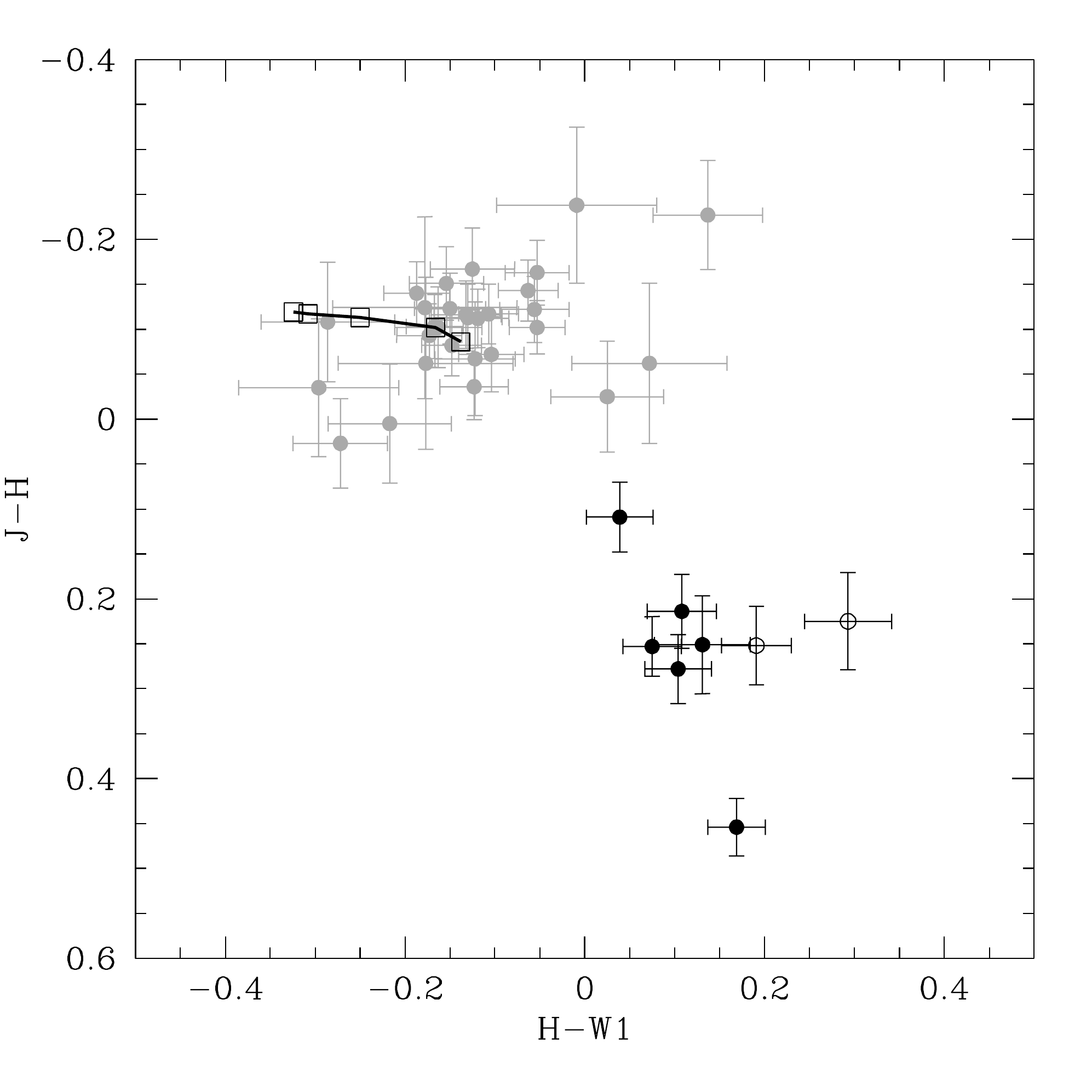}
\caption{Same as Fig.~\ref{fig_nmv_vmj} but showing $J-H$ versus $H-W1$. 
\label{fig_jmh_hmw1}}
\end{figure}

Fig.~\ref{fig_nmv_vmj} shows the $NUV-V$ versus $V-J$ colour diagram
for the sample of hot subdwarfs listed in Table~\ref{tbl_phot}. Fig.~\ref{fig_vmj_jmh} 
and Fig.~\ref{fig_jmh_hmw1} present the $V-J$ versus $J-H$ and
$J-H$ versus $H-W1$ diagrams, respectively.
The effect of interstellar extinction is evident in the $NUV-V$ colour of
some objects, but many are also affected by large systematic errors in the
$GALEX$ $NUV$ photometry (non-linearity).
For example, a larger extinction than observed in the interstellar line-of-sight is apparent toward GALEX~J1632+0759.
The International Ultraviolet Explorer ($IUE$) spectra that supplement uncertain $GALEX$ photometric measurements
indicate $E_{B-V}=0.4$, largely in excess of that found in the extinction map of \citet{sch1998}, $E_{B-V}=0.08$.
The effect on ultraviolet colours of an extinction coefficient $E_{B-V}=0.05$ (see Fig.~\ref{fig_nmv_vmj}) is relatively
modest, but coefficients in excess of 0.4 would displace colours from the upper left to the reddened, lower right corner 
in the vicinity of composite stars. 

Individual SEDs may be contaminated by the presence of a nearby star, either physically associated or unrelated to the hot subdwarf.
Inspection of the P82, P83, P85 and P89 EFOSC2 acquisition images obtained by \citet{ven2011} and \citet{nem2012} revealed the presence of nearby companions ($<3$~arcsec) to 
J0047$+$0337, J0657$-$7324, and J1632$+$0759 (see below).  Visual inspections of the guiding
images displayed at KPNO did not reveal the presence 
of a nearby companion to any other objects. An inspection of photographic plate material ({\tt http://surveys.roe.ac.uk/ssa/index.html}) helps locate other, more distant
objects ($>3$~arcsec) that may contaminate photometric measurements with large PSF (e.g., $WISE$). For example,
GALEX~J0751+0925 is accompanied by a faint ($\Delta m\sim$5 mag), nearby star $\sim$6~arcsec away at
a position angle of 200$^\circ$ (Epoch 1993). 

The composite nature of the spectra of GALEX~J0047+0337, J1411+7037, J1736+2806, J1753$-$5007, J2038$-$2657 \citep{nem2012} 
are confirmed by their IR photometric colours.
The flux contributions in the V band for these hot subdwarfs are offset by $\sim$0.7 (possibly contaminated), $\sim$0.0 (weak secondary detection in the optical), $\sim$1.2, $\sim$1.3, and $\sim$1.3 mag, respectively,
relative to their observed composite $V$ magnitudes.
Although evident in the IR colours of GALEX~J1356$-$4934 (Fig.~\ref{fig_sed2}), the presence of a companion was not detected by \citet{nem2012}. The flux contribution from the hot subdwarf in the V band is offset by
$\sim$0.4 mag relative to the observed composite $V$ magnitude. We re-examined the
spectra of GALEX~J1356$-$4934 and we found weak signatures of a cool main-sequence companion in the blue spectrum used by \citet{nem2012}, and stronger
spectral lines representative of a cool main-sequence star in the red spectra used in this paper. We present our spectral decomposition of this system
in Section 3.2.2.
The IR colours for GALEX~J0047$+$0337, J0657$-$7324, and J1632$+$0759 are almost certainly contaminated by their nearby, resolved
companions.

\subsubsection{Overview of the photometric time series}

Table~\ref{tbl_bin_phot} summarizes the photometric time series analyses.
We included objects showing significant radial velocity variations (e.g., GALEX~J2205$-$3141), 
objects with composite optical spectra (e.g., GALEX~J1736+2806), and, finally,
three objects with previously published analyses from the present survey: GALEX~J0321+4727 and GALEX~J2349+3844 \citep{kaw2010a}, and
GALEX~J1411$-$3053 \citep{ven2012}. Photometric variations observed in the sdB plus white dwarf system GALEX~J1411$-$3053 are an example of ellipsoidal variations in this class of objects.
Fourier transform calculations of available light curves (Section 2.3) uncovered
three objects with significant periodic variations. The light curves were analysed using
fast Fourier transform analysis from \citet{pre1992}. Both GALEX~J1736+2806 and  GALEX~J2038$-$2657 are binaries
comprising a hot subdwarf with a more luminous optical companion, while GALEX~J2205$-$3141 is composed of a hot
subdwarf and late-type companion. The photometric variations in the latter are clearly timed with the orbital 
period (see Section 3.2) and caused by reflection of the primary on the cool secondary. Variations in GALEX~J1736+2806
and GALEX~J2038$-$2657 may be caused by a spot on the surface of the secondary coupled to the rotation period.
The variable, double-peaked H$\alpha$ line profile of GALEX~J2038$-$2657 also implies the presence of surface inhomogeneities (see Section 3.3.1).

\begin{table*}
\centering
\begin{minipage}{\textwidth}
\caption{Photometric time series. \label{tbl_bin_phot}}
\renewcommand{\footnoterule}{\vspace*{-15pt}}
\renewcommand{\thefootnote}{\alph{footnote}}
\begin{tabular}{lccccccc}
\hline
Name         & Survey & HJD range & Number & Period range & Semi-amplitude & Average magnitude & Standard deviation \\
             &        & (2450000+)&        & (d)          & (mmag)    & (mag)        & (mmag)        \\
\hline
J0047+0337   & ASAS   & 1868-5168 & 378    & $>0.02$      & $4.4\pm7.0$ & 12.336     & 94.5          \\
             & NSVS   & 1382-1549 & 177    & $>0.01$      &$17.8\pm3.4$ & 12.676     & 32.2          \\
J0321+4727   & NSVS   & 1373-1630 & 173    & 0.26586 \footnotemark[1]\footnotetext[1]{Spectroscopic period.} &$61.3\pm3.9$ & 12.034 & 56.9 \\
             & SWASP  & 3196-4458 & 4575   & 0.26586 \footnotemark[1]  &$43.5\pm1.0$ & 11.490     & 49.4          \\
J0401$-$3223 & SWASP  & 3964-4485 & 14208  & $>0.01$      &$8.4\pm0.1$  & 11.268     & 12.4          \\
             &        &           &        & 1.85735 \footnotemark[2]\footnotetext[2]{Possible spectroscopic period.}  & $0.8\pm0.1$  & &  \\
J0507+0348 & Catalina & 3643-6592 & 347    & $>0.02$      & $14.4\pm1.8$ & 14.172  & 23.8  \\
             &        &           &        & 0.52813 $^a$  & $2.1\pm1.8$  &         &       \\
J0613+3420   & SWASP  & 3232-4573 & 4700   & $>0.03$      &$12.7\pm2.2$  & 13.958  & 106.4 \\
J0751+0925   & ASAS   & 2623-5131 & 198    & $>0.1$       &$80.3\pm20.3$ & 14.126  & 205.9 \\
             &        &           &        & 0.17832 $^a$  &$43.5\pm20.8$ &         &       \\
             & Catalina & 3466-6368 & 119  & $>0.1$       &$12.0\pm3.2$  & 14.168  & 18.7 \\
             &        &           &        & 0.17832 $^a$  &$3.6\pm3.0$   &         &      \\
J0805$-$1058 & ASAS   & 1868-5168 & 570    & $>0.1$       &$19.9\pm3.8$  & 12.270  & 65.6  \\
             &         &          &        & 0.17370 $^a$  &$11.7\pm4.8$  &         &       \\
             & NSVS   & 1488-1630 & 132    & $>0.01$      &$33.0\pm7.4$  & 12.812  & 59.8  \\
             &        &           &        & 0.17370 $^a$  &$10.9\pm7.2$  &         &        \\
J1356$-$4934 & ASAS   & 1900-5088 & 729    & $>0.04$      &$3.3\pm3.9$   & 12.269  & 74.2 \\
J1411$-$3053 & ASAS   & 1902-5088 & 1060   & 0.02449 \footnotemark[3]\footnotetext[3]{Ellipsoidal variations at half-spectroscopic period.} &$46.8\pm3.8$  & 12.342  &  88.2 \\
             & SWASP  & 3860-4614 & 13079  & 0.02449 \footnotemark[3] &$51.2\pm2.0$  & 12.723  & 165.6 \\
J1632+0759   & ASAS   & 2175-5106 & 399    & $>0.1$       &$23.9\pm5.1$  & 12.763  & 74.5 \\
             &        &           &        & 2.9515 $^a$   &$3.9\pm5.2$   &         &      \\
             & Catalina & 3466-6471 & 338  & $>0.02$      &$49.8\pm8.4$  & 12.587  & 121.6 \\
             &        &           &        & 2.9515 $^a$   &$7.5\pm9.5$   &         &       \\
             & NSVS   & 1275-1417 & 115    & $>0.01$      &$37.6\pm8.6$  & 13.248  & 69.3 \\
             &        &           &        & 2.9515 $^a$   &$17.9\pm8.9$  &         &      \\
J1731+0647   & ASAS   & 2727-5009 & 73     & $>0.1$       &$98.7\pm49.6$ & 13.799  & 299.1 \\
             &        &           &        & 1.17334 $^a$  &$44.5\pm51.5$ &         &       \\
             & Catalina & 3466-6457 & 105  & $>0.1$       &$74.9\pm4.7$  & 13.825  & 67.0  \\
             &        &           &        & 1.17334 $^a$  &$21.4\pm9.2$  &         &       \\
J1736+2806   & SWASP  & 3128-4325 & 9140   & 1.33320 \footnotemark[4]\footnotetext[4]{Photometric period.} & $10.2\pm0.2$ & 11.639  & 17.0 \\
J1753$-$5007 & ASAS   & 1947-5137 & 544    & $>0.1$       &$18.6\pm5.7$  & 12.955  & 94.2  \\
J1903$-$3528 & SWASP  & 3860-4551 & 7388   & $>0.01$      &$3.6\pm0.6$   & 13.089  & 35.2  \\
J2038$-$2657 & NSVS   & 1348-1483 & 42     & 1.860 $^d$    &$56.4\pm8.4$  & 11.950  & 51.1  \\
             & SWASP  & 3958-4614 & 8332   & 1.87022 $^d$  &$12.9\pm0.4$  & 11.856  & 28.2  \\
J2205$-$3141 & ASAS   & 1873-5166 &  521   & 0.34156 $^d$  &$40.0\pm5.0$  & 12.381  & 81.4 \\
             & Catalina & 3598-6217 & 252  & 0.34156 $^d$  &$46.2\pm6.2$  & 12.07   & 64.4 \\
             & SWASP  & 3862-4614 & 22731  & 0.34156 $^d$  &$26.7\pm1.0$  & 12.409  & 110.1 \\
J2254$-$5515 & ASAS   & 1869-5168 & 672    & $>0.02$      &$20.0\pm3.8$  & 12.113  & 70.9 \\
             &        &           &        & 1.22702 $^a$  &$4.7\pm3.9$   &         &      \\
             & Catalina & 3580-6076 & 78   & $>0.1$       &$98.4\pm16.8$ & 12.442  & 103.1 \\
             &        &           &        & 1.22702 $^a$  &$28.1\pm15.3$ &         &      \\
J2349+3844   & NSVS   & 1321-1579 & 261    & $>0.1$       &$19.9\pm3.9$  & 12.287  & 44.7 \\
             &        &           &        & 0.46252 $^a$  &$5.0\pm3.9$   &         &       \\
             & SWASP  & 3154-4669 & 12175  & $>0.01$      &$1.7\pm0.2$   & 11.640  & 15.6 \\
             &        &           &        & 0.46252 $^a$  &$1.1\pm0.2$   &         &      \\
\hline
\end{tabular}
\end{minipage}
\end{table*}

An analysis of time series helps constrain the nature of the companion \citep{max2002}. For example, a simple geometric
model suggests that the presence of a late-type companion generally leads to detectable photometric variations phased
on the orbital period. This reflection effect scales as $R_2^2$, where $R_2$ is the secondary radius, 
but only as $a^{-1/2}$, where $a$ is the orbital separation \citep{max2002}. Interestingly, an application 
of these simple relations confirms the results of detailed light curve modelling and most importantly that, 
for a given mass function, the effect of a lower binary inclination, which reduces the light contrast between inferior and
superior conjunctions as well as its intensity through increased binary separation, is compensated by the increased mass and radius of the secondary calculated using the 
mass function, hence increasing the fraction of intercepted light. Following \citet{max2002} and adding
a slight modification to account for the effect of inclination on the visibility of the exposed hemisphere at inferior
and superior conjunctions, the amplitude of the variations is given by:
\begin{displaymath}
\delta\, m =  2.5 \log\Big{(} \frac{f^+}{f^-}\Big{)},
\end{displaymath}
where
\begin{displaymath}
f^\pm = 1 + \Big{(}\frac{R_2}{R_1}\Big{)}^2 \Big{(}\frac{R_1}{\sqrt{2}\,a}\Big{)}^{1/2} \frac{1\pm\sin{i}}{2},
\end{displaymath}
where $f^+$ is the relative flux at superior conjunction, $f^-$ is at inferior conjunction, 
$i$ is the binary inclination and $R_1$ is the primary radius estimated from the measured surface gravity and
assumed mass (0.23 or 0.47\,\msun). The mass of the putative late-type secondary was estimated using the binary mass
function, the assumed primary mass and by varying the inclination angle: The radius is then estimated following the mass-radius
relation for late-type stars of \citet{cai1990}.
Applications of this approximate formula for $\delta\,m$ lead to an overestimation of the amplitude of a factor of $\approx 3$ when
applied to the well known case of GALEX~J0321+4727 (Table~\ref{tbl_bin_phot}): 
The semi-amplitude of the phased light curve of GALEX~J0321+4727 reaches 44 and 61~mmag in the SWASP and NSVS data sets, respectively,
and reveals the presence of an irradiated late-type companion. 
This simple model also shows that the amplitude of the variations is more
or less constant ($\pm30$ per cent) when varying the inclination as shown in the detailed models of \citet{max2002}.
Applying a factor of 0.3 to the amplitude calculated with the simple formula for $\delta\,m$ presented above should allow us to
confirm or rule out the presence of a late-type companion in the new binaries. For example, the companion to
GALEX~J2349+3844 \citep{kaw2010a} is almost certainly a white dwarf: The predicted semi-amplitude of variations due to a late-type companion
is $\approx 70$~mmag, while the observed variations are less than 1 and 5~mmag in the SWASP and NSVS phased light curves, respectively
(Table~\ref{tbl_bin_phot}). Based on this insight, the photometric times series will help constrain in the following Sections the nature of the companion in
the new binaries.

\subsubsection{Overview of the radial velocity data set}

We measured the radial velocities by fitting a Gaussian function to the deep and narrow
H$\alpha$ core for most red spectra, or He{\sc i}$\lambda$ 6678.15 in a few instances described below.
In the blue we used the H$\beta$ core,
He{\sc ii}$\lambda$4685.698, or He{\sc i}$\lambda$4471.48 if necessary (see below). All measured velocities are heliocentric corrected
and tabulated in Table~\ref{tbl_rad_vel} in Appendix~B. For each target Table~\ref{tbl_rad_vel} also includes the number of spectra, the average
velocity ($\bar{\varv}$) and velocity dispersion ($\sigma_\varv$).

\begin{figure}
\includegraphics[width=1.00\columnwidth]{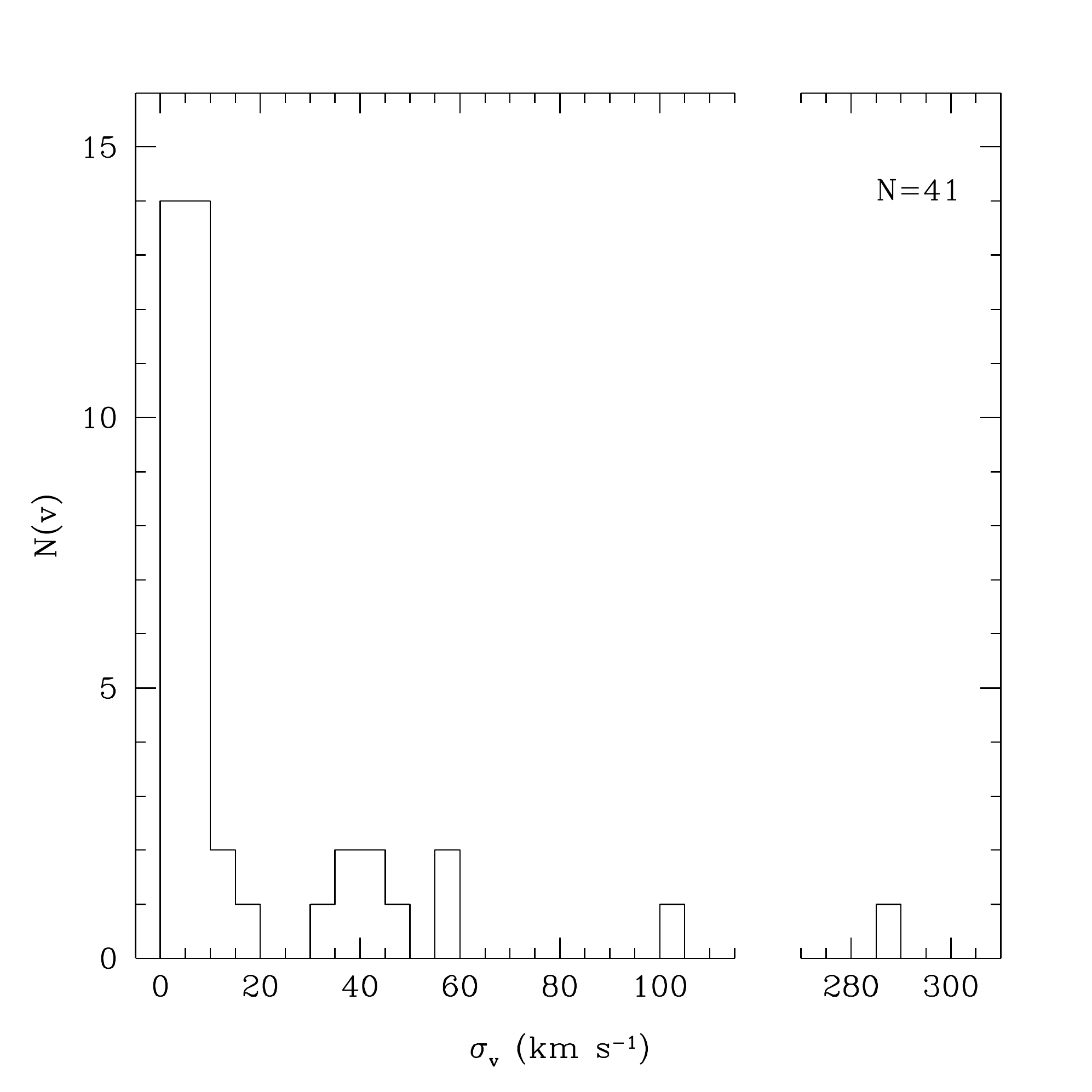}
\caption{Number of objects per velocity dispersion bin (width$=5$\,\kmps). The sample includes data
from Table~\ref{tbl_rad_vel} in Appendix~B and published results from the same survey \citep{kaw2010a,ven2012}.
\label{fig_stat}}
\end{figure}

Fig.~\ref{fig_stat} shows the distribution of the measured velocity dispersion. 
The sample includes three objects published earlier \citep{kaw2010a,ven2012} and seven new identifications described in the following Section. 
Three additional objects show significant radial velocity variations ($\sigma_\varv>10$\,\kmps), but with insufficient sampling
to determine the orbital parameters. Adding two likely close binaries identified through photometric variations 
(GALEX~J1736+2806 and J2038$-$2657, see Section 3.1.2) but for which
we only dispose of radial velocity measurements of the secondary, we estimate
that 15 out of 41 hot subdwarfs presently investigated are in close binaries, or a 37 per cent yield, lower than previously estimated \citep[e.g., ][]{cop2011}.
Our survey strategy aimed at short-period binaries would be insensitive
to long-period, low-amplitude variation ($<10$\,\kmps) systems. A detailed comparison with the sample of known hot subdwarf binaries
should allow us to secure a global estimate of binarity in this population (Section 4.1).

\subsection{Individual properties}

\begin{figure}
\includegraphics[width=1.00\columnwidth]{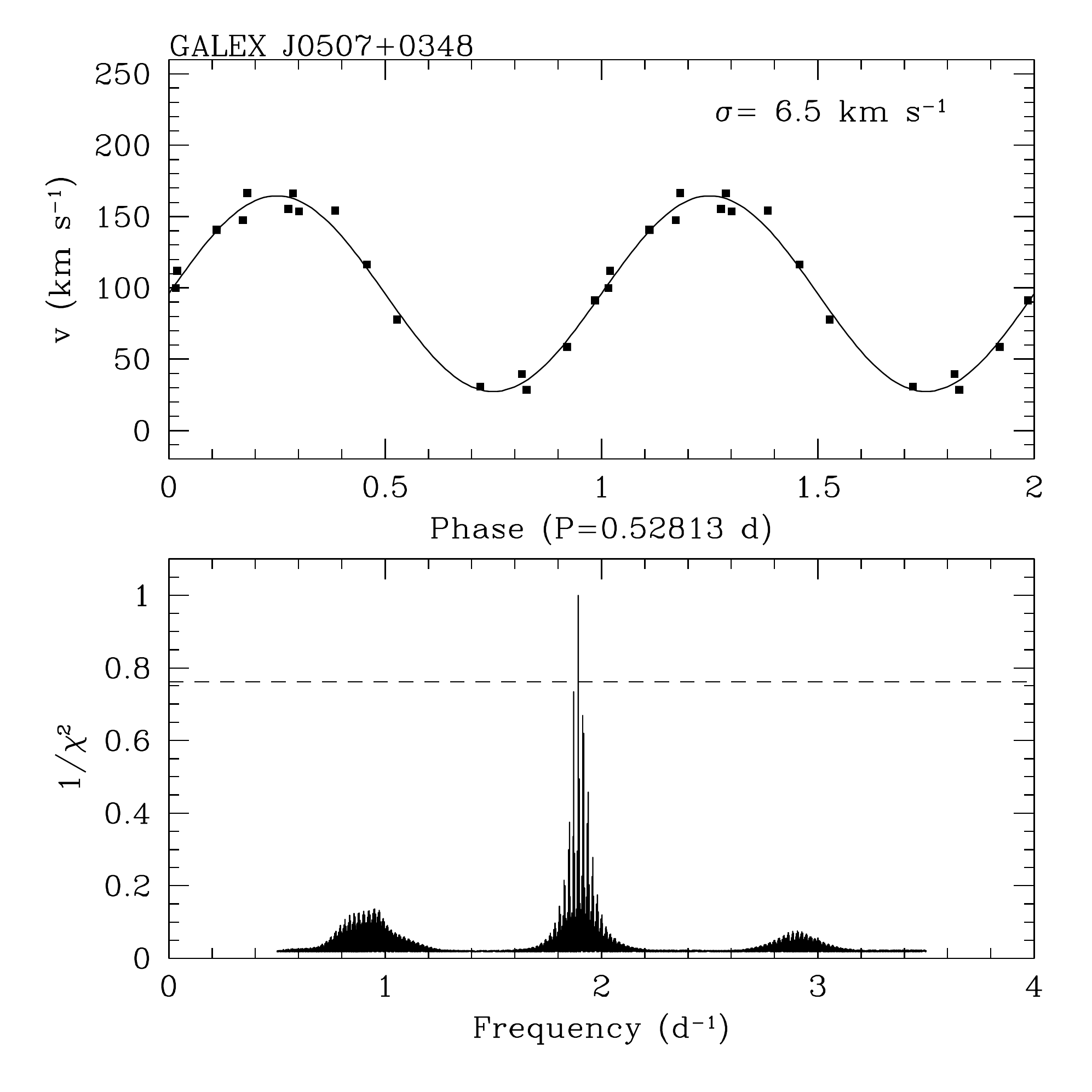}
\caption{(Bottom) period analysis of radial velocity measurements of GALEX~J0507$+$0348 (full line) and 66 per cent confidence level (dashed line). (Top) radial velocity measurements phased on the orbital period ($0.52813$\,d) and
and best-fitting sine curve (full line) with the dispersion in velocity residuals shown in upper-right. 
The initial epoch $T_0$ corresponds to inferior conjunction of the sdB star. Details are presented in Section 3.2.1.
\label{fig_0507}}
\end{figure}

\begin{figure}
\includegraphics[width=1.00\columnwidth]{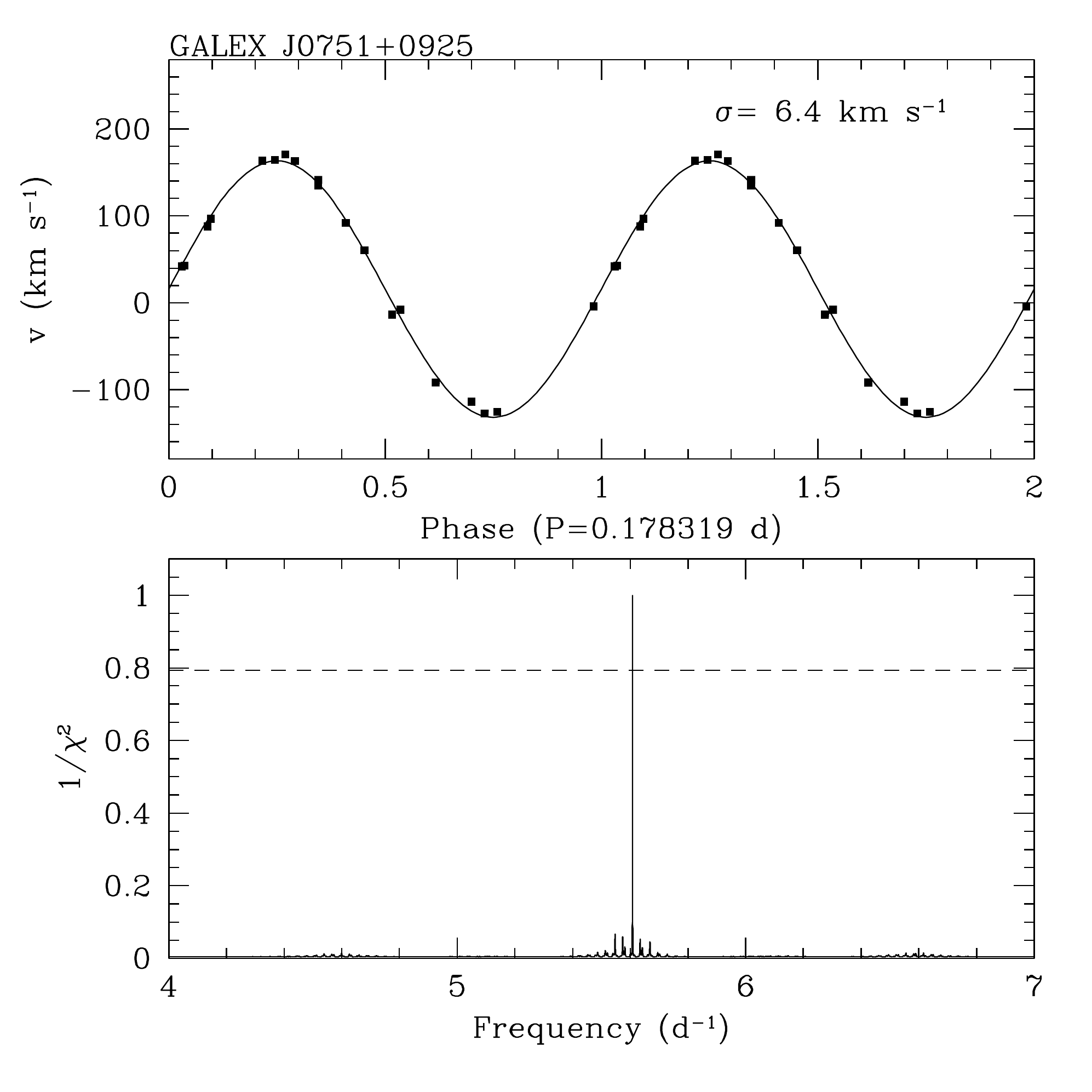}
\caption{Same as Fig.~\ref{fig_0507} but for GALEX~J0751$+$0925.
\label{fig_0751}}
\end{figure}

\begin{figure}
\includegraphics[width=1.00\columnwidth]{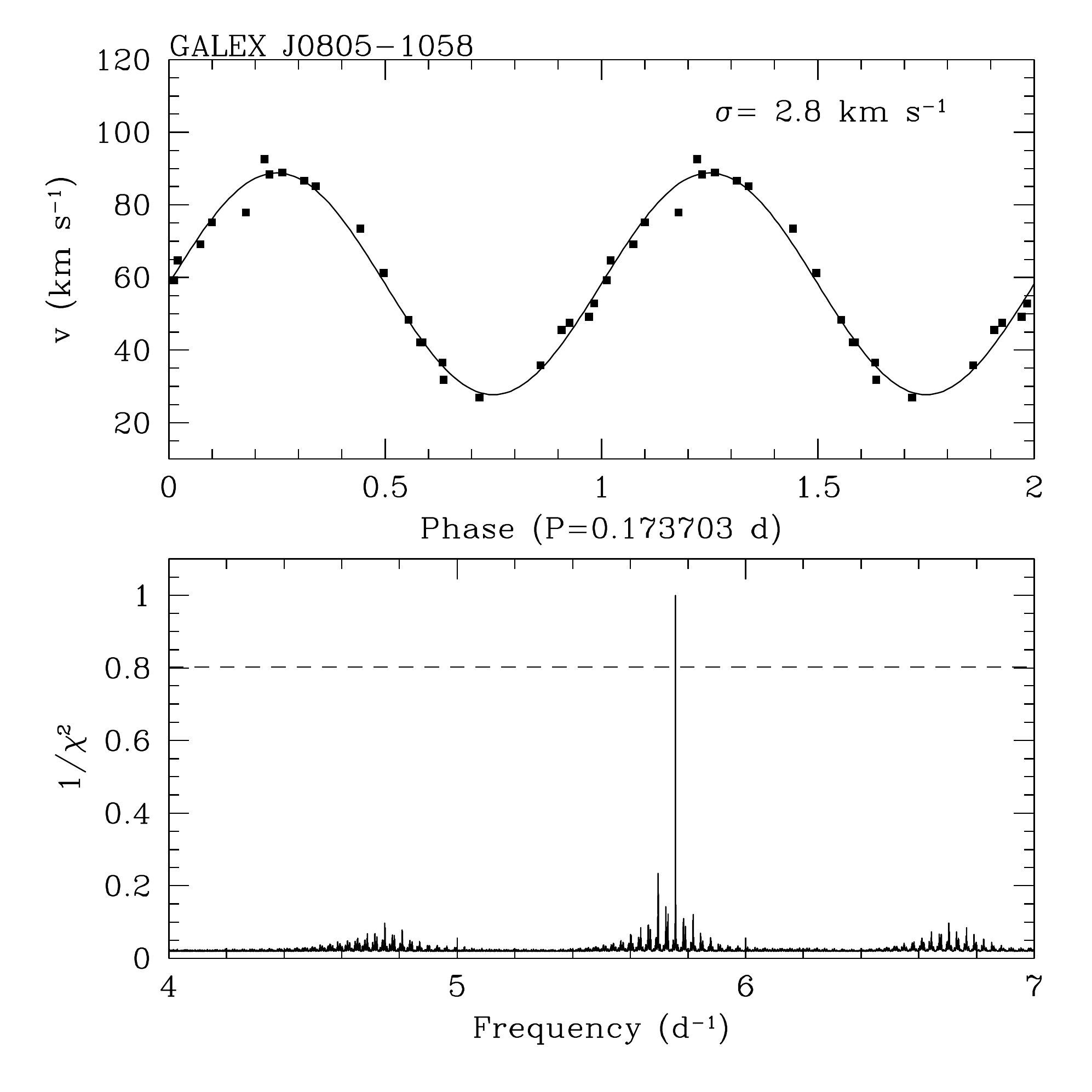}
\caption{Same as Fig.~\ref{fig_0507} but for GALEX~J0805$-$1058.
\label{fig_0805}}
\end{figure}

\begin{figure}
\includegraphics[width=1.00\columnwidth]{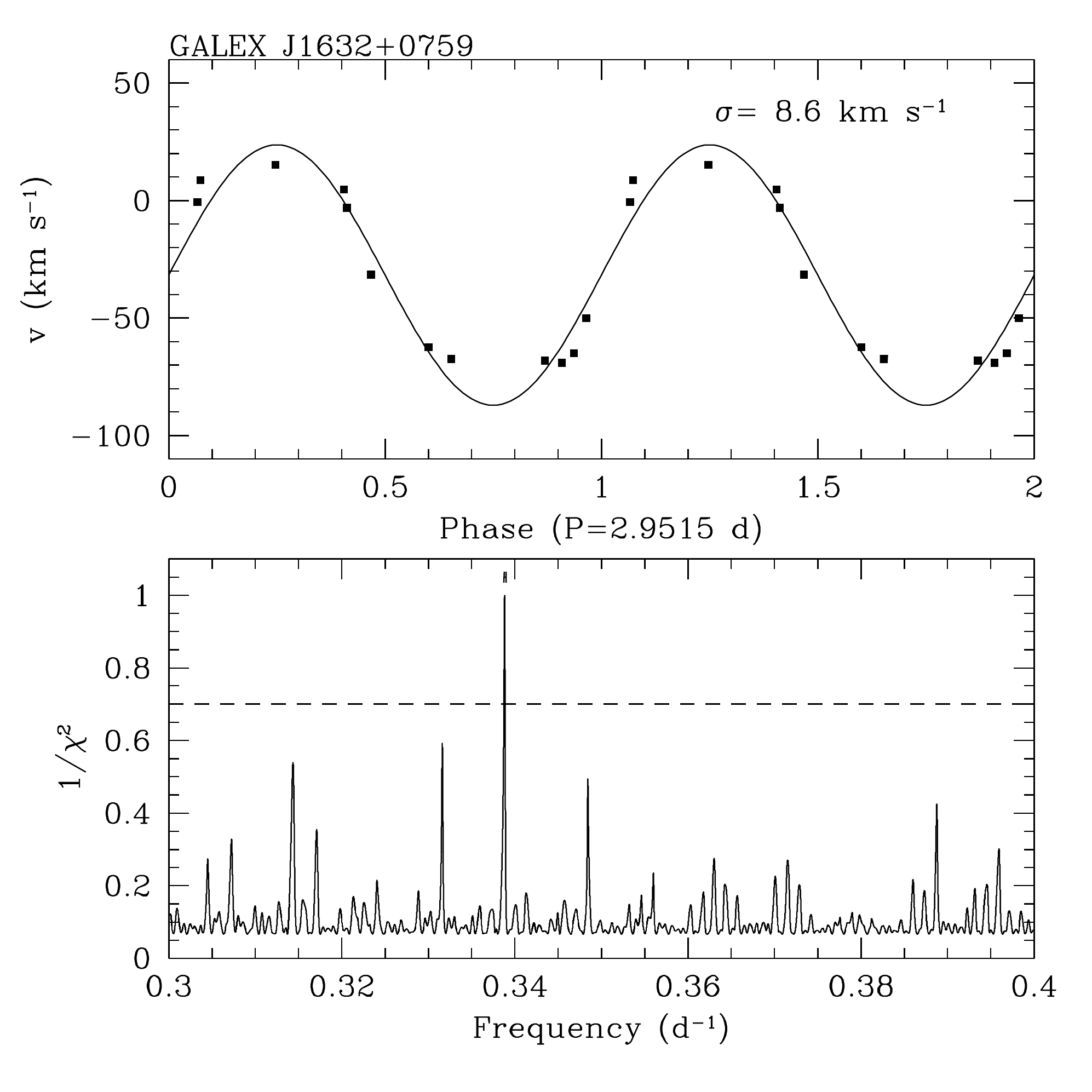}
\caption{Same as Fig.~\ref{fig_0507} but for GALEX~J1632$+$0759. The tick mark above the best-fitting period
indicates the results of the period analysis of \citet{bar2014}, 
$P=2.951\pm0.001$\,d.
\label{fig_1632}}
\end{figure}

\begin{figure}
\includegraphics[width=1.00\columnwidth]{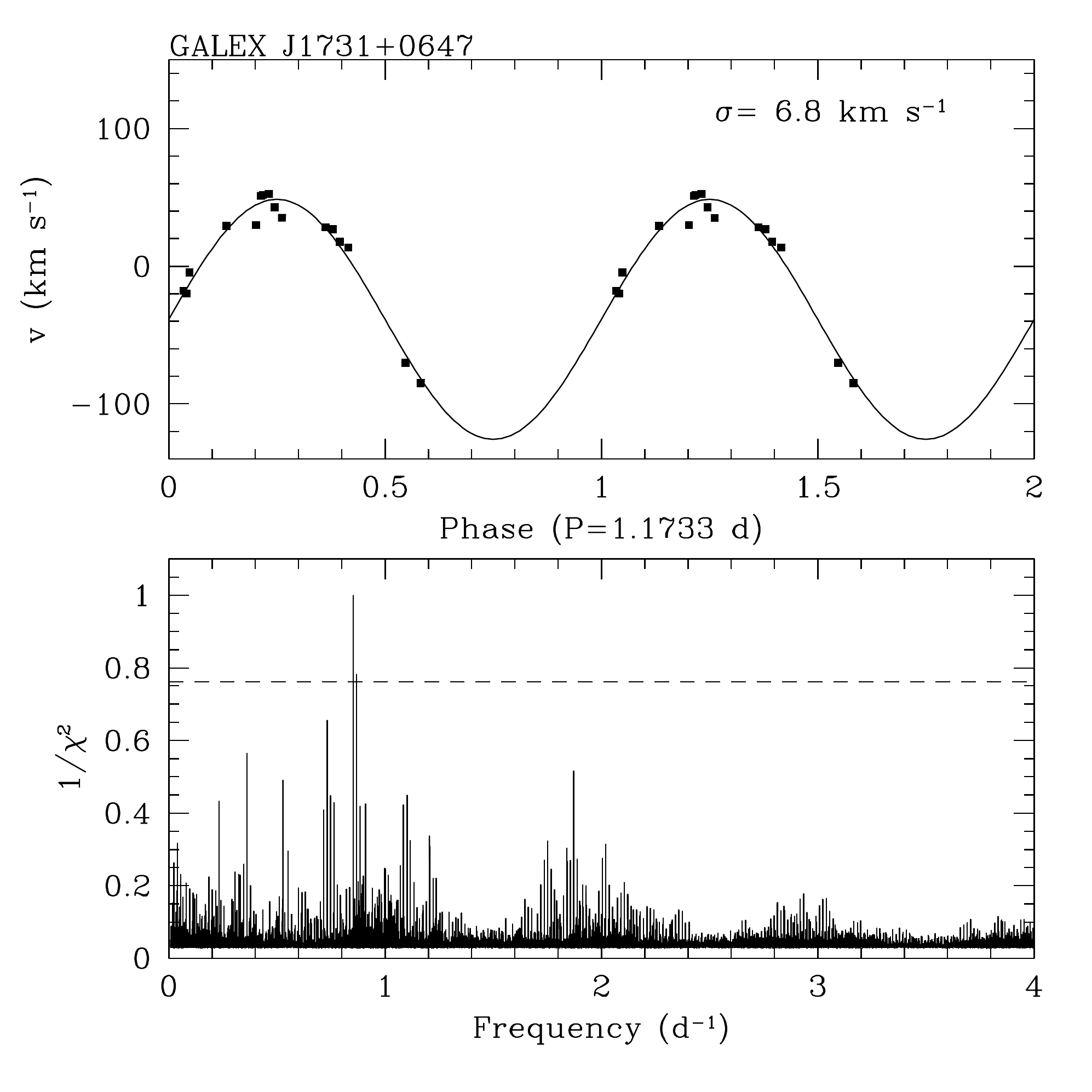}
\caption{Same as Fig.~\ref{fig_0507} but for GALEX~J1731$+$0647.
\label{fig_1731}}
\end{figure}

\begin{figure}
\includegraphics[width=1.00\columnwidth]{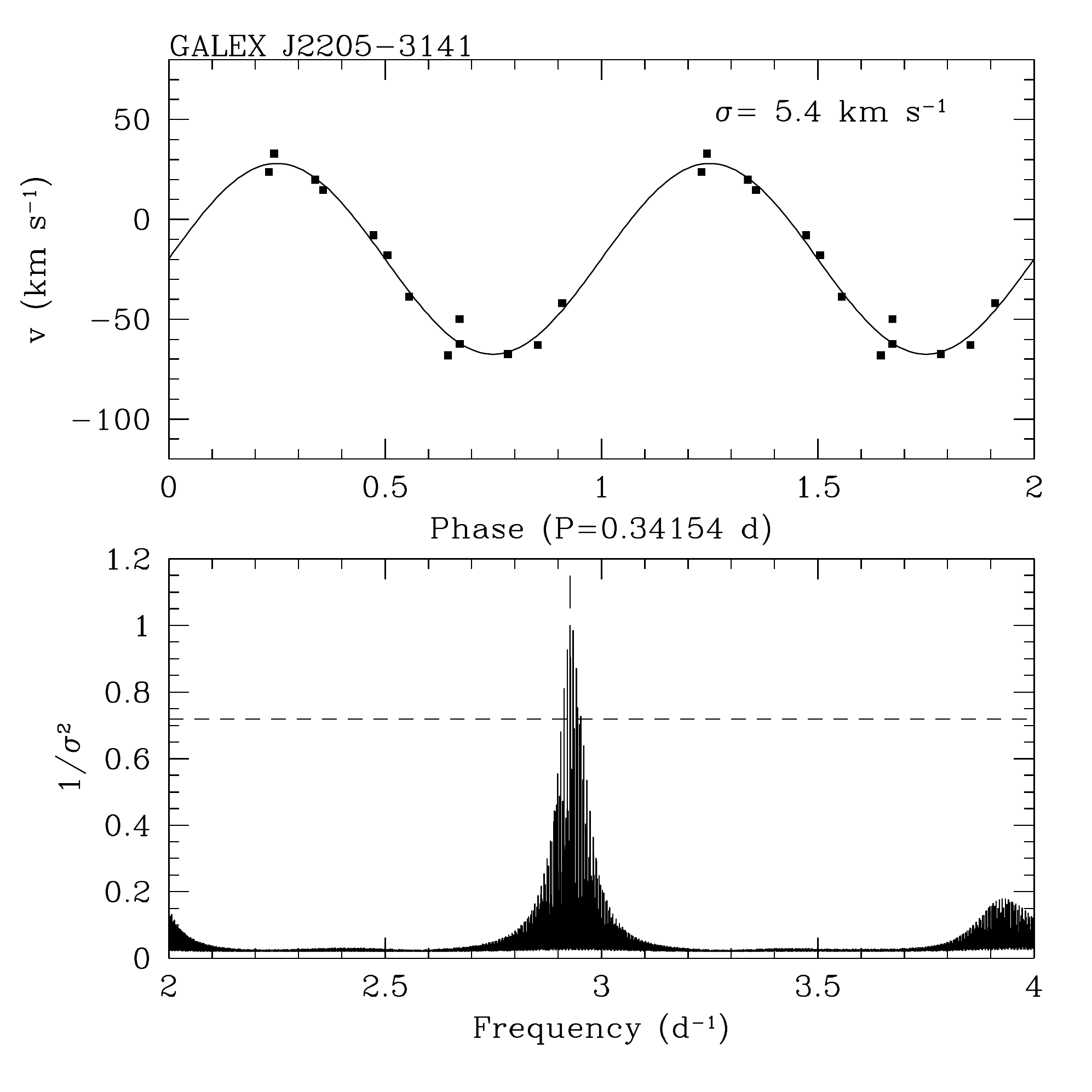}
\caption{Same as Fig.~\ref{fig_0507} but for GALEX~J2205$-$3141. The photometric period is marked
close to the peak frequency of the velocity periodogram.
\label{fig_2205}}
\end{figure}

\begin{figure}
\includegraphics[width=1.00\columnwidth]{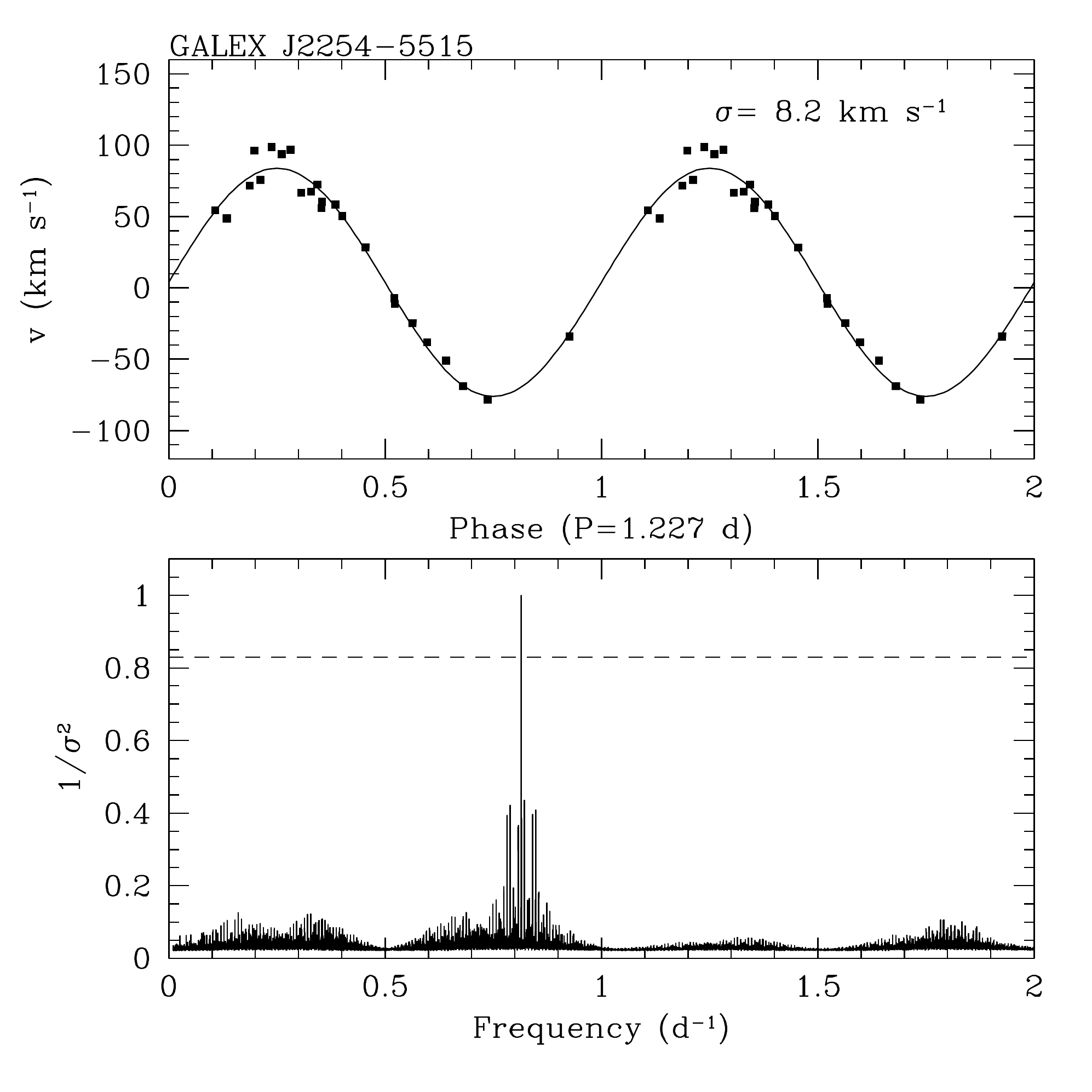}
\caption{Same as Fig.~\ref{fig_0507} but for GALEX~J2254$-$5515.
\label{fig_2254}}
\end{figure}

Section 3.2.1 describes objects with variable radial velocities
suggesting the presence of a close binary companion:
Fig.~\ref{fig_0507}, Fig.~\ref{fig_0751}, 
Fig.~\ref{fig_0805}, Fig.~\ref{fig_1632}, Fig.~\ref{fig_1731}, Fig.~\ref{fig_2205},
and Fig.~\ref{fig_2254}
show results of the period analysis ($1/\chi^2$ versus frequency) for this group of objects.
The confidence level is set at 1$\sigma$ (66 per cent) for
a four-parameter ($p=4$) analysis with the $\chi^2$ normalized on the best-fitting solution
and the radial velocity measurements phased on the best-fitting period.
Sections 3.2.2, 3.2.3, and 3.2.4 review the properties of the remaining systems, i.e., those with
composite spectra, unresolved radial velocity variations, or photometric variability, respectively, and the likelihood
that they might belong to the close binary population.
Finally, Section 3.3 presents known facts concerning the remaining objects.
In the following Section, the subscript ``1'' refers to the hot subdwarf and the 
subscript ``2'' refers to its companion. Similarly, the suffix ``B'' designates the companion.
The binary parameters are listed in Table~\ref{tbl_bin_param} including the number of spectra per object ($N$) 
and the dispersion in velocity residuals ($\sigma_{vr}$).

\subsubsection{Close binaries}

\begin{table*}
\centering
\begin{minipage}{\textwidth}
\caption{Spectroscopic binary parameters. \label{tbl_bin_param}}
\renewcommand{\footnoterule}{\vspace*{-15pt}}
\renewcommand{\thefootnote}{\alph{footnote}}
\begin{tabular}{lccccccc}
\hline
                      & J0507$+$0348    & J0751$+$0925    & J0805$-$1058               & J1632$+$0759    & J1731$+$0647    & J2205$-$3141 & J2254$-$5515 \\
Parameter             &                 &                 &                            &                 &                 &              &              \\
\hline
$P$ (d)               & $0.528127$      & $0.178319$      & $0.173703$                 & $2.9515$ \footnotemark[1]\footnotetext[1]{Based in part on the period obtained by \citet{bar2014}: $P=2.951\pm0.001$ d.}  & $1.17334$      & $0.341543$    & $1.22702$  \\
$\sigma_P$\,(d)       & $0.000013$      & $0.000005$      & $0.000002$                 & $0.0006$        & $0.00004$       & $0.000008$    &  $0.00005$ \\
$T_0$ (HJD)           & $2456315.349$   & $2455972.827$   & $2456299.0335$             & $2456150.701$   & $2456313.119$   & $2456313.3387$& $2456444.616$ \\
$\sigma\,(T_0)$       & $      0.015$   & $      0.001$   & $      0.0026$             & $      0.016$   & $      0.004$   & $      0.0005$& $      0.001$ \\
K (\kmps)             & $68.2\pm2.5$    & $147.7\pm2.2$   &  $29.2\pm1.3$              & $54.9\pm4.6$    & $87.7\pm4.1$    & $47.8\pm2.2$  & $79.7\pm2.6$ \\
$\gamma$ (\kmps)      & $96.2\pm1.8$    & $15.5\pm1.6$    & $58.2\pm0.9$               & $-31.6\pm2.7$   & $-39.1\pm3.0$   & $-19.4\pm1.7$ & $4.2\pm2.0$ \\
$f_{\rm sec}$ (\msun) & $0.017\pm0.002$ & $0.059\pm0.003$ & $(4.4\pm0.6)\times10^{-4}$ & $0.048\pm0.013$ & $0.080\pm0.012$ & $0.0037\pm0.0005$ & $0.063\pm0.006$ \\
$M_1$\ (\msun)        & (0.47)          &   (0.47)        &  (0.234)                   & (0.47)          &  (0.47)         & (0.47)        & (0.47)         \\
$M_2$\ (\msun)        & $>0.20$         &   $>0.34$       &  $>0.03$                   & $>0.31$         &  $>0.39$        & $>0.11$       & $>0.35$        \\
N                     &    16           & 19              & 23                         & 12              & 16              & 13            & 24             \\
$\sigma_{\varv}$ (\kmps) &    6.5       &   6.4           &   2.8                      &   8.6           &    6.8          &    5.4        &    8.2        \\
Notes                 &  probable WD    & probable WD     & low mass sdB,              & probable        &  probable       &    reflection, & probable      \\
                      &  secondary      & secondary       & possible BD                & WD secondary    & WD secondary    & dM secondary & WD secondary \\
	              &                 &                 & secondary                  &                 &                 &              &              \\
\hline
\end{tabular}
\end{minipage}
\end{table*}

The sdB star GALEX~J0507+0348 is part of a newly identified spectroscopic binary.
The star is close to the ZAEHB and may be a low-mass sdB star. 
The H$\alpha$ radial velocity measurements are phased on a period of $\sim$0.528\,d (Fig.~\ref{fig_0507}).
The SED of GALEX~J0507$+$0348 shows an infrared excess but in the $WISE$ W3 band only.
The nearby object ($sep.=17$~arcsec) visible on photographic plates is not likely to affect the $WISE$ measurements.
Also, low-dispersion spectra show weak CaH\&K lines with an equivalent width of
$E.W.$(CaK)$=270$\,m\AA. The calcium doublet could indicate the presence of a late-type companion. However, no radial
velocity measurements were obtained in that spectral region and we could not confirm variations in the
line position. Moreover, we could not confirm the presence of other late-type spectral signatures such as Mg{\sc i} lines,
and our composite spectral analysis \citep{nem2012} rejects the presence of a companion with 
a flux contribution above 1 percent in the optical range. 
A series of high dispersion spectra
is required to determine whether the CaH\&K lines originate from the companion, the interstellar medium (ISM), or in the circumstellar environment.
The mass function allows us to infer $M_2>0.20$\,\msun\ assuming $M_1=0.47$\,\msun, or 
$M_2>0.15$\,\msun\ assuming $M_1=0.30$\,\msun. A late-type (dM,dK) companion would satisfy these constraints.
However, the Catalina time series folded on the orbital period constrains the photometric variations to a semi-amplitude lower than 2~mmag. 
Reflection effect on a late-type star with a mass exceeding 0.2\,\msun\ would result in variations of a semi-amplitude of $\approx60$~mmag as observed in the
case of GALEX~J0321+4727 \citep{kaw2010a}.
We conclude that the companion is most likely a white dwarf: At a binary inclination $i<30^\circ$ the mass function implies a minimum mass of 0.51\,\msun\ that would be consistent with a normal white dwarf star.

The sdB star GALEX~J0751$+$0925 is part of a close binary with the largest velocity semi-amplitude 
measured in the present sample, $K\sim$148\,\kmps\ (Fig.~\ref{fig_0751}). 
The mass function implies the presence of a companion relatively more massive than
in other similar systems with $M_2>0.34$\,\msun\ assuming $M_1=0.47$\,\msun.
The SED of this system also appears to show an infrared excess in the W1, W2, and W3 bands 
which may be caused, in part, by a
nearby star only 6~arcsec away (particularly in the W3 band). 
The Catalina and ASAS light curves do not show significant variations at the orbital period. Again, a comparison with
the photometrically variable system GALEX~J0321+4727 indicates that the companion is probably
a white dwarf. 
The semi-amplitude of the phased light curve of GALEX~J0751$+$0925 is limited to 4~mmag in the Catalina observations although 
the minimum secondary mass in this shorter-period system
is larger than in GALEX~J0321+4727, i.e., 0.34 versus 0.13\,\msun, and the predicted semi-amplitude of variations
due to a late-type companion would be $\approx190$~mmag. Therefore, the absence of a reflection effect in GALEX~J0751$+$0925 
rules out the presence of a late-type companion leaving only the possibility of a $>0.34$\,\msun\ white dwarf companion,
or $>0.50$\,\msun\ if $i<50^\circ$.

The sdB GALEX~J0805$-$1058 clearly lies below the ZAEHB (Fig.~\ref{fig_sample}); following the evolutionary tracks of
\citet{dri1998} the mass of the subdwarf is estimated to be in the range 0.2-0.3\,\msun. The low velocity
amplitude (Fig.~\ref{fig_0805}) and small mass function imply
a very low mass for the companion, $M_2>0.03$\,\msun, assuming $M_1=0.234$\,\msun. 
At inclinations higher than $i = 26^\circ$, the secondary mass remains lower than 0.08\,\msun\ and the object is substellar.
Assuming a probability distribution for the inclination angle $i$ of the form $P_i\,di=\sin{i}\,di$, inclinations higher than
$26^\circ$ have $P(>26)=$89.9 per cent probability of occurring.
At lower inclinations ($10^\circ<i<26^\circ$, i.e., $P(10-26)=$8.6 per cent), the secondary mass does not exceed 0.3\,\msun\ and the object would be a low-mass M dwarf.
At very low inclinations ($i<7^\circ$, i.e., $P(<7)=$0.7 per cent) the secondary would be a normal white dwarf star ($M>0.5$\,\msun).
The SED shows a single, hot subdwarf star, i.e., as far as the $WISE$ W2 band, and the faint ($\Delta m\sim$6 mag), nearby ($sep.\sim$11~arcsec) object visible in photographic plates does not appear to contaminate
the SED. The ASAS and NSVS light curves do not show significant variations when folded on the orbital period, 
i.e., $\lesssim12$~mmag, while the predicted semi-amplitude of variations due to a substellar object 
\citep[$R_2\approx 0.1\,R_\odot$, see a review by][]{cha2009} would be $\approx20$~mmag.
Variations of the order of 10~mmag have been observed in brown dwarf plus hot subdwarf binaries \citep[see, e.g.,][]{sch2014c}
and such variations would be detectable in GALEX~J0805$-$1058 in quality photometric time series.
We conclude that the apparent lack of variations is a consequence of the
small radius of a substellar secondary.

\citet{bar2012} recorded radial velocity variations in spectra of the sdB star GALEX~J1632+0759. Their data 
suggested a period ranging from 2 to 11 days. 
Our measurements, based on H$\alpha$ and He\,{\sc i}5875.621 in the red,
and H$\beta$ and He\,{\sc i}4685.698 in the blue, also revealed large velocity variations (Fig.~\ref{fig_1632}).
Recently, \citet{bar2014} obtained new radial velocity
measurements and determined a period of $2.951\pm0.001$\,d.
We restricted our period analysis to frequencies between 0.3 and 0.4\,d$^{-1}$ and recovered
an identical period. The mass function implies a secondary mass $M_2>0.31$\,\msun, assuming $M_1=0.47$\,\msun.
The SED shows a large flux excess apparent in the 2MASS and $WISE$ bands as well as heavy extinction in
the ultraviolet range. The measured extinction coefficient ($E_{B-V}=0.4$) largely exceeds the coefficient
inferred from the maps of \citet{sch1998}, $E_{B-V}=0.08$. 
The additional extinction probably originates in the immediate, possibly dusty,
circumstellar environment of the system.
An inspection of our acquisition images of GALEX~J1632$+$0759 reveals the presence of
a nearby star; we measured a separation of 2.3~arcsec at a position angle of 225$^\circ$. 
The 2MASS and $WISE$ photometric measurements are likely
contaminated by this object. \citet{ost2005} also resolved GALEX~J1632$+$0759 
and the nearby star and measured a separation of 2.1~arcsec. In addition to GALEX~J1632$+$0759, \citet{bar2014} 
obtained radial velocity measurements of the nearby star which they classified
as a late G dwarf or early K dwarf that is itself in a close binary: The radial velocity varied with a period of $1.42\pm0.01$~d. 
\citet{bar2014} also found that 
the systems share the same systemic velocity
suggesting that this is a quadruple system. The ASAS, Catalina, and NSVS time series constrain photometric variations to
semi-amplitudes lower than 4, 8, and 18~mmag. The predicted semi-amplitude of photometric variations due to
the presence of a late-type companion would be $\approx60$~mmag. We conclude that the secondary star is most probably 
a white dwarf with a mass ranging from 1.3 to 0.5\,\msun\ assuming a low inclination ($24\lesssim i\lesssim 46^\circ$), 
or with a peculiar low mass ($0.3-0.5$\,\msun)
assuming $i\gtrsim 46^\circ$.

The new binary system GALEX~J1731$+$0647 (Fig.~\ref{fig_1731}) harbours the heaviest binary companion identified
in our sample. The mass function implies a mass $M_2>0.39$\,\msun, assuming $M_1=0.47$\,\msun. 
The field surrounding this subdwarf is relatively crowded but only two objects are found 
between 13 and 15~arcsec away and with photographic magnitude differentials 
$\Delta m\sim$3 and 5 mag. These objects would not affect the SED which shows a single hot subdwarf. 
The lack of photometric variations, $<45$~mmag in ASAS time series and $<21$~mmag in Catalina time series, compared
to expected variations of $\approx90$~mmag due to a relatively
large M or K dwarf suggests that
the companion is most likely a white dwarf. 
We infer a mass between 1.3 and 0.5\,\msun\ assuming an inclination of $29\lesssim i\lesssim 58^\circ$, 
or a peculiar low mass ($0.4-0.5$\,\msun)
assuming $i\gtrsim 58^\circ$.

\begin{figure*}
\includegraphics[width=1.00\textwidth]{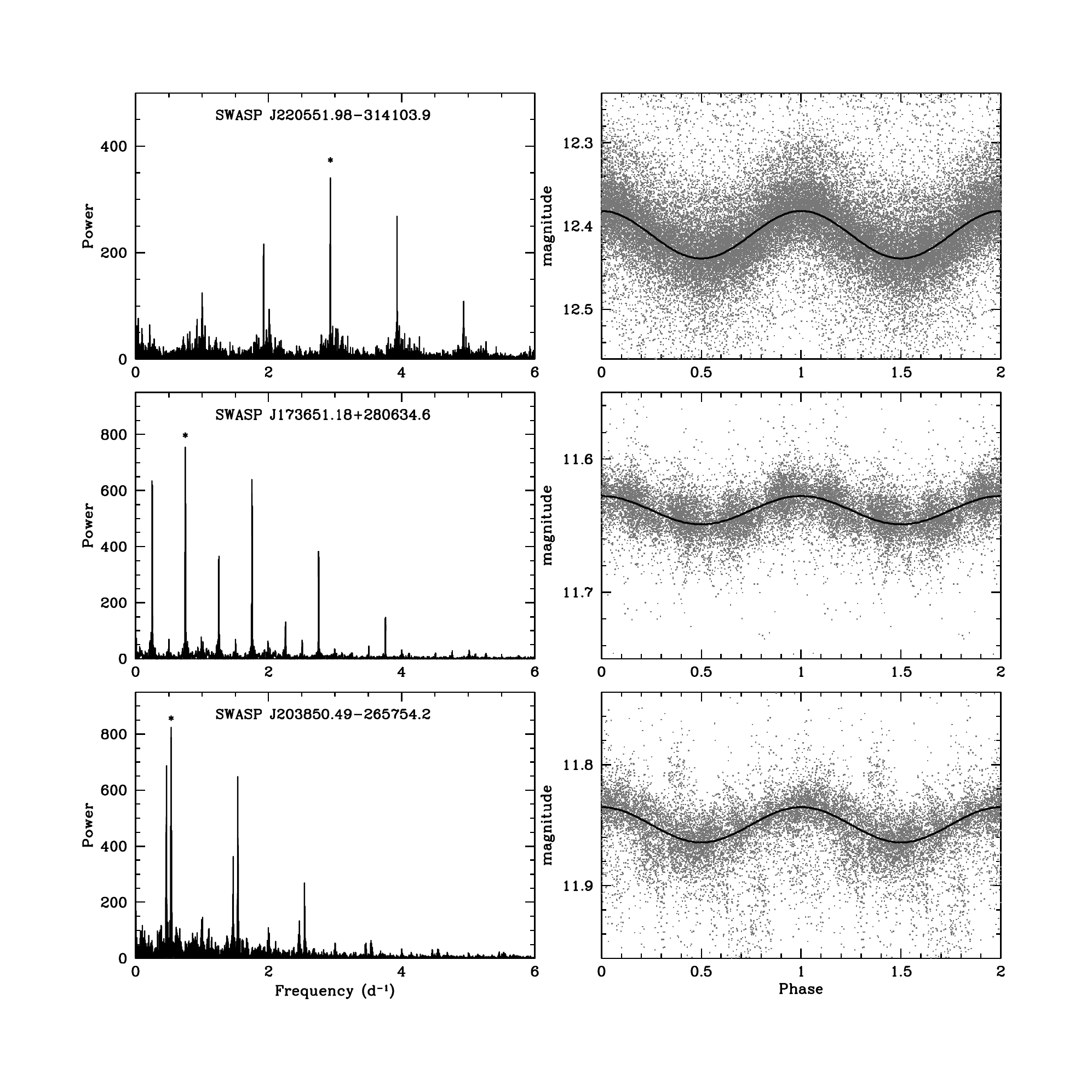}
\caption{(Left panels) Fourier transform analysis of the variable stars GALEX~J2205$-$3141 (top), J1736+2806 (middle) and J2038$-$2657 (bottom), and
phased light curves (right panels). The identification of the photometric period with the spectroscopic period clearly indicates that the light curve of GALEX~J2205$-$3141 
shows the effect of reflection of the bright primary on the secondary star.
The initial epoch $T_0$ in GALEX~J2205$-$3141 corresponds to the passage of the secondary star at superior conjunction. 
Without knowing their orbital periods, we can only state that the cool, secondary stars in both GALEX~J1736+2806 and J2038$-$2657 are variable.
The photometric periods are marked with star symbols: $P=0.341563$~d (J2205$-$3141), 1.333204~d (J1736+2806), and 1.870221~d (J2038$-$2657).
\label{fig_swasp}}
\end{figure*}

GALEX~J2205$-$3141 is a close binary with $P\approx 0.34$\,d (Fig.~\ref{fig_2205}) showing a reflection effect in the SWASP time series 
(semi-amplitude $\Delta m \approx 27$~mmag). Similar variations were also observed in the
ASAS and Catalina time series. The mass function is consistent with the presence of a late-type
companion ($M_2>0.11$\,\msun).
Photometric time series from the Catalina survey, SWASP (1SWASP~J220551.98$-$314103.9), and ASAS show
variability with mutually consistent periods of $0.341559\pm0.000003$, $0.341563\pm0.000002$, and
$0.341561\pm0.000002$\,d, respectively. The photometric periods are somewhat longer than
the spectroscopic orbital period $P=0.341543\pm0.000008$\,d.
The radial velocity measurements are based on the He{\sc i} 4471.48 and 6678.15 lines.
We noted that the Balmer H$\alpha$ and H$\beta$ lines cores are asymmetric and are possibly contaminated by emission from the
companion as noted in the case of AA~Dor \citep{vuc2008}.
Fig.~\ref{fig_swasp} shows the SWASP measurements phased on the photometric period.
We identify the initial epoch with the passage of the secondary star at superior conjunction 
corresponding to maximum reflected light.
Although the photometric variations are clearly caused by the reflection of the primary light
on a late-type dwarf companion, the phasing error between photometric and spectroscopic ephemeris is $\Delta \Phi \approx 0.1$.
We attribute this error to a large gap between the epoch of the spectroscopic observations and that of the
photometric observations. The SED shows a mild flux excess in the IR to mid-IR range possibly due to
the late-type companion. A star found 9~arcsec away and 4 mag fainter does not affect the SED. 
However, renormalizing on the $J$ band rather than the $V$ band nearly eradicates this excess.
Assuming a possible $K$ band contribution from the companion of 15 to 40 per 
cent, we estimated for the M dwarf companion $M_{K,2}\approx7.5$ to 6.5 if
$M_{K,1}\approx 5.5$. Bearing in mind that reprocessing of ultraviolet 
radiation from the hot primary into the cool secondary
atmosphere should contribute to this IR excess,
the absolute K magnitude of the secondary star corresponds to a spectral type later than M3-4, or a mass $M_2\lesssim 0.24$-0.4\,\msun\ which
requires an orbital inclination $i\gtrsim 20$-30$^\circ$.
We find possible evidence of extinction in the ultraviolet range in excess of the extinction expected
from the \cite{sch1998} map, although the $GALEX$ NUV dip may be the result of larger uncertainties than estimated. 
This system is the only confirmed binary in our sample comprised of a hot subdwarf
and late-type companion.

The sdB GALEX~J2254$-$5515 shows large radial velocity variations (Fig.~\ref{fig_2254}) although
the Catalina and ASAS time series indicate that the star is not photometrically variable with semi-amplitudes lower
than 28 and 5~mmag, respectively.
The minimum mass of the secondary, $M_2>0.35$\,\msun\ assuming $M_1=0.47$\,\msun, combined with the lack of photometric
variability when compared to expected variations of 150~mmag caused by a reflection effect on a putative late-type companion
imply that the
companion is a white dwarf.

We neglected the possible effect of orbital eccentricity in the period analysis.
The orbits of post-common envelope binaries is expected to be circular
due to the synchronization during the post-CE phase.
However, eccentric orbits in close binaries containing a subdwarf were reported by 
\citet{ede2005} and \citet{kaw2012a}. In these cases the eccentricity was 
small ($e < 0.1$). Larger eccentricities were reported for long
period binaries, such as BD$+20^\circ 3070$, BD$+34^\circ 1543$, Feige~87
\citep{vos2013} and PG~1449$+$653 \citep{bar2013a}.
Eccentric orbits may indicate the presence of a circumbinary disc 
\citep{art1991}.

Now, we summarize additional constraints on the properties of spectroscopic composites, other
likely systems showing radial velocity variations, and systems displaying photometric variability.

\subsubsection{Composite spectra}

Using spectral decomposition, \citet{nem2012} classified GALEX~J0047$+$0337 as a
binary consisting of a hot sdB and a main-sequence F star. 
The radial velocity measurements obtained for GALEX~J0047$+$0337B imply a constant velocity with standard deviation of 
only 6.3\,\kmps\ and include a single measurement
deviating from the average velocity by more than 10\,\kmps.
The ASAS and NSVS photometry
do not show evidence of significant variations: The ASAS data constrain potential variations to a
semi-amplitude lower than 4.4~mmag for all periods larger than 0.5 hr.
The EFOSC acquisition images revealed a nearby star approximately
$\sim$3~arcsec away at a position angle of 344$^\circ$. The object is about 
1.2~mag fainter in $R$ and the quoted $WISE$ and
2MASS magnitudes include both stars since they would not be resolved in either
surveys. Fortunately, our optical spectra were not contaminated by the nearby star and the composite nature of
the object is not affected.

GALEX~J1411+7037 and J1753$-$5007 are sdB stars with F-type companions.
Their SEDs are consistent with the presence of a luminous companion derived from the
spectral decomposition of \citet{nem2012}. The H$\alpha$ line profile in each star
is dominated by the main-sequence star and no significant radial
velocity variations have been found for these objects. 
The ASAS times series of GALEX~J1753$-$5007 constrain photometric variations 
to a semi-amplitude lower than 19~mmag.

The SED of GALEX~J1356$-$4934 shows significant
infrared excess. An inspection of the acquisition images did not reveal
a resolvable, nearby companion and
radial velocity measurements show only marginal variability with radial velocity
maxima reaching a span of 20\,\kmps. First, we performed a SED decomposition to estimate
the spectral type of the companion. We adopted the sdB parameters determined by \citet{nem2012} and
calculated sdB absolute magnitudes of $M_K = 5.47$ and $M_V = 4.46$. 
Adopting the apparent visual magnitude $V=12.3$ and 2MASS magnitude $K=11.633$ and using the
main-sequence colour and absolute magnitude relations from \citet{pec2013}, we determined the absolute
visual and infrared magnitudes of the late-type companion, $M_{K,2}=3.57$ and $M_{V,2}=5.36$, and a distance of 444 pc.  
Consequently, the companion mass is 0.94\,\msun\ corresponding to a G8V star. 
Next, we 
performed a spectral decomposition with XTGRID \citep{nem2012} making use of both the blue 
and red spectra of GALEX~J1356$-$4934. The spectral decomposition showed that the companion
contributes 27 per cent of the flux at 7000 \AA. The new parameters of the
sdB star are $T_{\rm eff} = 32370^{+230}_{-660}$ K, $\log{g} = 5.72^{+0.07}_{-0.16}$, $\log{\rm He/H} = -2.75^{+0.25}_{-0.43}$
and do not differ significantly from our earlier measurements. The parameters of the companion are
$T_{\rm eff} = 5470$, $\log{g} = 4.47$, [Fe/H] $=0.003$, also corresponding to a G8 main-sequence star.
These values supersede those of \citet{nem2012} for GALEX~J1356$-$4934. The ASAS time series limits the photometric
variations to a semi-amplitude of 3~mmag.

Optical spectra of subdwarf plus early-type F-stars are dominated in the red by the companion.
Because the mass ratio is $\gtrsim$3, high-dispersion spectroscopy is required to detect the
secondary star motion.

\subsubsection{Radial velocity variable}

Other objects, in addition to the confirmed binaries listed in Table~\ref{tbl_bin_param}, are likely close
systems. The measured radial velocity extrema suggest that these subdwarfs are in close orbit
with a companion, but the small number of spectra did not allow us to perform a period analysis.

We measured velocity extrema $\Delta\varv\approx80$\,\kmps\ for GALEX~J0613$+$3420, 
$\Delta\varv\approx28$\,\kmps\ for GALEX~J0812$+$1601, and
$\Delta\varv\approx35$\,\kmps\ for GALEX~J1903$-$3528. 
The SWASP time series for GALEX~J0613$+$3420 and GALEX~J1903$-$3528 constrain photometric variations
to maximum semi-amplitudes of 13 and 4~mmag, respectively, which exclude the presence of close late-type companions.
Further investigations are required to clarify their binary status.

The SED of each object does not reveal the presence of a companion, however the SED
of GALEX~J0613$+$3420 shows evidence of a large interstellar extinction \citep{sch1998},
and a possible excess ($E_{B-V}=0.64$) above the interstellar value ($E_{B-V}=0.36$).

\subsubsection{Photometrically variable}

The SWASP light curve of the
sdB plus F7V pair GALEX~J1736+2806 (1SWASP~J173651.18+280634.6) varies with a period $P=1.33$\,d and a semi-amplitude
of 11~mmag (Fig.~\ref{fig_swasp}). No significant variations were observed in a nearby comparison object (1SWASP~J173635.80+280902.2). 
The grouping of data points observed in the light curve are also observed in the light curve of the nearby object and, therefore,
it must be an artefact of data sampling.
Using the SED we found that the absolute $V$ magnitude of the companion is $\sim$0.72~mag brighter than the sdB star
consistent with a value of $\sim$0.81~mag obtained by \citet{nem2012}.
The absolute magnitude of a late F7 star, $M_V\sim$4, would imply for the hot subdwarf
$M_V\sim$4.7; The atmospheric parameters of the hot subdwarf are very uncertain \citep{nem2012}
but would be reconciled with the companion spectral type at the lowest acceptable temperature (30,000K at $\log{g}=5.7$).
The photometric variability may be caused by irradiation of the exposed hemisphere of the
F star although we failed to detect radial velocity variations at the same period.

\begin{figure}
\includegraphics[width=1.0\columnwidth]{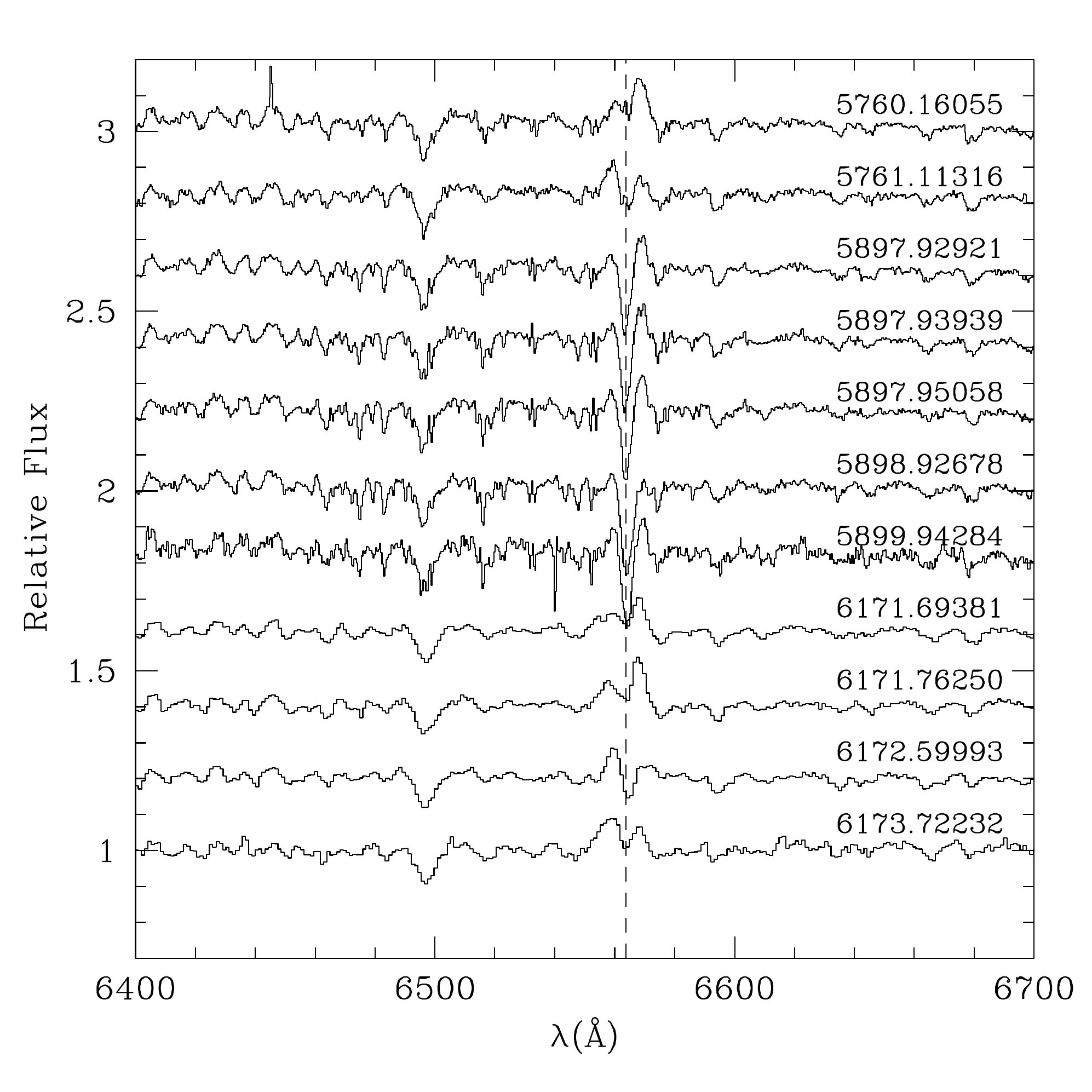}
\caption{Spectra of the photometrically variable sdO+G8III star GALEX~J2038$-$2657 obtained at SSO and La Silla and showing short-term variable H$\alpha$ emission. 
\label{fig_2038}}
\end{figure}

\begin{figure}
\includegraphics[width=1.0\columnwidth]{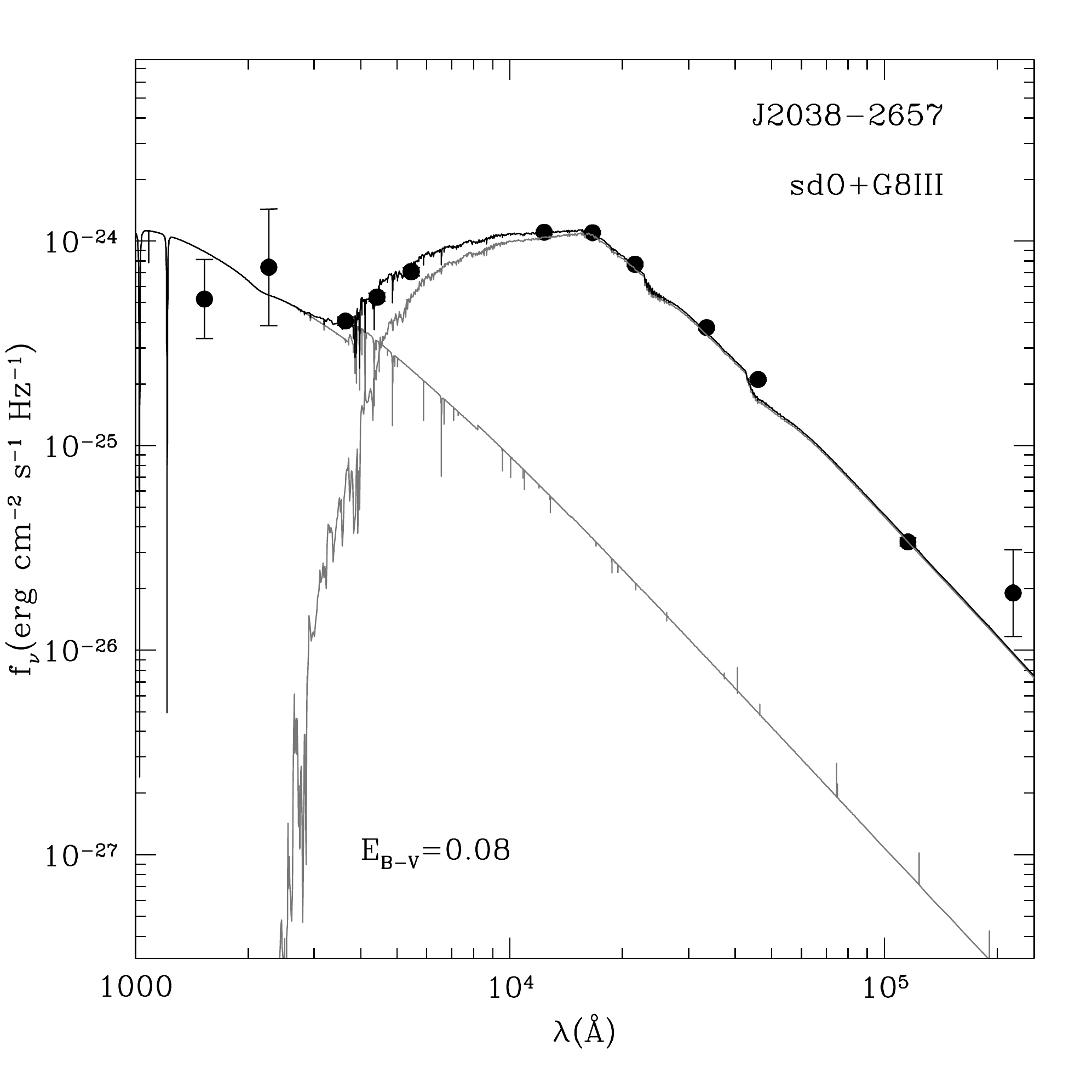}
\caption{Spectral energy distribution of GALEX~J2038$-$2657 combining a cool G8III secondary
and a sdO primary star (full line). Individual contributions are shown with grey lines. The effect of interstellar extinction ($E_{B-V}=0.08$) is included.
\label{fig_2038_sed}}
\end{figure}

GALEX~J2038$-$2657 is a relatively luminous hot sdO star with a G type companion \citep{nem2012}. 
Our spectroscopic observations revealed variability in the H$\alpha$ profile 
(Fig.~\ref{fig_2038}) on a time-scale of a day or less. However, cross-correlation
measurements in the spectral series dominated by the G companion show little variations with a dispersion $\sigma_v=7.8$\,\kmps\ 
comparable to the expected accuracy of the wavelength scale. The measurements imply
that the velocity semi-amplitude of the G8III star does not exceed $\approx16$\,\kmps.
Fig.~\ref{fig_2038_sed} shows the SED of 
GALEX~J2038$-$2657 where the ultraviolet range is dominated by the hot subdwarf
and the optical range by the red giant.
The SWASP time series (1SWASP~J203850.49$-$265754.2) reveals variations of 12~mmag semi-amplitude over a period of 1.87022~d (Fig.~\ref{fig_swasp}) that are confirmed by similar
variations in a short NSVS time series. Again, no significant variations were observed in a nearby comparison object
(1SWASP~J203851.22$-$265943.1).
These variations are most likely linked to the observed spectroscopic variability, 
but they cannot yet be clearly associated to a possible orbital period.

The system shares some properties with the sdB plus K\,III-IV system HD~185510 \citep{jef1992,fek1993} and 
the sdO plus K0\,III system FF~Aqr \citep{vac2003}. All three systems have an evolved secondary star, from
sub-giant to giant, and all three are photometrically variable. However, the hot subdwarf in HD~185510 is
possibly the progenitor of a low-mass, helium white dwarf \citep[0.3\,\msun, ][]{jef1997}. Photometric
variations in HD~185510 and FF~Aqr coincide with the orbital period and are caused by irradiation of the
exposed hemisphere of the secondary stars. Moreover, the orbital periods HD~185510 and FF~Aqr are 20.7 and 9.2~d, respectively,
with orbital separations of $\sim$43 and $\sim$25\,\rsun, respectively, and well outside the radius of a 
sub-giant or giant star ($R({\rm K0\,III})\approx 16$\,\rsun).

Without an estimate of the orbital period we can only set limits to the orbital parameters, such as the binary separation. The identification of the
late-type giant secondary is based on spectral decomposition \citep{nem2012}: The absolute $V$ magnitude of the hot sdO star 
is only about $\sim$1.0\,mag fainter than its companion. Adopting a G8\,III type from the spectral decomposition shown in Fig.~\ref{fig_2038_sed},
the absolute magnitude of the companion is $M_V$(G8\,III)$=0.9$, implying an absolute magnitude $M_V$(sdO)$=1.9$ for the primary in agreement
with the estimate of \citet{nem2012}, $M_V$(sdO)$=2.0$. The minimum orbital period for a systemic mass of 2-3\,\msun\ and an orbit outside the
G8\,III radius (15\,\rsun) is $P\gtrsim$3.1\,d.  
Adopting a radius of 15\,\rsun for the G8\,III star, the photometric period of 1.87~d implies a rotation velocity $\varv_{\rm rot}\approx350$\,\kmps. 
The narrowest features in the SSO spectra have a width of $\varv_{\rm rot}\sin{i}=130$\,\kmps, that would enforce
a low inclination $i\lesssim 22^\circ$. High dispersion spectroscopy is necessary to help determine the orbital parameters and
help clarify the origin of
the photometric variations. 
The most likely scenario is that the photometric variations are caused by a surface spot coupled to the rotation of the
star, and that the orbital period probably exceeds several days with a low velocity amplitude ($K\lesssim$20\,\kmps).

\subsection{Notes on other objects from this survey}

GALEX~J0047$+$0958 (HD 4539) is a well known hot sdB star \citep[see, e.g., ][]{kil1984}. Spectropolarimetric
measurements hint at the presence of a weak magnetic field 
\citep[$\sim 0.5$ kG][]{lan2012}. \citet{sch2007} reported line profile 
variations and radial velocity variations of a few \kmps\ that may be due 
to g-mode pulsations. \citet{lyn2012} obtained photometric series and measured
variations with a frequency of $9.285\pm0.003$~d$^{-1}$  and an amplitude of 
$0.0023\pm0.0003$ mag. This photometric frequency is consistent with one of the
frequencies ($9.2875\pm0.0003$~d$^{-1}$) determined from low-amplitude radial velocity 
variations, and both are possibly associated to stellar pulsation. 

The SED of GALEX~J0049+2056 (Fig.~\ref{fig_sed1}) shows an IR excess that could be attributed to a yet unidentified companion
or nearby object, or to a dusty environment.

The sdB GALEX~J0059$+$1544 (PHL~932) is 
embedded in an emission nebula. However, \citet{fre2010} have shown that
the association is only coincidental, but that PHL~932 does contribute and ionize a dense 
region of the ISM surrounding it. 
The SED of this object shows, as in the case of GALEX~J0049+2056, a considerable IR excess (Fig.~\ref{fig_sed1}).
Several radial velocity measurements of PHL~932 were reported in literature.
\citet{arp1967} measured $15\pm20$\,\kmps\ using two low-dispersion spectra.
\citet{ede2003a} measured $18\pm2$\,\kmps\ using echelle spectra. These
velocities are in agreement with our measurements ($\bar{\varv} = 16.7$, 
$\sigma = 3.1$\,\kmps) and, therefore, it does not appear that PHL~932
is in a close binary.
\citet{gei2012a} report a rotational velocity of $v_{\rm rot} = 9.0\pm1.3$\,\kmps.
\citet{lan2012} obtained spectropolarimetric measurements of PHL~932 but did not detect a magnetic field with an
upper limit of $\sim$300 G.

\citet{bro2008} classified GALEX~J0206$+$1438 (CHSS~3497) as a hot subdwarf. 
Our radial velocity measurements vary only marginally ($\sigma_\varv < 10$\,\kmps), and  we do not dispose of 
sufficient data to determine a period. A radial velocity measurement of 
$V_{r} = 7\pm16$\,\kmps\ was obtained by \citet{bro2008} and is 
consistent with our measurements ($\bar{\varv}=13.8,\ \sigma_\varv=7.5$\,\kmps). 

GALEX~J0232$+$4411 (FBS 0229$+$439) was classified as a sdB star in the First 
Byurakan Survey of blue stellar objects \citep{mic2008}.

\citet{cop2011} presented a set of radial velocity measurements for GALEX~J0401-3223 which suggest
that the sdB star is in a close binary system, but 
were unable to determine the orbital period with limited data. 
\citet{cop2011} measured an average velocity and dispersion of $\varv\pm\sigma_\varv=55.2\pm4.4$\,\kmps, consistent with our own measurements.
We have combined the \citet{cop2011} data with ours and conducted a period search. We found a best period of 1.8574~d, however
two significant aliases at $P = 0.64$ and 0.066~d cannot be ruled out. The velocity semi-amplitude at all three
periods does not exceed 10 km~s$^{-1}$ and excludes a white dwarf or late-type companion. The relatively short period and low velocity amplitude imply a minimum mass in the
substellar range, 0.01-0.04\,\msun. The SWASP data folded on the best period (1.8574~d) constrain photometric variations to a 
semi-amplitude of only 1~mmag,
or 8~mmag when folded on any periods larger than 0.01~d. The expected variations due to a substellar companion 
would be as low as 6~mmag at the two longest periods or 20~mmag at the shortest, but are all significantly larger than
the SWASP limit. It is not possible to describe the companion with present data, although a substellar companion
is a distinct possibility.

\citet{ost2010a} obtained series of photometric observations of 
GALEX~J0500$+$0912 in order to search for pulsations and concluded that 
it is not photometrically variable. Our limited radial velocity data set does not
indicate variability.

An inspection of the acquisition images of GALEX~J0657$-$7324 shows a nearby companion
and, therefore, the 2MASS and $WISE$ colours of the hot subdwarf are certainly contaminated.
\citet{hei1992} reported that GALEX~J0657$-$7324 (HEI 714)
is a visual double star with a separation of 1.9~arcsec, and our own acquisition image locates the companion
1.8~arcsec away at a position angle of 270$^\circ$. Also, our optical spectra do not
appear to be contaminated by this object and do not indicate variability.

GALEX~J1845$-$4138 is a relatively cool He-rich subdwarf displaying a strong He\,{\sc i}
line series and weaker Balmer lines. The velocity measurements based on He\,{\sc i}6678.154
($\varv=-59.7\pm3.3$\,\kmps) are consistent with the measurements based on H$\alpha$ and
do not suggest any variability.

GALEX~J1902$-$5130 is a helium sdO star. \citet{lan2012} obtained spectropolarimetry
of GALEX~J1902$-$5130 with a measurement that shows that this star does
not have a magnetic field down to a few hundred gauss.
Our radial velocity measurements suggest there may be long period, low-amplitude
variations. The measurements are based on He\,{\sc ii} 6560.088\AA.
The object is very hot and
our spectra display He\,{\sc i} emission.

GALEX~J1911$-$1406 is also a very hot He-rich subdwarf. The velocity measurements
are based on He\,{\sc ii}6560.088\AA.

\citet{gei2012a} report a rotational velocity of $v_{\rm rot} = 8.6\pm1.8$\,\kmps\ for GALEX~J2153-7003.
\citet{cop2011} obtained several radial velocity measurements of this star 
and found that it is not variable. Their average velocity and dispersion, $39.4\pm7.5$\,\kmps, 
are in a close agreement with our own measurements ($43.4\pm4.2$\,\kmps).

GALEX~J2344-3426 is a well known sdB star. \citet{ost2010a} obtained photometric series
of the star and found it to be non-variable. \citet{gei2012a} measured a rotational
velocity of $v_{\rm rot} = 7.3\pm1.0$\,\kmps. \citet{mat2012} and \citet{lan2012} obtained
spectropolarimetry of the star and constrained the longitudinal field to 261 G and $246\pm232$ G, respectively.

\section{Discussion}

The new binary identifications are placed into context with a compilation of all known
spectroscopically identified binaries (Appendix C).
Table~\ref{tbl_sdb_bin_all} lists the orbital parameters of these systems. The compilation includes hot subdwarfs
with an unseen companion and spectroscopically identified late- to early-type companions in a range
of periods from 0.05\, to 1363\,d. Table~\ref{tbl_sample_kine} lists the properties of the primary
as well as kinematical properties of the systems.

Throughout this discussion we assume for most subdwarfs a mass of 0.47\,\msun\ with
a few exceptions such as the ELM progenitors (0.23\,\msun) and hot sdO companions to early-type stars (1\,\msun).
Using pulsating properties of sdB stars and binary systems for which the sdB mass 
was measured \citet{fon2012} found that the average mass of a sdB star is 
0.47\,\msun\ with a standard deviation of 0.031\,\msun. \citet{zha2009}
found that most sdB stars have a mass between 0.42 and 0.54\,\msun\ and
an average mass of about 0.50\,\msun. \citet{zha2009} used evolutionary
models and the parameters of a sample of 164 sdB stars.
                   
\subsection{Properties of known binaries: Period and mass function}

\begin{figure}
\includegraphics[width=1.00\columnwidth]{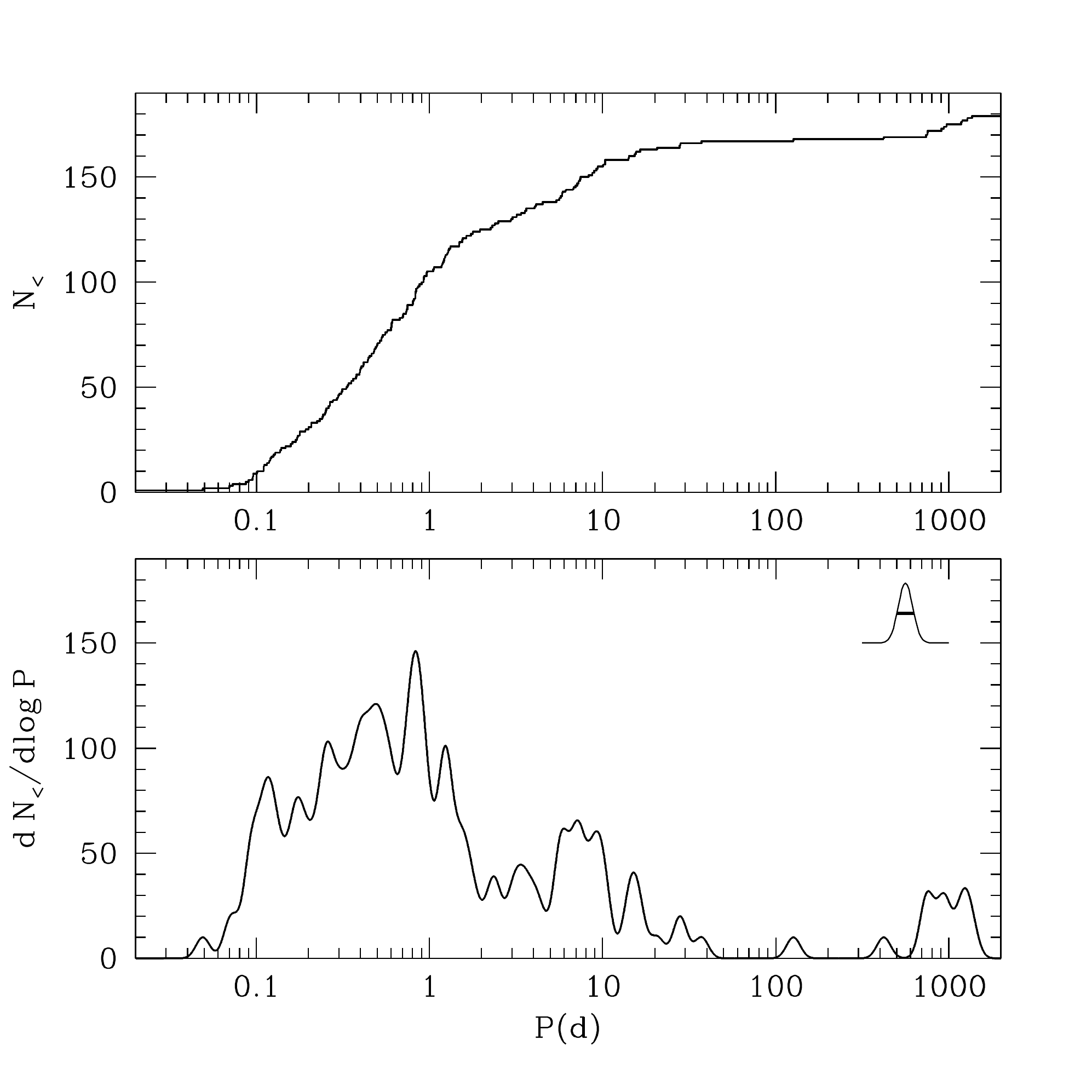}
\caption{Cumulative function of period (top), $N_<$ versus $\log{P}$, and its first derivative (bottom).
The derivative was smoothed with a Gaussian function ($FWHM=0.1$ dex), and shown in the
upper right corner.
\label{fig_cumul}}
\end{figure}

Fig.~\ref{fig_cumul} (top) shows the cumulative distribution of orbital periods in the
population of binaries with a hot subdwarf primary. The derivative of the function with
respect to the logarithm of the period provides an estimate of the period distribution (Fig.~\ref{fig_cumul}, bottom). 
Several peaks stand-out,
particularly at 0.1, 0.5-2.0, 10, and 1000 days. The last two peaks are clearly separated by
a gap within which few binaries are known: A few Be stars with a hot subdwarf companion populate the
gap and the distribution includes spectroscopically (UV) confirmed Be+sdO \citep{pet2008,pet2013}.
We noticed a hint of a hierarchy in the period distribution: The shortest periods coincide with dM 
companions emerging from a CE phase (low mass ratio $M_2/M_1 < 1/2$), 
white dwarfs ($M_2/M_1 \approx 1$), and,
at high mass ratio ($M_2/M_1>2$), subdwarfs with a subgiant/giant or Be companion, and, finally, 
subdwarfs with a FGK companion at the longest periods and emerging from a RLOF. 
Following a RLOF, the orbital separation increases the least for more massive companions. It is remarkable 
that main-sequence A-type star companions are still missing although they are predicted in population
syntheses \citep{han2003}. Subdwarfs with A-type main sequence companions should be detectable as UV excess
objects or as low-amplitude radial velocity variations similar to Be+subdwarf binaries.
In summary, binaries near the main peak are mostly white dwarfs plus subdwarfs after possibly two episodes of mass-transfer.
Note that the orbital parameters of many systems with large velocity amplitudes remain unresolved \citep[see, e.g., ][]{cop2011,gei2011a} and are not included in this analysis.

\begin{figure}
\includegraphics[width=1.00\columnwidth]{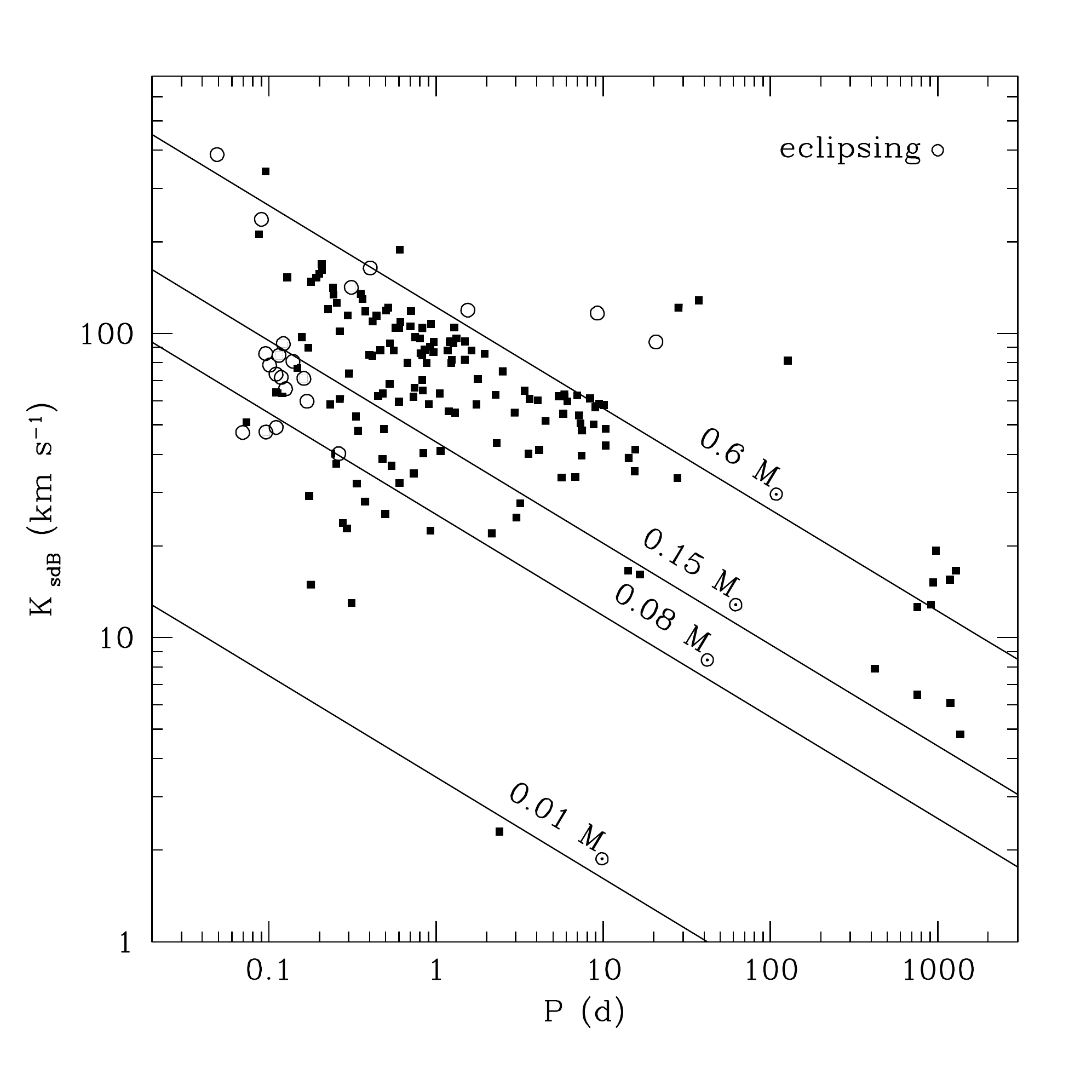}
\caption{Measured velocity amplitude versus period with a sub-sample of
eclipsing binaries shown with open circles. Full lines are labelled with
the mass of secondary stars which were computed for $i=90^\circ$.
\label{fig_eclipse}}
\end{figure}

Fig.~\ref{fig_eclipse} shows the sample of known binaries in the velocity amplitude versus
period plane. Most eclipsing systems have secondary masses between 0.08 to 0.15\,\msun. Systems
with known white dwarf secondaries have secondary masses close to or above 0.60\,\msun.

\begin{figure}
\includegraphics[width=1.00\columnwidth]{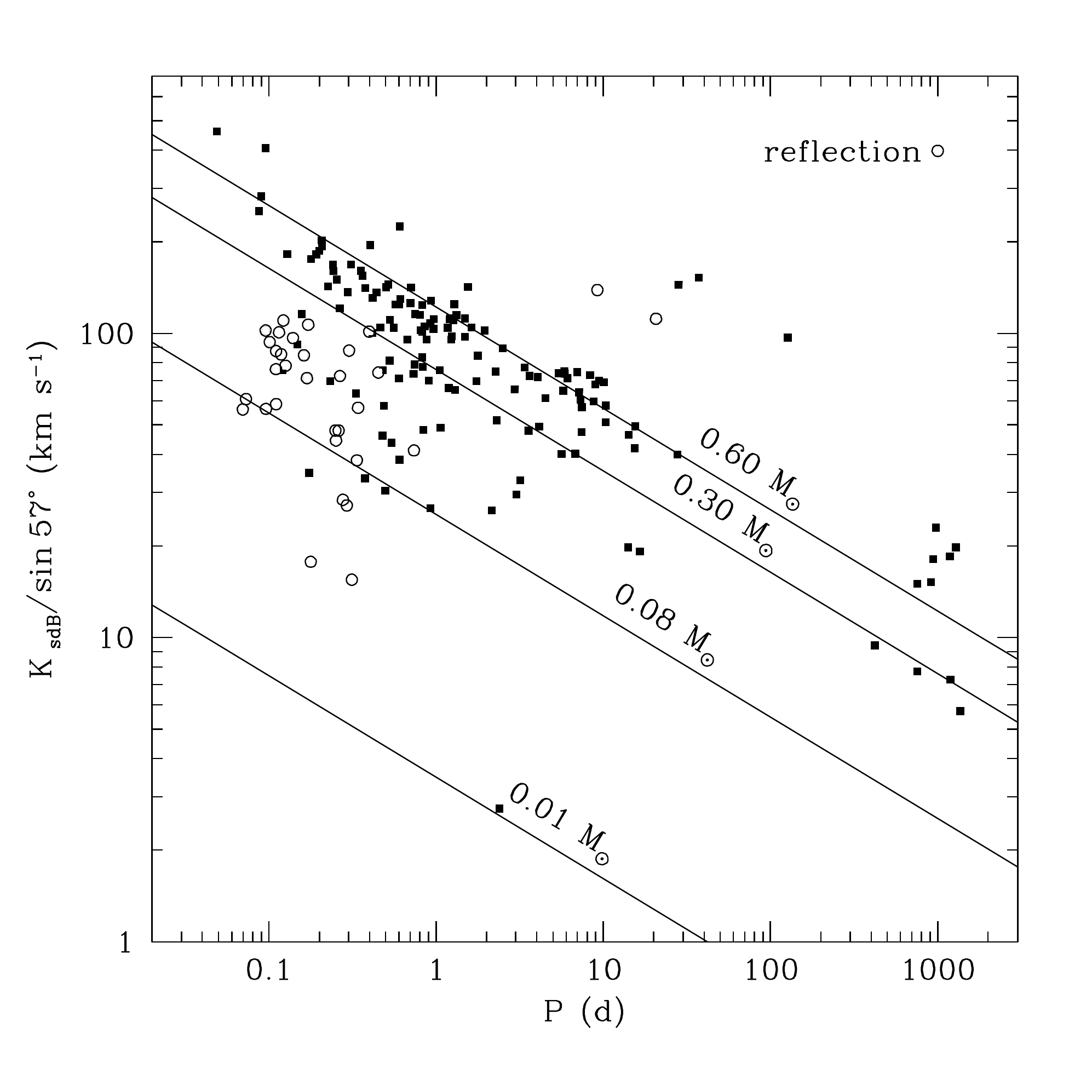}
\caption{Same as Fig.~\ref{fig_eclipse} but with
reflection binaries shown with open circles. The velocity scale is corrected for 
the effect of an average inclination of $57^\circ$. Secondary masses cluster
between 0.08 and 0.30\,\msun.
\label{fig_reflect}}
\end{figure}

\begin{figure}
\includegraphics[width=1.00\columnwidth]{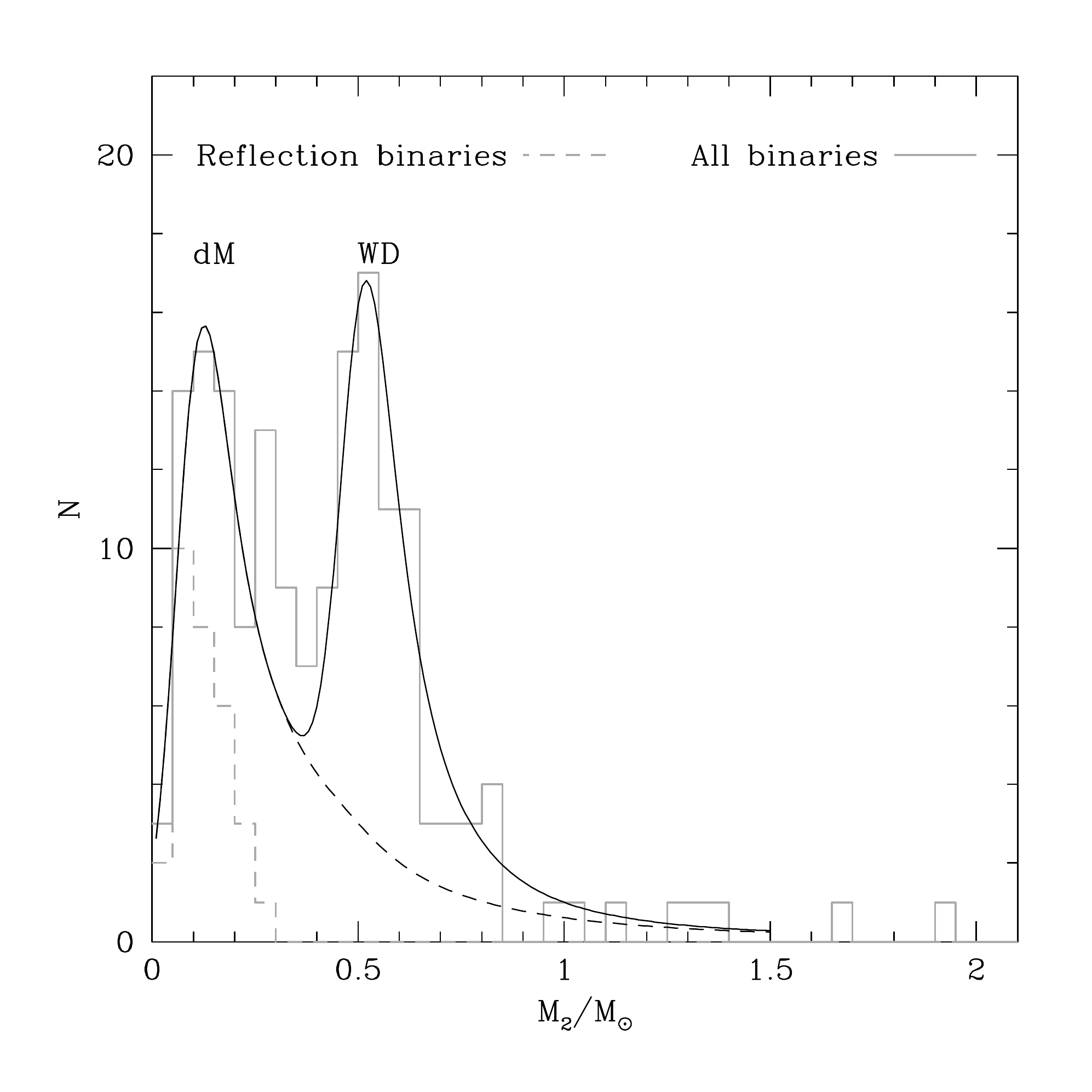}
\caption{Mass distribution of all known binaries with a hot subdwarf primary star as a function of the secondary mass, assuming
an average inclination of $57^\circ$ (full histogram). The peak distribution
of low-mass stars is marked ``dM'' and that of white dwarfs, ``WD''. Binaries showing reflection effect in their light curves
are shown with a dashed histogram. The full lines show synthetic distributions smoothed to two-bins width for a combination of late-type stars and white dwarfs (double-peaked full line), and
that excluding white dwarfs (dashed line).
\label{fig_mass_sec}}
\end{figure}

Fig.~\ref{fig_reflect} also shows the sample of known binaries in the velocity amplitude versus
period plane but with the velocity
scale corrected for an average inclination of $57^\circ$. The correction allows to draw class properties but should not be applied
to individual objects. Secondary masses for systems showing a reflection effect range, with the exception of FF~Aqr and HD~185510, from
0.08 to 0.30\,\msun. Remarkably, secondary masses for most non-reflecting systems cluster near 
0.60\,\msun\ and the unseen objects are probably white dwarfs. Secondary stars in the long-period range
and with masses in excess of 0.60\,\msun\ are identifiable as G and K stars.
All eclipsing systems with $K<100$\,\kmps, i.e., with an estimated $M_2<0.3$\,\msun, also show a reflection effect
indicative of a late-type secondary, while the remaining systems cluster at a higher secondary mass $M_2\approx0.6$\,\msun\ and
almost certainly harbour a white dwarf secondary.
Fig.~\ref{fig_mass_sec} shows secondary mass distribution assuming average system inclination of $57^\circ$.
This distribution may be described by a superposition of two power laws: A shallow distribution with $M_2>0.08$\,\msun, i.e., $\alpha=1.3$ between
0.08 and 0.5\,\msun\ and $\alpha=2.3$ above 0.5\,\msun\ following the initial mass function $\xi(m)\propto m^{-\alpha}$ of \citet{kro2001}, and
a steeper distribution ($\alpha\approx 6$) with $M_2>0.48$\,\msun. The former power law encompasses mostly M-type dwarfs, many
of them showing a reflection effect, while the latter encompasses white dwarfs 
in the 0.5-1.0\,\msun\ mass
range. This simplified white dwarf distribution represents well the peak and high-mass tail of the white dwarf mass distribution \citep[see, e.g., ][]{kep2007}
 but excludes possible low-mass white dwarfs ($<0.48$\,\msun).
On the other hand, the late-type stellar mass distribution follows the initial mass function and the expectation of a randomly drawn set of late-type stars. 
The reflection effect is common in short-period binaries ($P<0.5$~d) but is relatively rare at longer periods (see Fig.~\ref{fig_reflect}) due to increased binary separations and weaker
photometric variations: The actual late-type mass distribution appears as a scaled-up version of the secondary mass distribution in the
sub-sample of reflection binaries, but it also includes longer period binaries with an indiscernible reflection effect.
Note that a third narrow peak is possibly present at $\approx 0.3$ \msun. 
Since this peak is mostly made up of companions that do not show the reflection
effect, the origin of this peak maybe be low-mass white dwarfs. It may also be 
due to an incorrect mass estimate of the subdwarf, for example an ELM 
progenitor is assumed rather than a normal subdwarf.
Most objects (60-70 per cent) are low mass main sequence stars, while 30-40 per cent are white dwarfs.

Our own survey delivered a 37 per cent fraction of hot subdwarfs in close binaries (Section 3.1.3).
Our survey strategy was aimed at and successfully uncovered short-period binaries. Fig.~\ref{fig_reflect}
shows that setting our detection threshold at 10~\kmps\ would have allowed for the detection of any stellar companion with an
orbital period $\lesssim20$~d, any late-type companion with a mass of 0.3\,\msun\ and $P\lesssim600$~d, or just about any white dwarf companion.
However, a close examination of data sampling shows that 60 per cent were obtained with a span of 2 days or less
with another 30 per cent with a span of 100-400 days, i.e., during a subsequent observing run or season.
Systems with periods of 10-20 days or longer would have only been partially covered and most likely avoided detection. Note that
the longest period detected in our survey is only 3~d long. Setting our detection threshold at 10~d,
i.e., some 155 objects out of 179 known binaries, the total yield including longer period binaries could be $\approx$15 per cent larger for 
a total binary fraction of 43 per cent. The four additional hot subdwarfs with composite spectra (see Table~\ref{tbl_sum}) which are likely to have longer orbital
periods ($\gtrsim 10$~d) are such objects.

\subsection{Properties of known binaries: Kinematics}

We calculated the Galactic velocity components ($U,V,W$), which are relative to the local standard of rest (LSR),
of all known hot subdwarf binary systems (listed in 
Appendix~C) using their positions, systemic ($\gamma$) velocities, proper
motions and apparent magnitudes. 
We adopted the right-handed system for the velocity components, where $U$ is positive in the direction of the Galactic
centre, $V$ is positive in the direction of Galactic rotation and $W$ is positive toward the North Galactic Pole.  
We assumed that the solar motion relative to the LSR is ($U,V,W$) = (10.1,4.0,6.7) km~s$^{-1}$
as determined by \citet{hog2005}.
The distribution of systemic velocities, i.e., radial velocities, follows $N\approx e^{-\mid\gamma\mid/\sigma}$, where
$\sigma_r=41$\,\kmps. The $\sigma_r$ value is the one-dimension equivalent of the two-dimension transversal 
velocity dispersion $\sigma_T=59$\,\kmps\ measured by \citet{ven2011}, where $\sigma_T=\sqrt{2}\,\sigma_r$.
In their study of the kinematics of EHB stars,
\citet{alt2004} measured significantly larger radial velocities with a distribution following $\sigma_r=65$\,\kmps\ compared to our sample, but 
they measured a 
transversal velocity distribution consistent with the present one.
 
To calculate the Galactic velocity vectors, we employed the method outlined in
\citet{joh1987} using as an input the radial velocity, proper motion \citep{zac2013} and distance measurements. 
We determined the distance toward each 
subdwarf using the distance modulus $V-M_V=5\log{d}-5$, where the magnitude $V$ is listed in Table~\ref{tbl_sdb_bin_all}. 
We estimated the absolute magnitude $M_V$ from the measured stellar parameters, i.e.,
\begin{displaymath}
M_V=-2.5\log{(4\pi\Omega\,\bar{H}_V)},
\end{displaymath}
where $\Omega = r^2/d^2$, with $d=10$~pc and $r^2 = GM/g$. The Eddington flux is averaged ($\bar{H}_V$) over the Johnson $V$ transmission curve. 
We assumed $M=0.23$\,\msun\ for low-mass objects, 0.47\,\msun\ for normal objects and 1.0\,\msun\ for subdwarfs in massive binaries,
and the published surface gravity measurements were usually obtained using a spectroscopic method similar to that described in Section 3.1. 

\begin{figure}
\includegraphics[width=1.00\columnwidth]{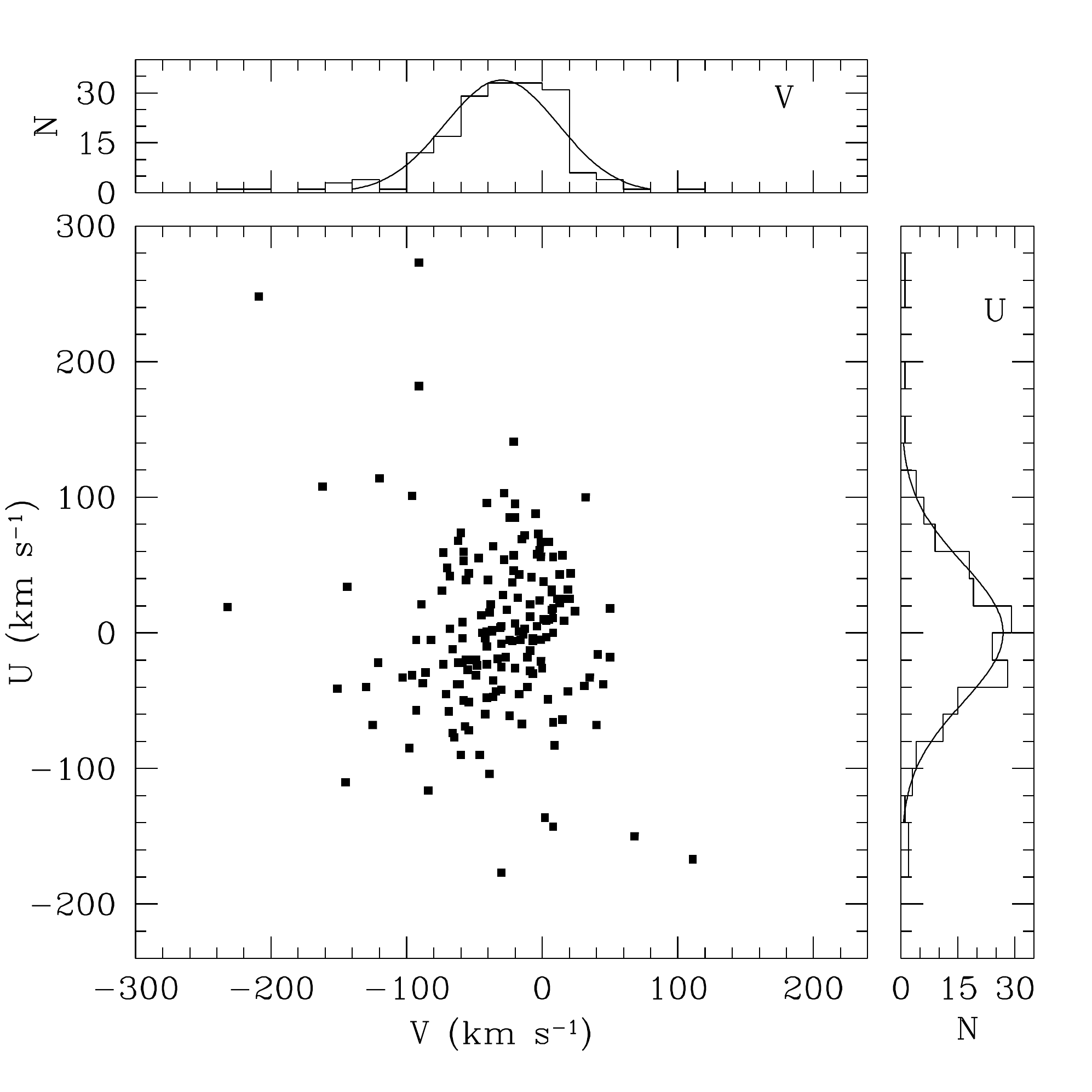}
\caption{Galactic velocity vectors $U$ and $V$ of all known binaries 
containing a hot subdwarf. The individual distributions are shown in the upper panel ($V$)
and right panel ($U$). Details of the measurements are shown in Table~\ref{tbl_kin} and
discussed in Section 4.2.
\label{fig_UV}}
\end{figure}

Fig.~\ref{fig_UV} shows the $U$ and $V$ velocity components for all
hot subdwarf binary systems. Table~\ref{tbl_kin} lists the ($U,V,W$) velocity
components and dispersions for the sample of known binaries. 
The distribution appears asymmetric with several objects trailing at large negative
Galactic $V$ velocity.
We computed the straight average and dispersion (``All'') but excluding the extreme case of
OGLE~BUL-SC16335. We also fitted the distributions with Gaussian functions excluding outliers, i.e., all bins with less than three members (`$N>3$`). The sample velocity dispersion is  
significantly smaller than that calculated by \citet{alt2004} for single hot subdwarf stars
($\sigma_U = 74$\,\kmps, $\sigma_V = 79$\,\kmps, $\sigma_W = 64$\,\kmps).
However, our velocity dispersion is in better agreement with the
dispersion ($\sigma_U = 62\pm8$\,\kmps, $\sigma_V = 52\pm7$\,\kmps, 
$\sigma_W = 59\pm8$\,\kmps) calculated by \citet{deb1997} based on
a sample of 41 hot subdwarf stars.
Possible explanations for the inflated Galactic velocities of \citet{alt2004} 
are that their sample included yet unidentified binaries or that lower dispersion
spectroscopy resulted in larger measurement errors.
In summary, the hot subdwarf population and the confirmed binaries among them are 
kinematically indistinguishable and drawn from the
same, general population of EHB stars.

Also, Table~\ref{tbl_kin} compares results for the hot subdwarf population
with that of field white dwarf stars.
\citet{pau2006} list 361 thin disc members
and 27 thick disc members, while \citet{kaw2012b} list 57 old disc white dwarfs, 
which is a mix of old thin-disk and thick-disk populations, and at least
one halo white dwarf.
A comparison of the velocity dispersions shows that the binary population may be an even older population than
field white dwarf stars with some population members belonging to the old disc or even the halo.

Four systems display peculiar kinematics. Three of these, SDSS~J1622$+$4730,
PHL~861 and SDSS~J1505+1108, have Galactic velocities that make them halo
candidates with a few additional objects lagging in the $V$ component making them thick-disk candidates. 
The fourth object, OGLE~BUL-SC16335, is in a crowded field and 
there is a possibility that the proper motion measurements are incorrect.

\citet{bar2013a} calculated kinematics for 5 long period systems and found
that two of these (PG~1449$+$653 and PG~1701$+$359) have kinematics suggesting 
that they belong either to the thick disc or halo. They also report that
there is a high probability that PG~1104$+$243 belongs to the thick disc. Our
calculated Galactic velocities are similar to those of \citet{bar2013a}.
Note that \citet{bar2013a} includes the disc rotation (220\,\kmps)
in their $V$ velocity.

A comparison of the velocity components of the hot subdwarf binary population
to that of \citet{kaw2012b} shows that the dispersion is larger for all
velocity components than that of the white dwarf population, suggesting that the
subdwarf population appears to be older than the white dwarf population. Finally, a comparison
with the work of \citet{pau2006} shows that the hot subdwarf binary population has
a velocity dispersion between the thin and thick disc dispersions for white
dwarfs.

\subsection{Low mass subdwarfs as progenitors of extremely low-mass white dwarfs}

The first known ELM white dwarf progenitor, HD~188112, was discovered by 
\citet{heb2003}. As part of their survey of ELM white dwarfs, \citet{kil2011}
found that SDSS~J1625$+$3632 is similar to HD 188112. Other recently 
discovered systems are KIC~6614501 \citep{sil2012} and NGC~6121-V46 \citep{oto2006a}. 
Our radial velocity survey adds one more object to the small sample of 
ELM white dwarf progenitors: \citet{ven2011} showed that GALEX~J0805$-$1058
has atmospheric properties representative of ELM white dwarf progenitors ($M\lesssim$0.3\,\msun) and,
therefore, it was selected for radial velocity follow-up measurements. New radial velocity measurements
proved that GALEX~J0805$-$1058 is in a close binary, 
and, because it also lies below the ZAEHB (see Section 3.1), we conclude that it is a genuine ELM white dwarf progenitor. It is also the first ELM white dwarf progenitor without
a more massive white dwarf companion, and this is likely to have significant implications for the origin and
evolution of ELM white dwarfs.
Two other objects from our sample lie below the ZAEHB, J1411+7037, which is paired with an F star, and J2153$-$7004, although we did
not detect significant radial velocity variations.

\begin{table}
\centering
\begin{minipage}{\textwidth}
\caption{Kinematical properties of the hot subdwarf binary population. \label{tbl_kin}}
\renewcommand{\footnoterule}{\vspace*{-15pt}}
\renewcommand{\thefootnote}{\alph{footnote}}
\begin{tabular}{lcccccc}
\hline
Vel. \footnotemark[1]\footnotetext[1]{All velocities expressed in \kmps.} & $N>3$\footnotemark[2]\footnotetext[2]{This work, but excluding outliers} & All \footnotemark[3]\footnotetext[3]{This work} & sdB \footnotemark[4]\footnotetext[4]{\citet{alt2004}} & WD \footnotemark[5]\footnotetext[5]{\citet{kaw2012b}} & WD$_{\rm thin}$ \footnotemark[6]\footnotetext[6]{\citet{pau2006}} & WD$_{\rm thick}$ \footnotemark[6] \\
\hline
$\bar{U}$   & $0\pm5$      & $2$   & $-8$  & $-7.8$  & ...  & ... \\
$\sigma_U$  & $52\pm3$     & 62    & $74$  &  $42.8$ & $34$ & $79$ \\
$\bar{V}$   & $-30\pm2$    & $-32$ & $-37$ & $-40.1$ & ...  & $-52$ \\
$\sigma_V$  & $42\pm2$     & 47    & $79$  & $31.9$  & $24$ & $36$ \\
$\bar{W}$   & $-5\pm3$     & $-6$  & $12$  & $-5.9$  & ...  & ... \\
$\sigma_W$  & $34\pm2$     & 41    & $64$  & $27.4$  & $18$ & $46$ \\
\hline
\end{tabular}
\end{minipage}
\end{table}

\subsection{Summary}

We presented an analysis of the orbital properties of seven new systems 
comprising a hot subdwarf primary. The secondary in one system is a late-type 
star showing a reflection effect (GALEX~J2205$-$3141), while we found evidence 
that the secondary star in GALEX~J0805$-$1058 is a very low mass M dwarf or 
possibly a substellar object. The mass function of the other objects implies the 
likely presence of a white dwarf companion.
The period of photometric variability of two additional systems is, in the case 
of GALEX~J1736+2806 probably coincident with the orbital period, and, in the 
case of J2038$-$2657, probably coincident with the rotation period of the giant 
companion. Our survey results, taking into account the survey strategy, imply 
an incidence of binarity of $\sim$43 per cent in the hot subdwarf population.

We have compiled a list of all known hot subdwarfs in binary systems 
and performed a binary population analysis. We found that systems showing
the reflection effect have components that are of a lower mass 
(0.08 to 0.30\,\msun) than those that do not show the reflection 
($\sim$0.6\,\msun). It is very likely that the companion to the hot subdwarf
in most of the systems not showing the reflection effect in the short period 
binaries ($P \la 1$ d) are white dwarfs.
The inferred secondary mass distribution is a superposition of two approximate
power laws, one low-mass power-law ($\gtrsim$0.1\,\msun) and composed of low-mass
main-sequence stars, and another, high-mass power-law ($\gtrsim$0.5\,\msun) and primarily
composed of white dwarfs with a few early-type main-sequence stars. 
White dwarfs constitute $\approx$30-40 per cent of all binary companions.

We have calculated the Galactic velocity components for all known hot subdwarfs
in binary system and showed that this population may be older than the field
white dwarf population. In this sample, we found three systems that possibly
belong to the halo.

Future work will involve high-dispersion spectroscopic follow-up of low-velocity amplitude
binary candidates, and of binaries comprising a hot subdwarf and an early-type main-sequence, or giant
companion. 

\section*{Acknowledgments}

A.K. and S.V. acknowledge support from the Grant Agency of the Czech Republic
(P209/12/0217 and 13-14581S) and Ministry of Education, Youth and Sports (LG14013). 
We wish to thank S. Ehlerov\'a and team members for their assistance with
observations obtained with the MPG~2.2-m telescope at La Silla.
We thank the referee for a thorough report and for stimulating further work on the paper.
This work was also supported by the project RVO:67985815 in the Czech Republic.
This paper uses observations made at the South African Astronomical Observatory.
This paper makes use of data obtained from the Isaac Newton Group Archive 
which is maintained as part of the CASU Astronomical Data Centre at the 
Institute of Astronomy, Cambridge.
This publication makes use of data products from the Wide-field Infrared Survey Explorer, 
which is a joint project of the University of California, Los Angeles, and the Jet 
Propulsion Laboratory/California Institute of Technology, funded by the National Aeronautics and Space Administration.
This publication makes use of data products from the Two Micron All Sky Survey, which 
is a joint project of the University of Massachusetts and the Infrared Processing and 
Analysis Center/California Institute of Technology, funded by the National Aeronautics 
and Space Administration and the National Science Foundation.

\appendix
\section{Photometry and spectral energy distributions.}

\begin{figure*}
\includegraphics[width=1.10\textwidth]{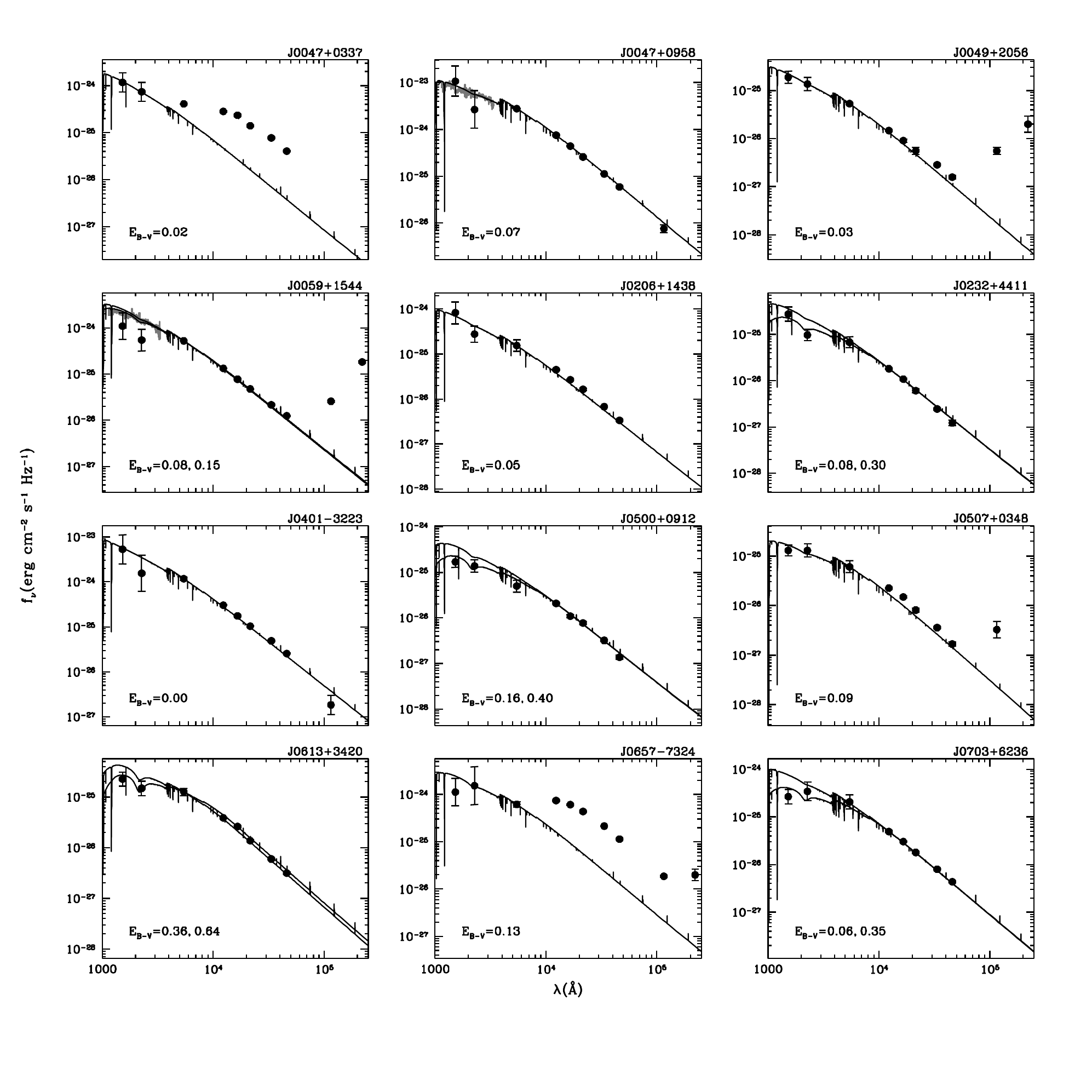}
\caption{Spectral energy distribution of the observed $GALEX$ sample.
\label{fig_sed1}}
\end{figure*}

\begin{figure*}
\includegraphics[width=1.10\textwidth]{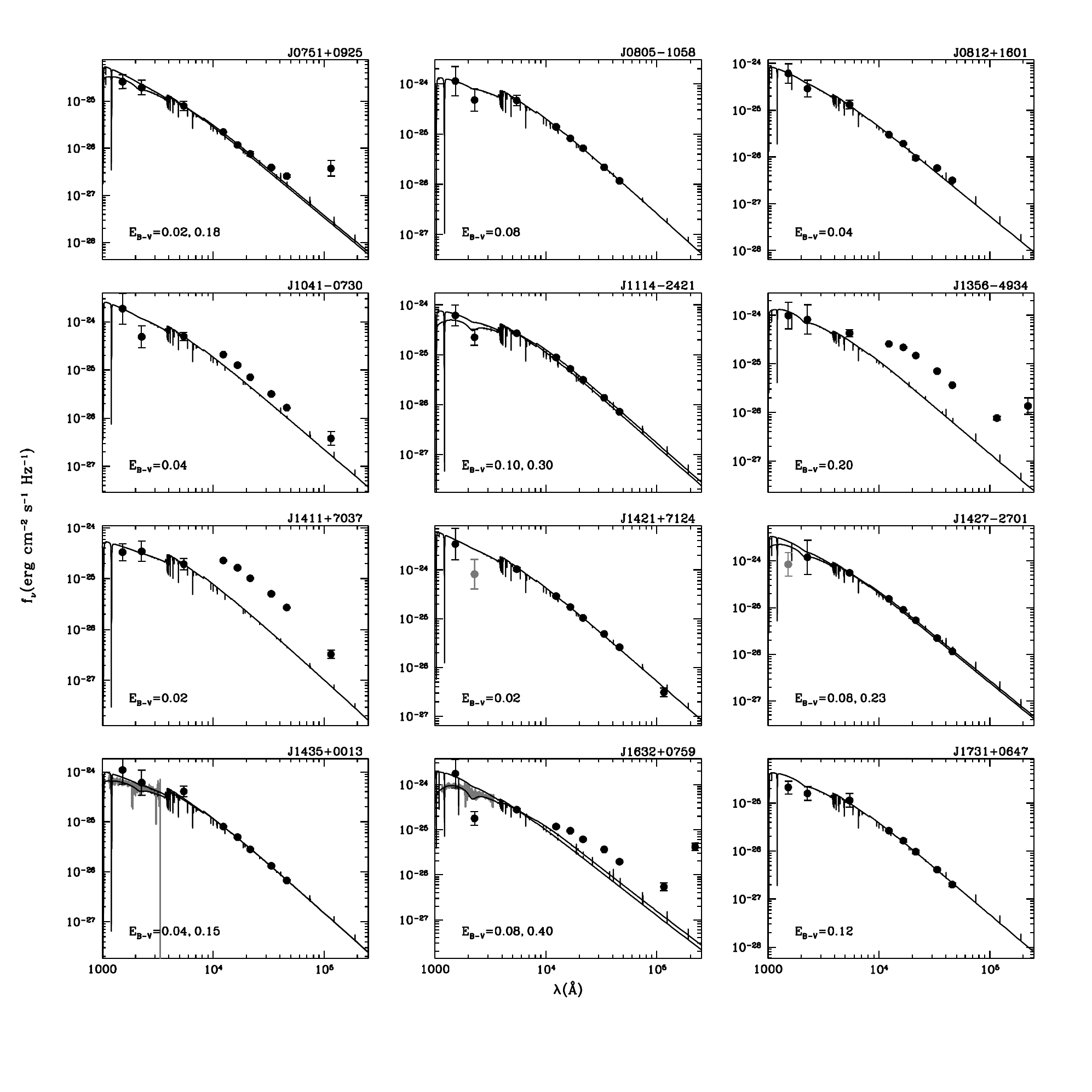}
\caption{Spectral energy distribution of the observed $GALEX$ sample.
\label{fig_sed2}}
\end{figure*}

\begin{figure*}
\includegraphics[width=1.10\textwidth]{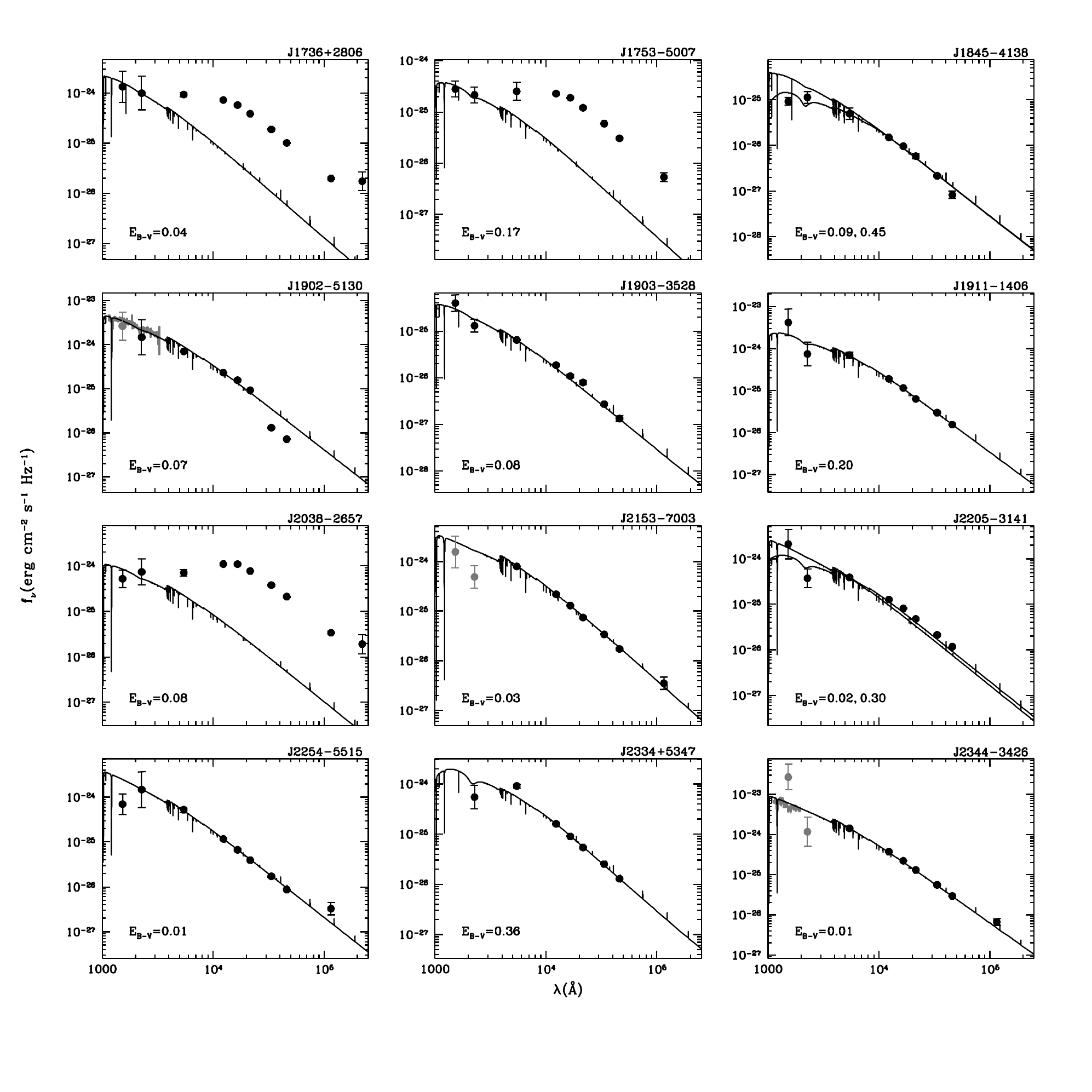}
\caption{Spectral energy distribution of the observed $GALEX$ sample.
\label{fig_sed3}}
\end{figure*}

Fig.~\ref{fig_sed1}, Fig.~\ref{fig_sed2} and Fig.~\ref{fig_sed3} show available photometry from Table~\ref{tbl_phot}
compared to model spectra. 
The extinction was determined by comparing the observed photometry with the
model spectrum. 
Extinction was measured toward several sdB stars \citep[e.g.,][]{moe1990,azn2001} 
and given that these stars are spread through out the Galaxy, i.e., at high 
latitudes and in the Galactic plane, the extinction coefficients $E(B-V)$ can 
vary from 0 to as much as 0.4. In these cases the extinction was measured
using International Ultraviolet Explorer ($IUE$) spectra combined with optical and 
infrared photometry. We used the parameterized extinction law as defined by 
\citet{car1989} with $R=3.2$.

We acquired from the Mikulski Archive for Space Telescopes (MAST) the following set of
$IUE$ spectra: 
\begin{itemize}
\item[] J0047+0958: swp26276mxlo and lwp07169mxlo,
\item[] J0059+1544: swp27142mxlo and lwp07144mxlo,
\item[] J1435+0013: swp23176mxlo and lwp03501mxlo,
\item[] J1632+0759: swp33790mxlo and lwp13481mxlo,
\item[] J1902-5130: swp17051mxlo and lwr13321mxlo,
\item[] J2344-3426: swp17981mxlo.
\end{itemize}

The model distributions include the effect of interstellar extinction with the line-of-sight
extinction coefficient obtained from \citet{sch1998}.
Also, we experimented with larger coefficients in an attempt to match SEDs showing possible
intrinsic absorption. For example, the SED of J1632+0759 reveals the presence of
an infrared flux excess as well as a larger extinction that revealed in \citet{sch1998}'s maps.

\begin{landscape}
\begin{table}
\hspace{-1.3cm}
\centering
\begin{minipage}{230mm}%
\renewcommand{\footnoterule}{\vspace*{-15pt}}
\caption{Photometric measurements. \label{tbl_phot}}
\vspace{-0.3cm}
\begin{tabular}{lcccccccccc}
\hline
                  & $F_{UV}$ & $N_{UV}$ & $V$ & $J$ & $H$ & $K$ & $W1$ & $W2$ & $W3$ & $W4$  \\
                  & 1528 \AA & 2271 \AA & 5455 \AA & $1.235\ \mu$m & $1.662\ \mu$m & $2.159\ \mu$m & $3.353\ \mu$m & $4.603\ \mu$m & $11.561\ \mu$m & $22.088\ \mu$m \\
                  & (mag)            & (mag)            & (mag)            &    (mag)           &   (mag)          &     (mag)        & (mag)            & (mag)              & (mag) & (mag) \\
GALEX~J           & & & & & & & & & & \\
\hline
004759.6$+$033742 & $11.24\pm 0.52$&$11.74 \pm0.40$     & $12.352\pm0.013$ & ${\it 11.880\pm0.036}^a$ & ${\it 11.629\pm0.041}^a$ & ${\it 11.697\pm0.038}^a$ & $11.498\pm0.034$ & $11.569\pm0.039$ & ... & ... \\
004729.4$+$095855 & $ 8.84\pm 0.80$&$10.34 \pm1.00$     & $10.272\pm0.004$ & $10.816\pm0.023$   & $10.939\pm0.032$ & $11.027\pm0.020$ & $11.089\pm0.023$ & $11.148\pm0.021$ & $11.449\pm0.209$ & ... \\
004917.2$+$205640 & $13.22\pm 0.32$&$13.56 \pm0.34$     & $14.559\pm0.009$ & $15.091\pm0.043$   & $15.153\pm0.078$ & $15.189\pm0.190$ & $15.081\pm0.037$ & $15.077\pm0.086$ & $11.796\pm0.194$ & $9.047\pm0.413$ \\
005956.7$+$154419 & $11.30\pm 0.71$&$12.06 \pm0.59$     & $12.076\pm0.004$ & $12.696\pm0.021$   & $12.818\pm0.030$ & $12.865\pm0.028$ & $12.874\pm0.024$ & $12.838\pm0.027$ & ... & ... \\
020656.1$+$143900 & $11.62\pm 0.62$&$12.80 \pm0.44$     & $13.41\pm0.31$   & $13.874\pm0.024$   & $13.976\pm0.038$ & $14.017\pm0.058$ & $14.139\pm0.030$ & $14.266\pm0.047$ & ... & ... \\
023251.9$+$441126 & $12.80\pm 0.38$&$13.93 \pm0.31$     & $14.30\pm0.29$   & $14.855\pm0.036$   & $14.963\pm0.056$ & $15.096\pm0.106$ & $15.249\pm0.049$ & $15.360\pm0.145$ & ... & ... \\
040105.3$-$322348 & $ 9.59\pm 0.80$&$10.92 \pm1.00$     & $11.20\pm0.08$   & $11.794\pm0.024$   & $11.937\pm0.024$ & $12.017\pm0.025$ & $12.000\pm0.023$ & $12.062\pm0.023$ & $12.992\pm0.525$ & ... \\
050018.9$+$091203 & $13.32\pm 0.30$&$13.55 \pm0.35$     & $14.63\pm0.33$   & $14.713\pm0.036$   & $14.951\pm0.079$ & $14.840\pm0.099$ & $14.960\pm0.041$ & $15.236\pm0.118$ & ... & ... \\
050735.7$+$034814 & $13.62\pm 0.26$&$13.62 \pm0.34$     & $14.42\pm0.30$   & $14.625\pm0.037$   & $14.620\pm0.055$ & $14.786\pm0.105$ & $14.837\pm0.041$ & $15.029\pm0.105$ & $12.374\pm0.407$ & ... \\
061325.3$+$342053 & $13.01\pm 0.35$&$13.48 \pm0.35$     & $13.63\pm0.18$   & $14.038\pm0.028$   & $14.011\pm0.041$ & $14.212\pm0.063$ & $14.283\pm0.033$ & $14.335\pm0.062$ & ... & ... \\
065736.7$-$732447 & $11.27\pm 0.72$&$10.93 \pm1.00$     & $11.90\pm0.14$   & $10.830\pm0.030$   & $10.578\pm0.032$ & $10.462\pm0.023$ & $10.387\pm0.022$ & $10.438\pm0.020$ & $10.477\pm0.052$ & $9.042\pm0.303$ \\
070331.5$+$623626 & $12.84\pm 0.37$&$12.56 \pm0.49$     & $13.10\pm0.37$   & $13.775\pm0.054$   & $13.842\pm0.033$ & $13.925\pm0.054$ & $13.964\pm0.030$ & $13.991\pm0.052$ & ... & ... \\
075147.1$+$092526 & $12.85\pm 0.37$&$13.17 \pm0.38$     & $14.12\pm0.26$   & $14.638\pm0.036$   & $14.865\pm0.049$ & $14.850\pm0.116$ & $14.728\pm0.036$ & $14.549\pm0.076$ & $12.221\pm0.414$ & ... \\
080510.9$-$105834 & $11.27\pm 0.72$&$12.21 \pm0.56$     & $12.21\pm0.27$   & $12.647\pm0.024$   & $12.764\pm0.023$ & $12.762\pm0.030$ & $12.871\pm0.024$ & $12.912\pm0.027$ & ... & ... \\
081233.6$+$160123 & $11.93\pm 0.52$&$12.74 \pm0.45$     & $13.57\pm0.22$   & $14.301\pm0.030$   & $14.326\pm0.054$ & $14.605\pm0.099$ & $14.301\pm0.032$ & $14.322\pm0.064$ & ... & ... \\
104148.6$-$073034 & $10.72\pm 0.80$&$12.17 \pm0.57$     & $12.14\pm0.21$   & $12.202\pm0.023$   & $12.295\pm0.027$ & $12.437\pm0.027$ & $12.468\pm0.024$ & $12.541\pm0.026$ & $12.211\pm0.363$ & ... \\
111422.0$-$242130 & $11.93 \pm0.52$&$13.03 \pm0.40$     & $12.80\pm0.01$   & $13.132\pm0.024$   & $13.248\pm0.029$ & $13.322\pm0.043$ & $13.380\pm0.026$ & $13.447\pm0.036$ & ... & ... \\
123723.5$+$250400$^b$ &     ...          & ...          & $10.509\pm0.002$ & $11.157\pm0.022$   & $11.270\pm0.030$ & $11.367\pm0.022$ & $11.400\pm0.023$ & $11.477\pm0.021$ & $11.591\pm0.184$ & ... \\
135629.2$-$493403 & $11.42\pm 0.67$&$11.62 \pm0.75$     & $12.30\pm0.17$   & $11.983\pm0.025$   & $11.705\pm0.029$ & $11.633\pm0.023$ & $11.601\pm0.023$ & $11.677\pm0.022$ & $11.443\pm0.105$ & $9.468\pm0.430$ \\
141133.3$+$703737 & $12.60\pm 0.41$&$12.55 \pm0.49$     & $13.17\pm0.28$   & $12.113\pm0.026$   & $12.004\pm0.029$ & $12.036\pm0.027$ & $11.965\pm0.023$ & $12.000\pm0.022$ & $12.369\pm0.203$ & ... \\
142126.5$+$712427 & $10.08 \pm0.80$&${\it 11.62\pm0.75}^a$  & $11.34\pm0.08$   & $11.848\pm0.021$   & $11.950\pm0.021$ & $12.017\pm0.028$ & $12.003\pm0.023$ & $12.047\pm0.022$ & $12.432\pm0.219$ & ... \\
142747.2$-$270108 & ${\it 11.58\pm0.62}^a$&$11.20 \pm0.92$  & $12.01\pm0.01$   & $12.529\pm0.022$   & $12.669\pm0.027$ & $12.732\pm0.029$ & $12.856\pm0.025$ & $12.926\pm0.027$ & ... & ... \\
143519.8$+$001352 & $11.30\pm 0.71$&$11.94 \pm0.63$     & $12.36\pm0.26$   & $13.244\pm0.032$   & $13.316\pm0.027$ & $13.436\pm0.044$ & $13.420\pm0.024$ & $13.521\pm0.032$ & ... & ... \\
160011.8$-$643330$^b$ &     ...          & ...          & $11.86\pm0.15$   & $12.568\pm0.028$   & $12.735\pm0.036$ & $12.786\pm0.036$ & $12.860\pm0.030$ & $12.951\pm0.037$ & ... & ... \\
163201.4$+$075940 & $10.79\pm 0.80$&$13.28 \pm0.37$     & $12.764\pm0.049$ & ${\it 12.836\pm0.034}^a$ & $12.611\pm0.042$ & $12.597\pm0.033$ & $12.318\pm0.024$ & $12.362\pm0.028$ & $11.823\pm0.220$ & $8.248\pm0.215$ \\
173153.7$+$064706 & $13.09\pm 0.34$&$13.40 \pm0.36$     & $13.74\pm0.36$   & $14.450\pm0.033$   & $14.512\pm0.090$ & $14.599\pm0.090$ & $14.689\pm0.037$ & $14.832\pm0.088$ & ... & ... \\
173651.2$+$280635 & $11.08\pm 0.78$&$11.40 \pm0.84$     & $11.44\pm0.10$   & $10.849\pm0.027$   & $10.635\pm0.031$ & $10.592\pm0.022$ & $10.527\pm0.023$ & $10.563\pm0.021$ & $10.410\pm0.057$ & $9.187\pm0.462$ \\
175340.5$-$500741 & $12.79 \pm0.38$&$13.07 \pm0.39$     & $12.88\pm0.43$   & $12.114\pm0.024$   & $11.861\pm0.023$ & $11.853\pm0.024$ & $11.786\pm0.023$ & $11.862\pm0.024$ & $11.839\pm0.206$ & ... \\
184559.8$-$413826 & $13.97\pm 0.20$&$13.76 \pm0.32$     & $14.63\pm0.32$   & $15.052\pm0.037$   & $15.087\pm0.067$ & $15.148\pm0.137$ & $15.383\pm0.059$ & $15.785\pm0.198$ & ... & ... \\
190211.7$-$513005 & ${\it 10.35\pm 0.80}^a$&$11.00 \pm1.00$     & $11.771\pm0.001$ & $12.103\pm0.020$   & $12.063\pm0.024$ & $12.158\pm0.023$ & $13.437\pm0.026$ & $13.451\pm0.034$ & ... & ... \\
190302.4$-$352828 & $12.40\pm 0.44$&$13.61 \pm0.34$     & $14.35\pm0.02$   & $14.834\pm0.044$   & $14.958\pm0.091$ & $14.813\pm0.109$ & $15.136\pm0.047$ & $15.274\pm0.150$ & ... & ... \\
191109.2$-$140651 & $ 9.84\pm 0.80$&$11.72 \pm0.71$     & $11.77\pm0.18$   & $12.314\pm0.024$   & $12.396\pm0.024$ & $12.561\pm0.031$ & $12.544\pm0.023$ & $12.616\pm0.031$ & ... & ... \\
203850.3$-$265750 & $12.11\pm 0.48$&$11.72 \pm0.71$     & $11.90\pm0.19$   & $10.398\pm0.024$   & $ 9.944\pm0.021$ & $ 9.843\pm0.022$ & $ 9.775\pm0.024$ & $ 9.772\pm0.020$ & ... & ... \\
215340.4$-$700430 &${\it 10.92\pm0.80}^a$&${\it 12.17\pm0.57}^a$ & $11.62\pm0.01$  & $12.152\pm0.021$   & $12.264\pm0.025$ & $12.375\pm0.027$ & $12.383\pm0.024$ & $12.486\pm0.027$ & $12.287\pm0.303$ & ... \\
220551.8$-$314105 & $10.61\pm 0.80$&$12.47 \pm0.51$     & $12.41\pm0.02$   & $12.747\pm0.024$   & $12.783\pm0.028$ & $12.863\pm0.034$ & $12.906\pm0.026$ & $12.903\pm0.030$ & ... & ... \\
225444.1$-$551505 & $11.79\pm 0.56$&$10.98 \pm1.00$     & $12.08\pm0.15$   & $12.825\pm0.024$   & $12.976\pm0.033$ & $13.068\pm0.031$ & $13.130\pm0.025$ & $13.238\pm0.031$ & $12.373\pm0.352$ & ... \\
233451.7$+$534701 & ... &$12.06 \pm0.59$                & $11.49\pm0.09$   & $12.502\pm0.025$   & $12.665\pm0.026$ & $12.726\pm0.021$ & $12.718\pm0.024$ & $12.812\pm0.027$ & ... & ...  \\
234421.6$-$342655 &${\it 7.82\pm0.80}^a$&${\it 11.23\pm0.91}^a$ & $10.982\pm0.006$ & $11.581\pm0.028$   & $11.684\pm0.022$ & $11.771\pm0.023$ & $11.851\pm0.023$ & $11.915\pm0.024$ & $11.604\pm0.201$ & ... \\
\hline
\end{tabular}\\
$^a$ Possibly unreliable.\\
$^b$ Not a $GALEX$ source.
\end{minipage}
\end{table}
\end{landscape}

\clearpage
\section{Radial velocity measurements.}

Table~\ref{tbl_rad_vel} lists the heliocentric-corrected velocity measurements ($\varv$), the
mid-exposure heliocentric julian dates (HJD) and data sources.
The instrument configurations and error estimates are described in Sections 2.1 and 2.2, respectively, and the measurement procedures are 
described in Section 3.1.

Stars labelled with the suffix ``B'' are the ``secondary'' components of each systems. The radial velocities are those of that secondary
component.

The label ``SSO'' refers to spectra obtained with the Wide Field Spectrograph (WiFeS) attached to the
2.3m telescope at Siding Spring Observatory; the label ``KPNO'' refers to spectra obtained
with the 4-m telescope and R.-C. Spectrograph at Kitt Peak National Observatory; the label ``NTT'' refers to spectra obtained
using EFOSC2 attached to the New Technology Telescope at La Silla; the label ``OND'' refers to spectra obtained
using the coud\'e spectrograph and the 2-m telescope at Ond\v{r}ejov Observatory; the label ``SAAO'' refers to spectra obtained at
the South African Astronomical Observatory using the 74-inch telescope and the Cassegrain spectrograph; {and new and} archival spectra
are labelled ``FEROS'' for the Fiber-fed Extended Range Optical Spectrograph on the MPG~2.2-m telescope at La Silla, 
and ``INT'' and ``WHT'' for spectrographs attached to the Isaac Newton Telescope and William Herschel Telescope at La Palma.

\begin{table*}
\centering%
\vspace{-0.3cm}
\begin{minipage}{\textwidth}%
\caption{Radial velocities. \label{tbl_rad_vel}}%
\vspace{-0.3cm}
\begin{tabular}{lcclcclcclcc}
\hline
\multicolumn{1}{c}{HJD} & $\varv$ & Source & \multicolumn{1}{c}{HJD} & $\varv$ & Source & \multicolumn{1}{c}{HJD} & $\varv$ & Source & \multicolumn{1}{c}{HJD} & $\varv$ & Source \\
(2450000+) & (\kmps) &  & (2450000+) & (\kmps) &  & (2450000+) & (\kmps) & & (2450000+) & (\kmps) &   \\
\hline
\multicolumn{3}{c}{GALEX~J0047+0337B}                                  & \multicolumn{3}{c}{$N= 4,\ \bar{\varv}= 45.3,\ \sigma_\varv= 4.4$}     & 5930.74878 &    31.3  & KPNO                                           & 6047.05680 &    29.3  & SSO                                           \\
5898.04458 &  $-$19.9 & SSO                                            & \multicolumn{3}{c}{GALEX J0401$-$3223}                                 & 5930.80712 &    18.8  & KPNO                                           & 6047.08829 &    31.4  & SSO                                           \\
5899.04507 &  $-$23.4 & SSO                                            & 3953.91980 &    53.0  & FEROS                                          & 5930.86267 &    13.6  & KPNO                                           & 6700.41812 &    22.9  & SAAO                                          \\
5930.57007 &  $-$22.1 & KPNO                                           & 4014.75477 &    52.1  & FEROS                                          & 5930.91715 &     5.7  & KPNO                                           & 6700.47565 &    34.4  & SAAO                                          \\
5930.57406 &  $-$23.5 & KPNO                                           & 5500.79501 &    49.3  & FEROS                                          & 5931.00340 &     0.5  & KPNO                                           & 6700.52813 &    30.5  & SAAO                                          \\
5930.66341 &  $-$23.2 & KPNO                                           & 5900.02546 &    45.2  & SSO                                            & 5932.71032 &    50.9  & KPNO                                           & \multicolumn{3}{c}{$N= 6,\ \bar{\varv}= 29.8,\ \sigma_\varv= 3.5$}    \\
5931.56622 &  $-$26.2 & KPNO                                           & 6171.78071 &    55.8  & NTT                                            & 5932.86851 &    42.3  & KPNO                                           & \multicolumn{3}{c}{GALEX J1114$-$2421}                                \\
5932.55868 &  $-$27.1 & KPNO                                           & 6171.83654 &    66.3  & NTT                                            & 5932.95043 &    36.1  & KPNO                                           & 6046.17066 &    20.2  & SSO                                           \\
6172.84455 &   $-$5.7 & NTT                                            & 6172.75132 &    56.8  & NTT                                            & 5933.00770 &    30.1  & KPNO                                           & 6047.00935 &    19.7  & SSO                                           \\
\multicolumn{3}{c}{$N= 8,\ \bar{\varv}= -21.4,\ \sigma_\varv= 6.3$}    & 6172.78790 &    52.3  & NTT                                            & \multicolumn{3}{c}{$N= 9,\ \bar{\varv}= 25.5,\ \sigma_\varv= 16.0$}    & 6047.04136 &    15.7  & SSO                                           \\
\multicolumn{3}{c}{GALEX J0047+0958 (HD~4539)}                         & 6173.74065 &    72.0  & NTT                                            & \multicolumn{3}{c}{GALEX J0751+0925}                                   & 6047.10305 &    15.9  & SSO                                           \\
3227.46321 &     3.7  & OND                                            & 6173.80204 &    63.0  & NTT                                            & 5898.15513 &   164.5  & SSO                                            & 6047.99465 &    16.9  & SSO                                           \\
3227.48556 &   $-$0.4 & OND                                            & 6173.88283 &    64.4  & NTT                                            & 5898.20329 &  $-$13.8 & SSO                                            & 6700.43054 &     7.6  & SAAO                                          \\
3227.50802 &   $-$1.4 & OND                                            & 6700.28962 &    54.0  & SAAO                                           & 5899.12739 & $-$113.9 & SSO                                            & 6700.48786 &     4.8  & SAAO                                          \\
3229.51947 &   $-$2.4 & OND                                            & 6700.34266 &    49.4  & SAAO                                           & 5899.22915 &   170.8  & SSO                                            & 6700.53977 &    16.3  & SAAO                                          \\
3229.54285 &   $-$0.2 & OND                                            & 6700.40330 &    46.9  & SAAO                                           & 5930.75957 &    87.8  & KPNO                                           & \multicolumn{3}{c}{$N= 8,\ \bar{\varv}= 14.6,\ \sigma_\varv= 5.2$}    \\
3255.46637 &   $-$0.5 & OND                                            & \multicolumn{3}{c}{$N= 14,\ \bar{\varv}= 55.8,\ \sigma_\varv= 7.6$}    & 5930.81729 &    92.2  & KPNO                                           & \multicolumn{3}{c}{Feige 66}                                          \\
3255.49221 &   $-$2.4 & OND                                            & \multicolumn{3}{c}{GALEX J0500+0912}                                   & 5930.87413 & $-$127.6 & KPNO                                           & 5311.46603 &   $-$5.5 & OND                                           \\
3255.54837 &   $-$3.0 & OND                                            & 5930.62297 &    48.5  & KPNO                                           & 5930.92836 &    42.6  & KPNO                                           & 5311.48899 &   $-$7.9 & OND                                           \\
3255.57093 &   $-$2.1 & OND                                            & 5930.73469 &    47.7  & KPNO                                           & 5930.93931 &    96.5  & KPNO                                           & 5311.51191 &   $-$4.2 & OND                                           \\
5499.73722 &   $-$3.2 & FEROS                                          & 5931.65783 &    50.2  & KPNO                                           & 5930.98374 &   135.2  & KPNO                                           & 5311.53487 &     5.8  & OND                                           \\
5500.56785 &   $-$3.7 & FEROS                                          & 5932.74346 &    45.0  & KPNO                                           & 5931.03215 &  $-$91.9 & KPNO                                           & 5311.55792 &   $-$7.1 & OND                                           \\
5500.70433 &   $-$2.7 & FEROS                                          & \multicolumn{3}{c}{$N= 4,\ \bar{\varv}= 47.8,\ \sigma_\varv= 1.9$}     & 5931.73038 &   $-$8.1 & KPNO                                           & 5312.46039 &   $-$3.2 & OND                                           \\
5500.78859 &   $-$3.2 & FEROS                                          & \multicolumn{3}{c}{GALEX~J0507+0348}                                   & 5931.87541 &   141.4  & KPNO                                           & 5312.48335 &   $-$5.0 & OND                                           \\
5930.55564 &   $-$5.0 & KPNO                                           & 5931.67914 &    77.7  & KPNO                                           & 5931.99730 &    41.9  & KPNO                                           & 5312.50638 &     0.0  & OND                                           \\
5930.55774 &     0.0  & KPNO                                           & 5931.83227 &    39.7  & KPNO                                           & 5932.75723 &   163.1  & KPNO                                           & 5312.52952 &   $-$4.2 & OND                                           \\
5930.65685 &   $-$3.4 & KPNO                                           & 5931.92094 &    91.2  & KPNO                                           & 5932.88029 &   $-$4.4 & KPNO                                           & 5312.55266 &   $-$4.7 & OND                                           \\
5931.57524 &   $-$3.7 & KPNO                                           & 5932.69831 &   116.5  & KPNO                                           & 5933.01902 & $-$125.4 & KPNO                                           & \multicolumn{3}{c}{$N= 10,\ \bar{\varv}= -3.6,\ \sigma_\varv= 3.7$}   \\
5932.55265 &   $-$6.4 & KPNO                                           & 5932.83768 &    30.7  & KPNO                                           & 6046.91015 &    60.7  & SSO                                            & \multicolumn{3}{c}{GALEX J1356$-$4934}                                \\
6172.83793 &     0.1  & NTT                                            & 5932.89420 &    28.6  & KPNO                                           & 6047.93820 &   163.7  & SSO                                            & 6046.15529 &     0.2  & SSO                                           \\
6173.77623 &   $-$1.4 & NTT                                            & 6171.78863 &   147.5  & NTT                                            & \multicolumn{3}{c}{$N=19,\ \bar{\varv}= 46.1,\ \sigma_\varv= 101.1$}   & 6046.27506 &    17.2  & SSO                                           \\
\multicolumn{3}{c}{$N= 20,\ \bar{\varv}= -2.1,\ \sigma_\varv= 2.1$}    & 6171.84440 &   155.4  & NTT                                            & \multicolumn{3}{c}{GALEX~J0805$-$1058}                                 & 6047.13503 &     4.8  & SSO                                           \\
\multicolumn{3}{c}{GALEX J0049+2056}                                   & 6171.85811 &   153.7  & NTT                                            & 5898.12316 &    52.8  & SSO                                            & 6172.47898 &    12.8  & NTT                                           \\
5930.59893 &    15.1  & KPNO                                           & 6171.90134 &   154.1  & NTT                                            & 5898.12983 &    64.7  & SSO                                            & 6172.49297 &     7.1  & NTT                                           \\
5930.68516 &    17.7  & KPNO                                           & 6172.81341 &   140.7  & NTT                                            & 5898.13875 &    69.2  & SSO                                            & 6172.51996 &    15.8  & NTT                                           \\
5931.58588 &    14.6  & KPNO                                           & 6172.90675 &   166.4  & NTT                                            & 5898.25067 &    26.9  & SSO                                            & 6172.54716 &     8.9  & NTT                                           \\
5932.57896 &    18.2  & KPNO                                           & 6173.81997 &    99.9  & NTT                                            & 5899.08066 &    61.3  & SSO                                            & 6700.49865 &    20.6  & SAAO                                          \\
\multicolumn{3}{c}{$N= 4,\ \bar{\varv}=  16.4,\ \sigma_\varv= 1.6$}    & 6173.90692 &   166.6  & NTT                                            & 5899.20870 &    88.4  & SSO                                            & 6700.56149 &     8.7  & SAAO                                          \\
\multicolumn{3}{c}{GALEX J0059+1544}                                   & 6700.31140 &    58.5  & SAAO                                           & 5900.24103 &    78.0  & SSO                                            & \multicolumn{3}{c}{$N= 9,\ \bar{\varv}= 10.7,\ \sigma_\varv= 6.1$}    \\
3955.82658 &    20.6  & FEROS                                          & 6700.36337 &   111.9  & SAAO                                           & 5930.76915 &    47.5  & KPNO                                           & \multicolumn{3}{c}{GALEX J1407+3103}                                  \\
3986.81551 &     9.6  & FEROS                                          & \multicolumn{3}{c}{$N=16,\ \bar{\varv}= 108.7,\ \sigma_\varv= 47.9$}   & 5930.82802 &    88.9  & KPNO                                           & 5932.03194 & $-$160.6 & KPNO                                          \\
5930.58194 &    15.5  & KPNO                                           & \multicolumn{3}{c}{GALEX J0613+3420}                                   & 5930.88403 &    42.1  & KPNO                                           & 5932.97687 & $-$157.3 & KPNO                                          \\
5930.58594 &    17.3  & KPNO                                           & 5931.74486 & $-$135.5 & KPNO                                           & 5930.97349 &    75.2  & KPNO                                           & 5933.05659 & $-$164.7 & KPNO                                          \\
5930.67212 &    20.7  & KPNO                                           & 5931.84915 & $-$139.9 & KPNO                                           & 5930.99445 &    92.6  & KPNO                                           & \multicolumn{3}{c}{$N= 3,\ \bar{\varv}= -160.9,\ \sigma_\varv= 3.0$}  \\
5931.60011 &    13.1  & KPNO                                           & 5931.93741 & $-$138.7 & KPNO                                           & 5931.75141 &    42.1  & KPNO                                           & \multicolumn{3}{c}{GALEX J1411+7037B}                                 \\
5932.56969 &    14.9  & KPNO                                           & 5932.72107 &  $-$40.9 & KPNO                                           & 5931.88363 &    85.1  & KPNO                                           & 5932.00693 &  $-$18.0 & KPNO                                          \\
5932.72419 &    18.3  & KPNO                                           & \multicolumn{3}{c}{$N= 4,\ \bar{\varv}= -113.7,\ \sigma_\varv= 42.1$}  & 5931.98199 &    45.6  & KPNO                                           & 5932.04748 &  $-$17.6 & KPNO                                          \\
6172.83077 &    12.4  & NTT                                            & \multicolumn{3}{c}{GALEX J0657$-$7324}                                 & 5932.86147 &    49.1  & KPNO                                           & 5932.99447 &  $-$14.9 & KPNO                                          \\
6173.78258 &    21.6  & NTT                                            & 6047.87067 &   $-$7.0 & SSO                                            & 5932.94370 &    73.5  & KPNO                                           & 5933.06557 &  $-$20.9 & KPNO                                          \\
\multicolumn{3}{c}{$N= 10,\ \bar{\varv}= 16.4,\ \sigma_\varv= 3.8$}    & 6171.82808 &  $-$11.3 & NTT                                            & 6046.92589 &    31.8  & SSO                                            & \multicolumn{3}{c}{$N= 4,\ \bar{\varv}= -17.9,\ \sigma_\varv=  2.1$}  \\
\multicolumn{3}{c}{GALEX J0206+1438}                                   & 6171.92035 &  $-$17.6 & NTT                                            & 6046.96533 &    36.0  & SSO                                            & \multicolumn{3}{c}{GALEX J1421+7124}                                  \\
5932.64012 &    19.8  & KPNO                                           & 6172.80377 &  $-$11.7 & NTT                                            & 6047.95401 &    48.4  & SSO                                            & 5461.35716 &  $-$24.2 & OND                                           \\
5932.76929 &    23.2  & KPNO                                           & 6172.91817 &     1.1  & NTT                                            & 6700.39670 &    36.6  & SAAO                                           & 5461.38072 &  $-$22.3 & OND                                           \\
6172.76462 &    19.9  & NTT                                            & 6172.92477 &  $-$12.6 & NTT                                            & 6700.46184 &    59.2  & SAAO                                           & 5461.40360 &  $-$25.5 & OND                                           \\
6172.79845 &     4.1  & NTT                                            & 6700.33004 &  $-$16.0 & SAAO                                           & 6700.51455 &    86.2  & SAAO                                           & 5462.34435 &  $-$29.5 & OND                                           \\
6172.89971 &     9.8  & NTT                                            & 6700.37998 &  $-$13.9 & SAAO                                           & \multicolumn{3}{c}{$N=23,\ \bar{\varv}= 60.1,\ \sigma_\varv= 20.0$}    & 5462.36727 &  $-$19.9 & OND                                           \\
6173.75136 &     1.6  & NTT                                            & 6700.44564 &  $-$19.7 & SAAO                                           & \multicolumn{3}{c}{GALEX J0812+1601}                                   & 5462.64138 &  $-$17.4 & OND                                           \\
6173.81315 &    12.5  & NTT                                            & \multicolumn{3}{c}{$N= 9,\ \bar{\varv}= -12.1,\ \sigma_\varv= 5.8$}    & 5898.18568 &  $-$26.7 & SSO                                            & 5463.32822 &  $-$21.2 & OND                                           \\
6173.92187 &    19.6  & NTT                                            & \multicolumn{3}{c}{GALEX J0703+6236}                                   & 5898.23480 &  $-$20.9 & SSO                                            & 5463.35791 &  $-$25.7 & OND                                           \\
\multicolumn{3}{c}{$N= 8,\ \bar{\varv}= 13.8,\ \sigma_\varv= 7.5$}     & 5931.04474 &    24.2  & KPNO                                           & 5932.85141 &     1.3  & KPNO                                           & 5464.32919 &  $-$23.3 & OND                                           \\
\multicolumn{3}{c}{GALEX J0232+4411}                                   & 5931.71772 &    18.3  & KPNO                                           & 5932.93361 &   $-$1.6 & KPNO                                           & 5464.35893 &  $-$24.2 & OND                                           \\
5931.61405 &    41.1  & KPNO                                           & 5931.85802 &    17.6  & KPNO                                           & 5933.02934 &   $-$4.1 & KPNO                                           & 5470.25157 &  $-$24.4 & OND                                           \\
5931.81305 &    52.1  & KPNO                                           & 5931.94502 &    19.8  & KPNO                                           & \multicolumn{3}{c}{$N= 5,\ \bar{\varv}= -10.4,\ \sigma_\varv= 11.2$}   & 5483.25104 &  $-$35.4 & OND                                           \\
5932.65498 &    46.2  & KPNO                                           & \multicolumn{3}{c}{$N= 4,\ \bar{\varv}= 20.0,\ \sigma_\varv= 2.6$}     & \multicolumn{3}{c}{GALEX J1041$-$0730}                                 & 5794.51260 &  $-$28.6 & OND                                           \\
5932.81723 &    41.8  & KPNO                                           & \multicolumn{3}{c}{GALEX J0716+2319}                                   & 6046.11527 &    30.0  & SSO                                            & \multicolumn{3}{c}{$N= 13,\ \bar{\varv}= -24.7,\ \sigma_\varv=  4.4$} \\
\hline
\end{tabular}
\end{minipage}
\end{table*}

\addtocounter{table}{-1}
\begin{table*}
\centering
\vspace{-0.3cm}
\begin{minipage}{\textwidth}
\caption{{\it continued}}
\vspace{-0.3cm}
\begin{tabular}{lcclcclcclcc}
\hline
\multicolumn{1}{c}{HJD} & $\varv$ & Source & \multicolumn{1}{c}{HJD} & $\varv$ & Source & \multicolumn{1}{c}{HJD} & $\varv$ & Source & \multicolumn{1}{c}{HJD} & $\varv$ & Source \\
(2450000+) & (\kmps) &  & (2450000+) & (\kmps) &  & (2450000+) & (\kmps) & & (2450000+) & (\kmps) &   \\
\hline
\multicolumn{3}{c}{GALEX J1427$-$2701}                                 & 5757.01406 &   $-$4.4 & SSO                                            & 5757.09665 &     7.1  & SSO                                            & 5898.96627 &    66.8  & SSO                                           \\
6048.11285 &   $-$6.6 & SSO                                            & 6048.21613 &    52.3  & SSO                                            & 5757.10547 &    14.4  & SSO                                            & 5898.99478 &    67.3  & SSO                                           \\
6171.47061 &    10.9  & NTT                                            & 6171.57012 &    28.2  & NTT                                            & 5760.12396 &    14.0  & SSO                                            & 5899.02602 &    60.4  & SSO                                           \\
6171.48306 &     6.8  & NTT                                            & 6171.59018 &    27.1  & NTT                                            & 5761.03781 &    20.1  & SSO                                            & 5899.98304 &    48.7  & SSO                                           \\
6171.49578 &     1.1  & NTT                                            & 6171.60924 &    17.8  & NTT                                            & 5761.09468 &     8.6  & SSO                                            & 6171.72238 &  $-$38.2 & NTT                                           \\
6171.50953 &     3.5  & NTT                                            & 6171.63315 &    13.7  & NTT                                            & 6171.68438 &    10.4  & NTT                                            & 6171.77618 &  $-$50.8 & NTT                                           \\
6171.52366 &     5.2  & NTT                                            & 6172.56912 &    51.2  & NTT                                            & 6171.74097 &    10.7  & NTT                                            & 6171.82488 &  $-$69.0 & NTT                                           \\
6171.53666 &   $-$0.5 & NTT                                            & 6172.60585 &    42.9  & NTT                                            & 6172.65633 &    18.1  & NTT                                            & 6171.89529 &  $-$78.4 & NTT                                           \\
6172.50574 &     0.6  & NTT                                            & 6172.62568 &    35.3  & NTT                                            & 6172.72634 &     4.9  & NTT                                            & 6172.63823 &    72.5  & NTT                                           \\
6172.53333 &   $-$7.6 & NTT                                            & 6361.86864 &  $-$70.0 & NTT                                            & \multicolumn{3}{c}{$N=9,\ \bar{\varv}= 12.0,\ \sigma_\varv= 4.7$}      & 6172.70799 &    50.3  & NTT                                           \\
6700.57282 &  $-$12.0 & SAAO                                           & 6361.90960 &  $-$84.9 & NTT                                            & \multicolumn{3}{c}{GALEX J1903$-$3528}                                 & 6172.77525 &    28.4  & NTT                                           \\
6700.60630 &   $-$5.0 & SAAO                                           & 6429.53360 &    51.7  & INT                                            & 6171.70234 &  $-$29.7 & NTT                                            & 6172.85901 &  $-$11.3 & NTT                                           \\
\multicolumn{3}{c}{$N= 11,\ \bar{\varv}= -0.3,\ \sigma_\varv= 6.6$}    & 6870.49571 &  $-$17.8 & NTT                                            & 6171.75097 &  $-$37.6 & NTT                                            & 6173.67386 &    71.5  & NTT                                           \\
\multicolumn{3}{c}{GALEX J1435+0013}                                   & 6870.61105 &    29.4  & NTT                                            & 6172.67409 &  $-$17.3 & NTT                                            & 6173.73606 &    98.9  & NTT                                           \\
3829.84393 &   $-$5.5 & FEROS                                          & 6870.69156 &    29.8  & NTT                                            & 6172.73670 &  $-$25.6 & NTT                                            & 6173.79014 &    96.8  & NTT                                           \\
4137.79323 &   $-$3.2 & FEROS                                          & \multicolumn{3}{c}{$N=16,\ \bar{\varv}= 11.4,\ \sigma_\varv= 40.0$}    & 6173.70238 &   $-$2.9 & NTT                                            & 6173.87812 &    55.8  & NTT                                           \\
5756.90232 &  $-$16.0 & SSO                                            & \multicolumn{3}{c}{GALEX J1736+2806B}                                  & \multicolumn{3}{c}{$N= 5,\ \bar{\varv}= -22.6,\ \sigma_\varv= 11.8$}   & 6870.63480 &    96.2  & NTT                                           \\
6046.18707 &   $-$3.7 & SSO                                            & 5045.37797 &  $-$14.2 & OND                                            & \multicolumn{3}{c}{GALEX J1911$-$1406}                                 & 6870.71233 &    93.8  & NTT                                           \\
6047.22908 &   $-$6.5 & SSO                                            & 5045.39926 &  $-$16.5 & OND                                            & 5760.14357 & $-$156.8 & SSO                                            & 6870.86537 &    58.6  & NTT                                           \\
6047.26718 &     2.2  & SSO                                            & 5310.44503 &  $-$19.6 & OND                                            & 5761.05696 & $-$147.0 & SSO                                            & 6989.54548 &    54.3  & FEROS                                         \\
6048.04105 &     3.6  & SSO                                            & 5310.46808 &   $-$8.2 & OND                                            & 5761.12734 & $-$151.3 & SSO                                            & 6990.55028 &  $-$34.2 & FEROS                                         \\
6048.07142 &   $-$5.9 & SSO                                            & 5310.49159 &  $-$16.9 & OND                                            & \multicolumn{3}{c}{$N= 3,\ \bar{\varv}= -151.7,\ \sigma_\varv= 4.0$}   & 6994.58156 &    75.6  & FEROS                                         \\
6048.09997 &   $-$4.6 & SSO                                            & 5310.51449 &   $-$7.4 & OND                                            & \multicolumn{3}{c}{GALEX J2153$-$7003}                                 & \multicolumn{3}{c}{$N=24,\ \bar{\varv}= 32.6,\ \sigma_\varv= 55.0$}   \\
6700.58473 &  $-$12.6 & SAAO                                           & 5310.53754 &  $-$14.7 & OND                                            & 3918.75562 &    43.4  & FEROS                                          & \multicolumn{3}{c}{GALEX J2334+5347}                                  \\
6700.62607 &   $-$6.0 & SAAO                                           & 5310.56068 &   $-$9.3 & OND                                            & 3956.76278 &    42.5  & FEROS                                          & 5461.43461 &    40.3  & OND                                           \\
\multicolumn{3}{c}{$N= 11,\ \bar{\varv}= -5.3,\ \sigma_\varv= 5.3$}    & 5310.58391 &  $-$13.9 & OND                                            & 5757.17717 &    41.8  & SSO                                            & 5461.48740 &    36.1  & OND                                           \\
\multicolumn{3}{c}{J1600$-$6433}                                       & 5311.58102 &  $-$13.2 & OND                                            & 5757.18600 &    43.1  & SSO                                            & 5461.55799 &    31.5  & OND                                           \\
5756.93811 &    55.6  & SSO                                            & 5311.60395 &  $-$18.7 & OND                                            & 5761.21680 &    44.2  & SSO                                            & 5461.58084 &    44.7  & OND                                           \\
5756.94716 &    55.1  & SSO                                            & 5312.57540 &  $-$11.5 & OND                                            & 5897.97485 &    41.4  & SSO                                            & 5462.39976 &    46.5  & OND                                           \\
5757.03853 &    48.6  & SSO                                            & 5312.59830 &   $-$7.9 & OND                                            & 5897.98518 &    36.9  & SSO                                            & 5462.42295 &    41.0  & OND                                           \\
5757.04736 &    53.6  & SSO                                            & 5377.40465 &  $-$11.8 & OND                                            & 5898.94745 &    39.8  & SSO                                            & 5462.59518 &    31.6  & OND                                           \\
5757.96919 &    65.3  & SSO                                            & 5377.42598 &   $-$6.9 & OND                                            & 5898.95287 &    42.1  & SSO                                            & 5462.61797 &    42.5  & OND                                           \\
5757.97801 &    56.6  & SSO                                            & 5430.35500 &   $-$9.3 & OND                                            & 5898.97969 &    40.3  & SSO                                            & 5463.41196 &    37.9  & OND                                           \\
5760.98197 &    41.6  & SSO                                            & 5430.37835 &  $-$10.7 & OND                                            & 5899.96199 &    41.4  & SSO                                            & 5463.44187 &    29.6  & OND                                           \\
6171.62174 &    37.8  & NTT                                            & 5483.27549 &   $-$7.8 & OND                                            & 6171.80243 &    50.5  & NTT                                            & 5463.47292 &    26.0  & OND                                           \\
6171.66370 &    45.9  & NTT                                            & 5675.52280 &  $-$16.0 & OND                                            & 6171.87909 &    47.0  & NTT                                            & 5464.41610 &    37.2  & OND                                           \\
6172.58713 &    56.0  & NTT                                            & 5766.38926 &   $-$7.9 & OND                                            & 6172.68364 &    42.9  & NTT                                            & 5464.44582 &    50.4  & OND                                           \\
6172.66055 &    40.6  & NTT                                            & 5794.39180 &   $-$6.5 & OND                                            & 6173.79603 &    54.2  & NTT                                            & 5483.37624 &    28.3  & OND                                           \\
6700.59333 &    43.1  & SAAO                                           & \multicolumn{3}{c}{$N= 21,\ \bar{\varv}= -11.9,\ \sigma_\varv= 4.0$}   & \multicolumn{3}{c}{$N=15,\ \bar{\varv}= 43.4,\ \sigma_\varv= 4.2$}     & 5483.56518 &    29.9  & OND                                           \\
6700.63352 &    37.7  & SAAO                                           & \multicolumn{3}{c}{GALEX J1753$-$5007B}                                & \multicolumn{3}{c}{GALEX J2205$-$3141}                                 & 5794.58849 &    34.8  & OND                                           \\
\multicolumn{3}{c}{$N=13,\ \bar{\varv}= 49.0,\ \sigma_\varv= 8.3$}     & 5757.06955 &  $-$56.9 & SSO                                            & 5757.23298 &  $-$67.4 & SSO                                            & 5834.29578 &    29.0  & OND                                           \\
\multicolumn{3}{c}{GALEX J1632+0759}                                   & 5757.07837 &  $-$56.4 & SSO                                            & 6171.77057 &  $-$17.9 & NTT                                            & \multicolumn{3}{c}{$N=17,\ \bar{\varv}= 36.3,\ \sigma_\varv=6.9$}     \\
5435.41483 &$-$67.4   &  WHT                                           & 5761.02112 &  $-$65.7 & SSO                                            & 6171.81889 &  $-$68.1 & NTT                                            & \multicolumn{3}{c}{GALEX J2344$-$3426}                                \\
5756.97348 &$-$62.5   &  SSO                                           & 5761.07352 &  $-$60.3 & SSO                                            & 6171.88946 &  $-$62.9 & NTT                                            & 3679.56424 &    21.5  & FEROS                                         \\
5761.00101 &$-$50.0   &  SSO                                           & 6171.64403 &  $-$56.2 & NTT                                            & 6172.70173 &    23.7  & NTT                                            & 3957.88130 &    21.5  & FEROS                                         \\
6048.12788 &  15.2    &  SSO                                           & 6171.71118 &  $-$73.0 & NTT                                            & 6172.78429 &   $-$8.0 & NTT                                            & 5757.29066 &    12.9  & SSO                                           \\
6171.55953 & $-$0.7   &  NTT                                           & 6172.61645 &  $-$75.6 & NTT                                            & 6172.85277 &  $-$62.3 & NTT                                            & 5898.02545 &    21.5  & SSO                                           \\
6171.57889 &   8.6    &  NTT                                           & 6173.71129 &  $-$74.9 & NTT                                            & 6173.73046 &    32.9  & NTT                                            & 5899.01277 &    19.2  & SSO                                           \\
6172.55735 &   4.8    &  NTT                                           & \multicolumn{3}{c}{$N= 8,\ \bar{\varv}= -64.9,\ \sigma_\varv= 8.0$}    & 6173.76269 &    19.9  & NTT                                            & 5899.06162 &    21.0  & SSO                                           \\
6172.57774 & $-$3.1   &  NTT                                           & \multicolumn{3}{c}{GALEX J1845$-$4138}                                 & 6173.83761 &  $-$38.8 & NTT                                            & 6171.81182 &    29.0  & NTT                                           \\
6432.47442 &$-$31.5   &  INT                                           & 6171.65588 &  $-$52.3 & NTT                                            & 6870.62358 &  $-$49.9 & NTT                                            & 6171.91775 &    26.4  & NTT                                           \\
6870.48215 &$-$68.0   &  NTT                                           & 6171.73152 &  $-$52.3 & NTT                                            & 6870.70432 &  $-$42.0 & NTT                                            & 6172.69358 &    31.2  & NTT                                           \\
6870.59970 &$-$69.0   &  NTT                                           & 6172.64693 &  $-$57.2 & NTT                                            & 6870.85814 &    14.7  & NTT                                            & 6173.76988 &    34.9  & NTT                                           \\
6870.68037 &$-$65.0   &  NTT                                           & 6172.71683 &  $-$69.1 & NTT                                            & \multicolumn{3}{c}{$N=13,\ \bar{\varv}= -25.1,\ \sigma_\varv= 36.4$}   & \multicolumn{3}{c}{$N=10,\ \bar{\varv}= 23.9,\ \sigma_\varv= 6.1$}    \\
\multicolumn{3}{c}{$N= 12,\ \bar{\varv}= -32.4,\ \sigma_\varv= 33.3$}  & 6173.68314 &  $-$56.9 & NTT                                            & \multicolumn{3}{c}{GALEX J2254$-$5515}                                 & \\
\multicolumn{3}{c}{GALEX J1731+0647}                                   & \multicolumn{3}{c}{$N= 5,\ \bar{\varv}= -57.6,\ \sigma_\varv= 6.1$}    & 5898.00341 &   $-$7.3 & SSO                                            & \\
5757.00524 &  $-$19.9 & SSO                                            & \multicolumn{3}{c}{GALEX J1902$-$5130}                                 & 5898.05579 &  $-$24.7 & SSO                                            & \\
\hline
\end{tabular}
\end{minipage}
\end{table*}

\section{Hot subdwarf binary systems}

We have compiled a list of all known hot subdwarf binary systems. Table~\ref{tbl_sdb_bin_all}
lists the name of the system, its coordinates, proper motion, $V$ magnitudes, orbital period in days, the
systemic velocity ($\gamma$) and the velocity semiamplitude of the hot subdwarf ($K$). The 
table also lists the eccentricity ($e$) of the system if the orbit of the binary was found
to be eccentric. If the companion to the hot subdwarf is known it is listed and also if the
system is photometrically variable, this is also noted in the table. The proper motions and $V$ magnitudes for 
the objects are from The fourth U.S. Naval Observatory CCD Astrograph Catalog \citep[UCAC4][]{zac2013}. For 
those objects that are not in the UCAC4 and has SDSS photometry, we calculated $V$ magnitudes using the
transformation equation of \citet{jes2005}:
\begin{displaymath}
V = g-0.58(g-r)-0.01
\end{displaymath}
The references for the binary system properties are provided in the final column.
GALEX~J1736+2806 and J2038$-$2657 are not included in Table~\ref{tbl_sdb_bin_all} because the photometric
variations are not clearly associated to orbital periodicities.
New Kepler identifications (KIC) are added to the list despite current lack of radial velocity data
because of the higher quality of light curve analysis and the timeliness of the results.
Table~\ref{tbl_sample_kine} lists published stellar parameters (with relevant references) and calculated
absolute $V$ magnitude and Galactic velocity vectors ($U,V,W$). The mass of sdO companions to Be stars
is assumed to be 1\,\msun\ and that of low-mass sdB stars is assumed to be 0.23\,\msun.

\begin{landscape}
\begin{table}
\begin{center}
\vspace{-0.3cm}
\begin{minipage}{230mm}%
\renewcommand{\footnoterule}{\vspace*{-15pt}}
\caption{Properties of hot subdwarf binary systems. \label{tbl_sdb_bin_all}}%
\vspace{-0.3cm}
\begin{tabular}{lcccccccccll}
\hline
Object            & R.A.      & Dec.        & $\mu_{\alpha}\cos{\delta}\ ,\mu_{\delta}$ & $V$ & Period & $\gamma$      & $K$           & $e$ & Sec. & Variable & Ref. \\
                  & (J2000)     & (J2000)   & (mas yr$^{-1}$)                           & (mag) & (d)  & (\kmps) & (\kmps) &     & type    &          &           \\
\hline
CD-30 11223      & 14 11 16.0 & $-$30 53 07 &   7.4$\pm$1.0, -6.4$\pm$1.8 & 12.34 & 0.04897906$\pm$0.00000004    & 31.5$\pm$1.3  & 386.9$\pm$1.9 &                 & WD & ell,ecl  & 1,2 \\
SDSSJ1622+4730   & 16 22 56.7 & $+$47 30 51 & -10.7$\pm$7.4,-29.6$\pm$8.0 & 16.19 & 0.06969$\pm$0.00003          & -54.7$\pm$1.5 &  47.2$\pm$2.0 &                 & bd & refl,ecl & 3 \\
PG1017-086       & 10 20 14.5 & $-$08 53 46 &  -5.0$\pm$2.5, 11.7$\pm$4.0 & 14.42 & 0.0729938$\pm$0.0000003      & -9.1$\pm$1.3  &  51.0$\pm$1.7 &                 & dM & refl     & 4 \\
NGC6121-V46      & 16 23 47.1 & $-$26 31 56 &     ...                     & 18.51 & 0.087159                     & 31.3$\pm$1.6  & 211.6$\pm$2.3 &                 & WD & ell      & 5 \\
KPD0422+5421     & 04 26 06.9 & $+$54 28 17 &   3.5$\pm$3.7, -5.0$\pm$4.3 & 14.66 & 0.09017945$\pm$0.00000012    & -57$\pm$12    & 237$\pm$18    &                 & WD & ecl,ell  & 6,7 \\
KPD1930+2752     & 19 32 14.8 & $+$27 58 35 &  -0.7$\pm$10.1,15.1$\pm$10.7& 13.82 & 0.0950933$\pm$0.0000015      & 5$\pm$1       & 341$\pm$1     &                 & WD & ell,puls & 8,9,10 \\
HS0705+6700      & 07 10 42.1 & $+$66 55 44 &  -3.2$\pm$1.6,-3.5$\pm$1.8  & 14.92 & 0.09564665$\pm$0.00000039    & -36.4$\pm$2.9 & 85.8$\pm$3.7  &                 & dM & refl,ecl & 11 \\
SDSSJ08205+0008  & 08 20 53.5 & $+$00 08 43 &  2.5$\pm$4.6, -7.1$\pm$4.6  & 15.17 & 0.096$\pm$0.001              & 9.5$\pm$1.3   & 47.4$\pm$1.9  &                 & bd & refl,ecl & 12 \\
PG1336-018       & 13 38 48.1 & $-$02 01 49 & -6.5$\pm$2.0, -12.5$\pm$2.1 & 13.75 & 0.101015999$\pm$0.000000002  & ...           & 78.6$\pm$0.6  &                 & dM & refl,ecl,puls & 13,14 \\
NSVS14256825     & 20 20 00.5 & $+$04 37 56 &  5.1$\pm$2.7, -2.0$\pm$3.0  & 13.23 & 0.110374230$\pm$0.000000002  & -12.1$\pm$1.5 & 73.4$\pm$2.0  &                 & dM & refl,ecl & 15 \\
HS2231+2441      & 22 34 21.5 & $+$24 56 57 & 13.9$\pm$1.4,-20.7$\pm$1.7  & 14.15 & 0.1105880$\pm$0.0000005      & ...           & 49.1$\pm$3.2  &                 & dM & refl,ecl & 16 \\
UVEXJ0328+5035   & 03 28 55.2 & $+$50 35 30 & -4.7$\pm$4.6, -1.2$\pm$2.2  & 14.26 & 0.11017$\pm$0.00011          & 44.9$\pm$0.7  & 64.0$\pm$1.5  &                 & dM & refl     & 17 \\
HW Vir           & 12 44 20.2 & $-$08 40 17 &  9.5$\pm$1.5, -16.0$\pm$1.6 & 10.69 & 0.115$\pm$0.008              & -13.0$\pm$0.8 & 84.6$\pm$1.1  &                 & dM & refl,ecl & 18 \\
EC10246-2707     & 10 26 56.6 & $-$27 22 59 & -4.4$\pm$2.9, -7.2$\pm$2.1  & 14.38 & 0.1185079936$\pm$0.0000000009 & ...          & 71.6$\pm$1.7  &                 & dM & refl,ecl & 19 \\
PG1043+760       & 10 47 05.0 & $+$75 44 23 & 5.1$\pm$1.5, 8.9$\pm$1.9    & 13.53 & 0.1201506$\pm$0.0000003      & 24.8$\pm$1.4  & 63.6$\pm$1.4  &                 & dM &          & 20 \\
OGLE BUL-SC16335 & 18 09 48.2 & $-$26 41 49 & 60.0$\pm$5.3, 9.0$\pm$6.6   & 16.5  & 0.122                        & 36.4$\pm$19.6 & 92.5$\pm$26.2 &                 & dM & refl,ecl & 21,22 \\
2M1938+4603      & 19 38 32.6 & $+$46 03 59 & 2.8$\pm$0.6, -2.7$\pm$0.7   & 12.06 & 0.125765300$\pm$0.000000021 & 20.1$\pm$0.3 & 65.7$\pm$0.6 &                & dM & refl,ecl,puls & 23 \\
EC00404-4429     & 00 42 48.4 & $-$44 13 25 & 21.6$\pm$1.3, 9.8$\pm$1.4   & 13.67 & 0.12834$\pm$0.00004          & 33.0$\pm$2.9  & 152.8$\pm$3.4 &                 &    &          & 24 \\
KIC7335517       & 18 43 06.8 & $+$42 59 18 & -2.8$\pm$4.9, -8.7$\pm$4.4  & 15.60 & 0.13729$\pm$0.00002          & ...           & ...           &                 & dM & refl     & 25,26 \\
SDSSJ0830+4751   & 08 30 06.2 & $+$47 51 50 & 1.1$\pm$3.8, -10.0$\pm$4.2  & 16.04 & 0.14780$\pm$0.00007          & 49.9$\pm$0.9  & 77.0$\pm$1.7  &                 & WD &          & 27 \\
ASAS102322-3737  & 10 23 21.9 & $-$37 37 00 & -27.4$\pm$1.3,-20.9$\pm$0.9 & 11.58 & 0.13926940$\pm$0.00000004    & ...           & 81$\pm$3      &                 & dM & ecl,refl & 28 \\
KIC6614501       & 19 36 50.0 & $+$42 01 44 & 42.8$\pm$8.6, 12.8$\pm$9.1  & 16.09 & 0.15749747$\pm$0.00000025    & -6.5$\pm$1.5  & 97.2$\pm$2.0  &                 & WD & ell,dop  & 29 \\
2M1533+3759      & 15 33 49.4 & $+$37 59 28 & -0.6$\pm$1.3, -19.9$\pm$3.1 & 13.61 & 0.16177042$\pm$0.00000001    & -3.4$\pm$5.2  & 71.1$\pm$1.0  &                 & dM & refl,ecl & 30 \\
SDSSJ1920+3722   & 19 20 59.8 & $+$37 22 20 & 1.7$\pm$5.6, 4.5$\pm$6.5    & 15.74 & 0.168876$\pm$0.00035         & 16.8$\pm$2.0  & 59.8$\pm$2.5  &                 & dM & refl,ecl & 31 \\
HS2333+3927      & 23 35 42.5 & $+$39 44 27 &  0.6$\pm$4.7, 3.9$\pm$4.8   & 14.79 & 0.1718023$\pm$0.0000009      & -31.4$\pm$2.1 & 89.6$\pm$3.2  &                 & dM & refl     & 32 \\
GALEX~J0805-1058 & 08 05 10.9 & $-$10 58 34 & -23.2$\pm$1.8, -27.3$\pm$1.5& 12.25 & 0.173703$\pm$0.000002        & 58.2$\pm$0.9  & 29.2$\pm$1.3  &                 & dM, bd?       & 33 \\
BPS CS 22169-0001& 03 56 23.3 & $-$15 09 19 & -3.3$\pm$1.7, 2.2$\pm$1.6   & 12.85 & 0.1780$\pm$0.0003            & 2.8$\pm$0.3   & 14.9$\pm$0.4  &                 &    & refl?    & 34 \\
GALEX~J0751+0925 & 07 51 47.1 & $+$09 25 26 & -9.6$\pm$1.6, -9.9$\pm$2.0  & 14.09 & 0.178319$\pm$0.000005        & 15.5$\pm$1.6  & 147.7$\pm$2.2 &                 & WD &          & 33 \\
HE1415-0309      & 14 18 20.9 & $-$03 22 54 &         ...                 & 15.14 & 0.192$\pm$0.004              & 104.7$\pm$9.5 & 152.4$\pm$11.2 &                & WD &          & 22 \\
HS1741+2133      & 17 43 19.0 & $+$21 32 38 & -12.1$\pm$2.3, 4.2$\pm$2.5  & 13.99 & 0.20$\pm$0.01                & -112.8$\pm$2.7& 157.0$\pm$3.4 &                 & WD &          & 17 \\
SDSSJ0823+1136   & 08 23 32.1 & $+$11 36 41 &      ...                    & 16.65 & 0.20707$\pm$0.00002          & 135.1$\pm$2.0 & 169.4$\pm$2.5 &                 &    &          & 27 \\
SDSSJ1138-0035   & 11 38 40.7 & $-$00 35 32 & -10.2$\pm$2.7, -27.2$\pm$3.3& 14.47 & 0.207536$\pm$0.000002        &  23.3$\pm$3.7 & 162.0$\pm$3.8 &                 & WD &          & 35 \\
PG1432+159       & 14 35 18.9 & $+$15 40 14 & 2.9$\pm$1.4, -25.7$\pm$1.9  & 13.90 & 0.22489$\pm$0.00032          & -16.0$\pm$1.1 & 120.0$\pm$1.4 &                 & WD &          & 4,36 \\
SDSSJ1625+3632   & 16 25 42.1 & $+$36 32 19 &       ...                   & 19.36 & 0.2324$\pm$0.0396            & -95.0$\pm$2.1 & 58.4$\pm$2.7  &                 &    &          & 37 \\
PG2345+318       & 23 48 07.5 & $+$32 04 48 & 5.4$\pm$1.8, -3.1$\pm$2.1   & 14.16 & 0.2409458$\pm$0.0000083      & -10.6$\pm$1.4 & 141.2$\pm$1.1 &                 &    &          & 36 \\
SDSSJ2046-0454   & 20 46 13.4 & $-$04 54 19 &10.9$\pm$13.4, -11.4$\pm$10.9& 16.32 & 0.24311$\pm$0.00001          & 87.6$\pm$5.7  & 134.3$\pm$7.8 &                 &    &          & 35 \\
PG1329+159       & 13 31 53.6 & $+$15 41 18 & -18.1$\pm$1.2, -8.6$\pm$1.7 & 13.51 & 0.249699$\pm$0.000002        & -22.0$\pm$1.2 & 40.2$\pm$1.1  &                 & dM & refl     & 20 \\
FBS0117+396      & 01 20 22.9 & $+$39 50 59 & -4.5$\pm$5.1, -1.7$\pm$5.7  & 15.34 & 0.252013$\pm$0.000013        & -47.3$\pm$1.3 & 37.3$\pm$2.8  &                 & dM & puls,refl& 38 \\
SDSSJ1654+3037   & 16 54 04.2 & $+$30 37 02 & 7.4$\pm$5.3, -8.4$\pm$3.3   & 15.41 & 0.25357$\pm$0.00001          & 40.5$\pm$2.2  & 126.1$\pm$2.6 &                 &    &          & 35 \\
AA Dor           & 05 31 40.4 & $-$69 53 02 & -13.6$\pm$1.2, 52.2$\pm$1.3 & 11.14 & 0.2614$\pm$0.0002            & 1.6$\pm$0.1   & 40.2$\pm$0.1  &                 & dM/bd & refl,ecl & 39 \\
HE0532-4503      & 05 33 40.5 & $-$45 01 35 & 6.7$\pm$4.8, -14.7$\pm$4.8  & 16.08 & 0.2656$\pm$0.0001            & 8.5$\pm$0.1   & 101.5$\pm$0.2 &                 &    &          & 40 \\
GALEXJ0321+4727  & 03 21 39.6 & $+$47 27 19 & 60.1$\pm$1.6, -8.5$\pm$0.9  & 11.72 & 0.265856$\pm$0.000003        & 69.6$\pm$2.2  & 60.8$\pm$4.5  &                 & dM & refl,puls& 41,42 \\
CPD-64$^\circ$481& 05 47 59.3 & $-$64 23 03 & -1.9$\pm$1.0, -30.1$\pm$0.9 & 11.29 & 0.27726315$\pm$0.00000008    & 93.5$\pm$0.1  & 23.8$\pm$0.1  &                 & bd & refl     & 43 \\
KBS13            & 19 26 09.4 & $+$37 20 08 & 4.0$\pm$1.7, -9.9$\pm$2.1   & 13.63 & 0.2923$\pm$0.0004            & 7.5$\pm$0.1   & 22.8$\pm$0.2  &                 & dM & refl     & 44 \\
SDSSJ1021+3010   & 10 21 51.6 & $+$30 10 11 &     ...                     & 18.22 & 0.2966$\pm$0.0001            & -28.4$\pm$4.8 & 114.5$\pm$5.2 &                 &    &          & 27 \\
HS2043+0615      & 20 46 20.8 & $+$06 26 25 & 2.5$\pm$5.9, -9.4$\pm$6.3   & 15.42 & 0.3015$\pm$0.0003            & -43.5$\pm$3.4 & 73.7$\pm$4.3  &                 & dM & refl     & 22 \\
PG0941+280       & 09 43 54.6 & $+$27 46 59 & -14.7$\pm$1.1, -40.1$\pm$2.8& 13.26 & 0.311                        & 73.0$\pm$4.9  & 141.7$\pm$19.4 &                & WD & ecl      & 22 \\
PHL457           & 23 19 24.5 & $-$08 52 37 & -11.2$\pm$2.3, -10.9$\pm$2.4& 12.95 & 0.3131$\pm$0.0002            & 20.7$\pm$0.2  & 13.0$\pm$0.2  &                 & bd & refl,puls& 43 \\
PG1528+104       & 15 31 10.4 & $+$10 15 01 & -18.9$\pm$1.4, -7.2$\pm$1.7 & 13.38 & 0.331$\pm$0.001              & -49.3$\pm$1.0 & 53.3$\pm$1.6  &                 & WD &          & 23 \\
\hline
\end{tabular}
\end{minipage}
\end{center}
\end{table}
\end{landscape}

\addtocounter{table}{-1}
\begin{landscape}
\begin{table}
\begin{center}
\vspace{-0.3cm}
\begin{minipage}{230mm}%
\renewcommand{\footnoterule}{\vspace*{-15pt}}
\caption{{\it continued}}
\vspace{-0.3cm}
\begin{tabular}{lcccccccccll}
\hline
Object            & R.A.      & Dec.        & $\mu_{\alpha}\cos{\delta}\ ,\mu_{\delta}$ & $V$ & Period & $\gamma$      & $K$           & $e$ & Sec. & Variable & Ref. \\
                  & (J2000)     & (J2000)   & (mas yr$^{-1}$)                           & (mag) & (d)  & (\kmps) & (\kmps) &     & type    &          &           \\
\hline
PG1438-029       & 14 40 52.8 & $-$03 08 53 & 7.8$\pm$1.5, -42.9$\pm$1.8  & 13.79 & 0.336                        & ...           & 32.1          &                 &    & refl     & 45 \\
GALEX~J2205-3141 & 22 05 51.8 & $-$31 41 05 & 22.3$\pm$0.9, -2.1$\pm$1.6  & 12.30 & 0.341543$\pm$0.000008        & -19.4$\pm$1.7 & 47.8$\pm$2.2  &                 & dM & refl     & 33 \\
PG1101+249       & 11 04 31.7 & $+$24 39 43 & -29.9$\pm$1.9, 19.2$\pm$2.3 & 12.78 & 0.35386$\pm$0.00014          & -0.8$\pm$0.9  & 134.6$\pm$1.3 &                 & WD &          & 36 \\
PG1232-136       & 12 35 18.7 & $-$13 55 09 & -44.0$\pm$1.7, 5.4$\pm1.7$  & 13.27 & 0.3630$\pm$0.0003            & 4.1$\pm$0.3   & 129.6$\pm$0.4 & 0.060$\pm$0.005 &    &          & 34 \\
Feige 48         & 11 47 14.5 & $+$61 15 32 & -28.1$\pm$3.2, -6.3$\pm2.7$ & 13.42 & 0.376$\pm$0.003              & -47.9$\pm$0.1 & 28.0$\pm$0.2  &                 & WD & puls     & 46 \\
GD 687           & 01 10 18.5 & $-$34 00 26 & -1.3$\pm$1.6, -16.1$\pm1.6$ & 14.08 & 0.37765$\pm$0.00002          & 32.3$\pm$3.0  & 118.3$\pm$3.4 &                 & WD &          & 47 \\
KIC11179657      & 19 02 22.0 & $+$48 50 53 &               ...           &  ...  & 0.3945$\pm$0.0002            & ...           & ...           &                 & dM & refl,puls& 26,48 \\
V 1405 Ori       & 04 44 56.9 & $+$14 21 50 & 3.6$\pm$4.3, -10.9$\pm4.7$  & 15.14 & 0.398                        & -33.6$\pm$5.5 & 85.1$\pm$8.6  &                 & dM & refl,puls& 22,49 \\
KIC2438324       & 19 21 12.9 & $+$37 45 51 &             ...             & ...   & 0.3984944$\pm$0.0000035      &   ...         &   ...         &                 & dM & ref,puls & 50 \\
KPD1946+4340     & 19 47 42.9 & $+$43 47 31 & -9.2$\pm$2.7, -1.4$\pm3.1$  & 14.28 & 0.4037503$\pm$0.0000002      & -5.5$\pm$1.0  & 164.0$\pm$1.9 &                 & WD & ell,ecl,dop & 20,51 \\
SDSSJ0951+0347   & 09 51 01.3 & $+$03 47 57 & -4.8$\pm$4.0, -9.5$\pm$3.6  & 15.90 & 0.4159$\pm$0.0007            & 111.1$\pm$2.5 & 84.4$\pm$4.2  &                 &    &           & 27 \\
$[$CW83$]$ 1419-09& 14 22 40.3 & $-$09 17 22& -6.0$\pm$0.9, -36.9$\pm1.0$ & 12.12 & 0.4178$\pm$0.0002            & 42.3$\pm$0.3  & 109.6$\pm$0.4 & 0.039$\pm$0.005 &    &           & 34 \\
HE0929-0424      & 09 32 02.1 & $-$04 37 37 & -3.1$\pm$4.6, -6.4$\pm$4.4  & 16.16 & 0.4400$\pm$0.0002            & 41.4$\pm$1.0  & 114.3$\pm$1.4 &                 &    &           & 40 \\
KIC2991403       & 19 27 15.9 & $+$38 08 08 &           ...               &  ...  & 0.44312$\pm$0.00008          & ...           & ...           &                 & dM & refl,puls & 52,53 \\
HE0230-4323      & 02 32 54.7 & $-$43 10 28 & -8.5$\pm$1.5, -0.9$\pm$1.5  & 13.77 & 0.4515$\pm$0.0002            & 16.6$\pm$1.0  & 62.4$\pm$1.6  &                 & dM & refl,puls & 34,54 \\ 
GALEXJ2349+3844  & 23 49 47.6 & $+$38 44 42 & -4.3$\pm$1.6, -2.5$\pm$1.1  & 11.72 & 0.462516$\pm$0.000005        & 2.0$\pm$1.0   & 87.9$\pm$2.2  & 0.06$\pm0$.02   & WD & puls      & 41,42 \\
KUV16256+4034    & 16 27 16.5 & $+$40 27 29 & -19.7$\pm$0.9, -13.6$\pm$0.6& 12.49 & 0.4776$\pm$0.0008            & -90.9$\pm$0.9 & 38.7$\pm$1.2  &                 & WD &           & 24 \\
BPS CS 22879-149 & 20 57 15.3 & $-$38 11 51 & 11.9$\pm$2.4, -10.8$\pm$2.4 & 14.24 & 0.478\footnote{Alternate P = 0.964 d.} & 21.9$\pm$2.5 & 63.5$\pm$2.8 &         &    &           & 22 \\
HE1318-2111      & 13 21 15.6 & $-$21 27 18 & 4.5$\pm$1.7, -0.7$\pm$2.0   & 14.77 & 0.487502                     & 48.9$\pm$0.7  & 48.5$\pm$1.2  &                 &    &           & 55,56 \\
PG1544+488       & 15 46 11.7 & $+$48 38 37 & -47.5$\pm$1.4, 32.7$\pm$1.1 & 12.79 & 0.496$\pm$0.002  & 86.6$\pm$0.5/95.0$\pm$0.4 & -25.5$\pm$0.4 &                 & He-sdB &       & 57 \\
SDSSJ1726+2744   & 17 26 24.1 & $+$27 44 19 & 8.0$\pm$3.7, -9.1$\pm$3.8   & 15.99 & 0.50198$\pm$0.00005          & -36.7$\pm$4.8 & 118.9$\pm$3.7 &                 &    &           & 35 \\
PG1743+477       & 17 44 26.4 & $+$47 41 47 & 0.7$\pm$1.2, 12.4$\pm$1.3   & 13.79 & 0.515561$\pm$0.000002        & -65.8$\pm$0.8 & 121.4$\pm$1.0 &                 &    &           & 20 \\
GALEX~J0507+0348 & 05 07 35.7 & $+$03 48 14 & 11.7$\pm$3.0, -4.0$\pm$3.3  & 14.24 & 0.528127$\pm$0.000013        & 96.2$\pm$1.8  & 68.2$\pm$2.5  &                 & WD &           & 33 \\
PG0001+275       & 00 03 55.6 & $+$27 48 37 & 3.3$\pm$1.9, -20.4$\pm$1.2  & 13.32 & 0.529842$\pm$0.000005        & -44.7$\pm$0.5 & 92.8$\pm$0.7  &                 &    &           & 34 \\
PG1519+640       & 15 20 31.4 & $+$63 52 08 & 24.5$\pm$1.0, 33.8$\pm$1.3  & 12.39 & 0.539$\pm$0.003              & 0.9$\pm$0.8   & 36.7$\pm$1.2  &                 & WD &           & 24 \\
HE1059-2735      & 11 01 24.8 & $-$27 51 42 & -11.4$\pm$2.4, 2.0$\pm$2.4  & 15.56 & 0.555624                     & -44.7$\pm$0.6 & 87.7$\pm$0.8  &                 &    &           & 55,56 \\
PG0101+039       & 01 04 21.7 & $+$04 13 37 & 11.7$\pm$0.7, -29.3$\pm$1.0 & 12.06 & 0.569899$\pm$0.000001        & 7.3$\pm$0.2   & 104.5$\pm$0.3 &                 & WD & ell,puls  & 58 \\
EC20182-6534     & 20 22 51.3 & $-$65 25 20 & -12.6$\pm$1.3, -9.2$\pm$2.3 & 13.29 & 0.598819$\pm$0.000006        & 13.5$\pm$1.9  & 59.7$\pm$3.2  &                 &    &           & 24 \\
PG1725+252       & 17 27 57.4 & $+$25 08 36 & -20.1$\pm$1.4, 7.4$\pm$1.2  & 13.06 & 0.601507$\pm$0.000003        & -60.0$\pm$0.6 & 104.5$\pm$0.7 &                 &    &           & 20 \\
PG1247+554       & 12 50 04.3 & $+$55 06 03 & -76.5$\pm$3.6, -7.3$\pm$2.0 & 12.26 & 0.602740$\pm$0.000006        & 13.8$\pm$0.6  & 32.2$\pm$1.0  &                 &    &           & 59 \\
HD188112         & 19 54 31.4 & $-$28 20 21 & 33.7$\pm$0.7, 22.5$\pm$1.2  & 10.18 & 0.6065812$\pm$0.0000005      & 26.7$\pm$0.2  & 188.4$\pm$0.2 &                 & WD &           & 34 \\
PG1648+536       & 16 49 59.9 & $+$53 31 32 & 0.9$\pm$1.3, -15.4$\pm$2.1  & 14.09 & 0.6109107$\pm$0.0000004      & -69.9$\pm$0.9 & 109.0$\pm$1.3 &                 & WD &           & 24 \\
SDSSJ1522-0130   & 15 22 22.1 & $-$01 30 18 &    ...                      & 17.81 & 0.67162$\pm$0.00003          & -79.5$\pm$2.7 & 80.1$\pm$3.5  &                 &    &           & 27 \\
SDSSJ2256+0656   & 22 56 38.3 & $+$06 56 51 & -2.0$\pm$3.7, -1.1$\pm$4.3  & 15.31 & 0.7004$\pm$0.0001            & -7.3$\pm$2.1  & 105.3$\pm$3.4 &                 &    &           & 35 \\
EC22202-1834     & 22 22 58.1 & $-$18 19 10 & 10.3$\pm$1.8, -15.7$\pm$1.7 & 13.80 & 0.70471$\pm$0.00005          & -5.5$\pm$3.9  & 118.6$\pm$5.8 &                 &    &           & 24 \\
PG1248+164       & 12 50 50.3 & $+$16 10 03 & 11.6$\pm$2.0, -8.8$\pm$2.2  & 14.46 & 0.73232$\pm$0.00002          & -16.2$\pm$1.3 & 61.8$\pm$1.1  &                 &    &           & 20 \\
JL82             & 21 36 01.3 & $-$72 48 27 & 15.0$\pm$1.2, -17.6$\pm$0.9 & 12.37 & 0.7371$\pm$0.0005            & -1.6$\pm$0.8  & 34.6$\pm$1.0  &                 & dM & refl,puls & 34,60 \\
PG0849+319       & 08 52 54.6 & $+$31 43 37 & -10.8$\pm$1.6, -9.6$\pm$1.8 & 14.58 & 0.74507$\pm$0.00001          & 64.0$\pm$1.5  & 66.3$\pm$2.1  &                 &    &           & 20 \\
SDSSJ1505+1108   & 15 05 13.5 & $+$11 08 37 & -17.7$\pm$8.3, -29.4$\pm$8.0& 15.38 & 0.74773$\pm$0.00005          & -77.1$\pm$1.2 & 97.2$\pm$1.8  &                 &    &           & 35 \\
EQ Psc           & 23 34 34.6 & $-$01 19 37 & -8.7$\pm$1.8, -40.4$\pm$1.3 & 12.78 & 0.801\footnote{based on Kepler light curves} &  ...  & ...   &                 & dM & refl,puls & 61 \\
EC02200-2338     & 02 22 19.8 & $-$23 24 56 & 34.2$\pm$1.6, -12.3$\pm$1.0 & 12.01 & 0.8022$\pm$0.0007            & 20.7$\pm$2.3  & 96.4$\pm$1.4  &                 &    &           & 24 \\
KPD2215+5037     & 22 17 29.7 & $+$50 52 59 & 9.7$\pm$4.4, 14.4$\pm$2.8   & 13.64 & 0.809146$\pm$0.000002        & -7.2$\pm$1.0  & 86.0$\pm$1.5  &                 &    &           & 24 \\
Ton S 183        & 01 01 17.6 & $-$33 42 45 & -7.1$\pm$1.2, -15.2$\pm$1.0 & 12.60 & 0.8277$\pm$0.0002            & 50.5$\pm$0.8  & 84.8$\pm$1.0  &                 &    &           & 34 \\
EC13332-1424     & 13 35 53.5 & $-$14 40 13 & -12.6$\pm$1.8, 16.6$\pm$2.0 & 13.40 & 0.82794$\pm$0.00001          & -53.2$\pm$1.8 & 104.1$\pm$3.0 &                 &    &           & 24 \\
PG1627+017       & 16 29 35.3 & $+$01 38 19 & -2.0$\pm$2.1, -9.7$\pm$4.4  & 12.94 & 0.8292056$\pm$0.0000014      & -54.2$\pm$0.3 & 70.1$\pm$0.1  &                 & WD & puls      & 62 \\
EC21556-5552     & 21 59 00.7 & $-$55 38 04 & 3.4$\pm$1.3, 6.4$\pm$1.3    & 13.09 & 0.8340$\pm$0.0007            & 31.4$\pm$2.0  & 65.0$\pm$3.4  &                 &    &           & 24 \\
\hline
\end{tabular}
\end{minipage}
\end{center}
\end{table}
\end{landscape}

\addtocounter{table}{-1}
\begin{landscape}
\begin{table}
\begin{center}
\vspace{-0.3cm}
\begin{minipage}{230mm}%
\renewcommand{\footnoterule}{\vspace*{-15pt}}
\caption{{\it continued}}
\vspace{-0.3cm}
\begin{tabular}{lcccccccccll}
\hline
Object            & R.A.      & Dec.        & $\mu_{\alpha}\cos{\delta}\ ,\mu_{\delta}$ & $V$ & Period & $\gamma$      & $K$           & $e$ & Sec. & Variable & Ref. \\
                  & (J2000)     & (J2000)   & (mas yr$^{-1}$)                           & (mag) & (d)  & (\kmps) & (\kmps) &     & type    &          &           \\
\hline
PG1230+052       & 12 33 12.6 & $+$04 57 38 & -10.6$\pm$2.4, -17.6$\pm$2.3& 13.24 & 0.837177$\pm$0.000003        & -43.1$\pm$0.7 & 40.4$\pm$1.2  &                 & WD &           & 24 \\
PG1116+301       & 11 19 04.8 & $+$29 51 53 & -13.0$\pm$3.3, -7.7$\pm$4.1 & 14.37 & 0.85621$\pm$0.00003          & -0.2$\pm$1.1  & 88.5$\pm$2.1  &                 &    &           & 20 \\
PG0918+029       & 09 21 28.2 & $+$02 46 02 & -23.9$\pm$2.4, -22.0$\pm$1.4& 13.30 & 0.87679$\pm$0.00002          & 104.4$\pm$1.7 & 80.0$\pm$2.6  &                 &    &           & 20 \\
EC12408-1427     & 12 43 30.0 & $-$14 43 49 & -24.1$\pm$1.6, 6.2$\pm$1.9  & 12.83 & 0.90243$\pm$0.00001          & -52.2$\pm$1.2 & 58.6$\pm$1.5  &                 &    &           & 24 \\
HE2135-3749      & 21 38 44.2 & $-$37 36 15 & 26.5$\pm$1.4, -0.8$\pm$1.2  & 13.90 & 0.9240$\pm$0.0003            & 45.0$\pm$0.5  & 90.5$\pm$0.6  &                 &    &           & 40 \\
PB5333           & 23 19 55.3 & $+$04 52 35 & 28.7$\pm$3.3, -26.6$\pm$2.2 & 12.81 & 0.92560$\pm$0.00012          & -95.3$\pm$1.3 & 22.4$\pm$0.8  &                 &    &           & 63 \\
HS2359+1942      & 00 02 08.5 & $+$19 59 13 & -12.1$\pm$2.8, -1.7$\pm$3.9 & 15.64 & 0.93261$\pm$0.00005          & -96.1$\pm$6.0 & 107.4$\pm$8.3 &                 & WD &           & 22 \\
PG1452+198       & 14 54 39.8 & $+$19 37 01 & 0.8$\pm$1.8, -17.9$\pm$1.3  & 12.48 & 0.96498$\pm$0.00004          & -9.1$\pm$2.1  & 86.8$\pm$1.9  &                 &    &           & 24 \\
SDSSJ1508+4940   & 15 08 29.0 & $+$49 40 50 &  ...                        & 17.52 & 0.967164$\pm$0.000009        & -60.0$\pm$10.7& 93.6$\pm$5.8  &                 &    &           & 27 \\
PG1000+408       & 10 03 54.3 & $+$40 34 18 & -1.7$\pm$1.9, -15.9$\pm$1.6 & 13.29 & 1.049343$\pm$0.000005        & 56.6$\pm$3.4  & 63.5$\pm$3.0  &                 & WD &           & 24 \\
SDSSJ1132-0636   & 11 32 41.6 & $-$06 36 52 &  ...                        & 16.27 & 1.06$\pm$0.02                & 8.3$\pm$2.2   & 41.1$\pm$4.0  &                 &    &           & 27 \\
GALEX~J1731+0647 & 17 31 53.7 & $+$06 47 06 & -18.9$\pm$2.0, -1.3$\pm$2.1 & 14.09 & 1.17334$\pm$0.00004          & -39.1$\pm$3.0 & 87.7$\pm$3.0  &                 & WD &           & 33 \\
HE1421-1206      & 14 24 08.8 & $-$12 20 20 & -7.8$\pm$2.8, -6.8$\pm$2.2  & 15.51 & 1.188                        & -86.2$\pm$1.1 & 55.5$\pm$2.0  &                 &    &           & 55 \\
PG2331+038       & 23 33 58.2 & $+$04 03 56 & -10.2$\pm$2.8, -16.7$\pm$3.3& 14.63 & 1.204964$\pm$0.000003        & -9.5$\pm$1.1  & 93.5$\pm$1.9  &                 &    &           & 24 \\
HE1047-0436      & 10 50 26.9 & $-$04 52 36 & -6.4$\pm$2.5, 0.1$\pm$2.7   & 14.95 & 1.213253                     & 25            & 94            &                 & WD &           & 64 \\
GALEX~J2254-5515 & 22 54 44.1 & $-$55 15 05 & 29.7$\pm$1.3, 6.2$\pm$1.5   & 12.12 & 1.22702$\pm$0.00005          & 4.2$\pm$2.0   & 79.7$\pm$2.6  &                 & WD &           & 33 \\
PG0133+114       & 01 36 26.2 & $+$11 39 32 & 22.0$\pm$1.7, -20.3$\pm$1.7 & 12.30 & 1.23787$\pm$0.00003          & -0.3$\pm$0.2  & 82.0$\pm$0.3  & 0.025$\pm$0.005 &    &           & 34 \\
PG1512+244       & 15 14 32.5 & $+$24 10 41 & -41.9$\pm$1.1, 3.0$\pm$0.9  & 13.18 & 1.26978$\pm$0.00002          & -2.9$\pm$1.0  & 92.7$\pm$1.5  &                 &    &           & 20 \\
$[$CW83$]$ 1735+22 & 17 37 26.4 & $+$22 08 58 & -24.2$\pm$0.8, 0.1$\pm$1.6& 11.86 & 1.280$\pm$0.006              & 20.6$\pm$0.4  & 104.6$\pm$0.5 &                 & WD &           & 34 \\
SDSSJ0118-0025   & 01 18 57.2 & $-$00 25 46 & 5.3$\pm$3.8, -9.6$\pm$4.2   & 14.80 & 1.30$\pm$0.02                & 37.7$\pm$1.8  & 54.8$\pm$2.9  &                 &    &           & 27 \\
HE2150-0238      & 21 52 35.8 & $-$02 24 32 & ...                         & 16.08 & 1.3209$\pm$0.0050            & -32.5$\pm$0.9 & 96.3$\pm$1.4  &                 &    &           & 40 \\
KPD2040+3955     & 20 42 33.9 & $+$40 05 42 & -12.9$\pm$2.6, -14.1$\pm$3.1& 14.48 & 1.482860$\pm$0.000004        & -16.4$\pm$1.0 & 94.0$\pm$1.5  &                 & WD &           & 24 \\
SDSSJ0023-0029   & 00 23 24.0 & $-$00 29 53 & 24.1$\pm$2.8, 8.5$\pm$7.1   & 15.58 & 1.4876$\pm$0.0001            & 16.4$\pm$2.1  & 81.8$\pm$2.9  &                 &    &           & 35 \\
HD49798          & 06 48 04.7 & $-$44 18 58 & -5.1$\pm$1.0, 6.0$\pm$1.0   &  8.29 & 1.547671$\pm$0.000011        & 13.5$\pm$2.2  & 119.2$\pm$3.2 &                 & WD/N& ecl,X-ray& 65,66 \\
KIC7664467       & 18 56 07.1 & $+$43 19 19 &  ...                        & ...   & 1.559110                     &  ...          &  ...          &                 &    & puls      & 48,53 \\
HD171858         & 18 37 56.7 & $-$23 11 35 & -16.8$\pm$1.4, -21.2$\pm$1.4&  9.86 & 1.63280$\pm$0.00005          & 62.5$\pm$0.1  & 87.8$\pm$0.2  &                 & WD &           & 34 \\
PG1403+316       & 14 05 59.8 & $+$31 24 37 & -25.2$\pm$2.0, 2.7$\pm$1.8  & 13.50 & 1.73846$\pm$0.00001          & -2.1$\pm$0.9  & 58.5$\pm$1.8  &                 &    &           & 24 \\
PG1716+426       & 17 18 03.9 & $+$42 34 13 & 8.2$\pm$1.3, -19.4$\pm$1.6  & 13.93 & 1.77732$\pm$0.00005          & -3.9$\pm$0.8  & 70.8$\pm$1.0  &                 &    & puls      & 20,67 \\
SDSSJ1346+2817   & 13 46 32.6 & $+$28 17 22 & -14.0$\pm$3.9, -7.4$\pm$4.2 & 14.91 & 1.96$\pm$0.03                & 1.2$\pm$1.2   & 85.6$\pm$3.4  &                 &    &           & 27\\
NGC188/II-91     & 00 47 52.3 & $+$85 19 08 & -5.7$\pm$6.7, -1.2$\pm$6.4  & 16.07 & 2.15                         &  ...          & 22.0:         &                 &    &           & 68\\
PG1300+279       & 13 02 41.8 & $+$27 40 42 & -7.8$\pm$1.5, -7.8$\pm$1.9  & 14.26 & 2.2593$\pm$0.0001            & -3.1$\pm$0.9  & 62.8$\pm$1.6  &                 &    &           & 20 \\
CPD-20$^\circ$ 1123 & 06 06 13.4 & $-$20 21 07 & 9.3$\pm$1.4, -15.4$\pm$1.3 & 12.17 & 2.3098$\pm$0.0003          & -6.3$\pm$1.2  & 43.5$\pm$0.9  &                 &    &           & 69 \\
HD149382         & 16 34 23.3 & $-$04 00 52 & -8.7$\pm$0.8, -1.9$\pm$0.5  &  8.94 & 2.391$\pm$0.002              & 25.3$\pm$0.1  & 2.3$\pm$0.1   &                 & bd &           & 70 \\
PG1538+269       & 15 40 23.4 & $+$26 48 30 & 7.8$\pm$1.4, -5.1$\pm$1.7   & 13.86 & 2.500                        & ...           & 75            &                 & WD &           & 71,72 \\
GALEX~J1632+0759 & 16 32 01.4 & $+$07 59 40 & 7.0$\pm$1.1, -2.7$\pm$1.3   & 12.76 & 2.9516$\pm$0.0006            & -31.6$\pm$2.7 & 54.9$\pm$4.6  &                 & WD &           & 33,73 \\
PG1253+284       & 12 56 04.9 & $+$28 07 19 & -11.8$\pm$1.2, 0.5$\pm$1.5  & 12.76 & 3.01634$\pm$0.00005          & 17.8$\pm$0.6  & 24.8$\pm$0.9  &                 &    &           & 24 \\
PG0958-073       & 10 00 47.3 & $-$07 33 31 & -43.1$\pm$1.8, -2.2$\pm$3.2 & 13.56 & 3.18095$\pm$0.00007          & 90.5$\pm$0.8  & 27.6$\pm$1.4  &                 &    &           & 24 \\
KIC10553698A     & 19 53 08.4 & $+$47 43 00 & 36.1$\pm$4.5, 11.5$\pm$4.3  & 14.90 & 3.387$\pm$0.014              & 52.1$\pm$1.5  & 64.8$\pm$2.2  &                 & WD & puls,dop  & 74 \\
KPD0025+5402     & 00 28 29.0 & $+$54 19 15 & -8.9$\pm$7.6, -8.3$\pm$3.8  & 13.91 & 3.571$\pm$0.001              & -7.8$\pm$0.7  & 40.2$\pm$1.1  &                 &    &           & 20 \\
PB7352           & 22 55 43.2 & $-$06 59 40 & -2.0$\pm$1.0, 2.1$\pm$1.1   & 12.26 & 3.62166$\pm$0.00005          & -2.1$\pm$0.3  & 60.8$\pm$0.3  &                 &    &           & 34 \\
PG0934+186       & 09 37 16.3 & $+$18 25 11 & -14.8$\pm$1.3, -9.5$\pm$0.8 & 13.13 & 4.050$\pm$0.01               & 7.7$\pm$3.2   & 60.3$\pm$2.4  &                 &    &           & 24 \\
Ton S 135        & 00 03 22.1 & $-$23 38 58 & 4.2$\pm$2.5, -17.6$\pm$1.7  & 13.28 & 4.122$\pm$0.008              & -3.7$\pm$1.1  & 41.4$\pm$1.5  &                 &    &           & 34 \\
EC20369-1804     & 20 39 46.5 & $-$17 54 04 & 9.2$\pm$1.3, -8.3$\pm$1.6   & 13.35 & 4.5095$\pm$0.0004            & 7.2$\pm$1.6   & 51.5$\pm$2.3  &                 &    &           & 24 \\
SDSSJ1832+6309   & 18 32 49.0 & $+$63 09 10 & 2.1$\pm$4.6, 6.1$\pm$4.5    & 15.70 & 5.4$\pm$0.2                  & -32.5$\pm$2.1 & 62.1$\pm$3.3  &                 &    &           & 27\\
PG0839+399       & 08 43 12.7 & $+$39 44 50 & -3.5$\pm$1.9, -10.7$\pm$2.2 & 14.34 & 5.622$\pm$0.002              & 23.2$\pm$1.1  & 33.6$\pm$1.5  &                 &    &           & 20 \\
PG1244+113       & 12 47 06.6 & $+$11 03 14 & 6.6$\pm$1.7, -0.7$\pm$3.2   & 14.14 & 5.75211$\pm$0.00009          & 7.4$\pm$0.8   & 54.4$\pm$1.4  &                 &    &           & 24 \\
CD-24$^\circ$ 731& 01 43 48.5 & $-$24 05 10 & 84.3$\pm$2.0, -48.6$\pm$1.2 & 11.72 & 5.85$\pm$0.30                & 20$\pm$5      & 63$\pm$3      &                 & WD &           & 34 \\
HE1115-0631      & 11 18 11.6 & $-$06 47 32 & -13.8$\pm$2.7, -10.3$\pm$3.2& 15.08 & 5.870                        & 87.1$\pm$1.3  & 61.9$\pm$1.1  &                 &    &           & 55,56 \\
PG0907+123       & 09 10 25.4 & $+$12 08 27 & -9.1$\pm$1.4, -3.6$\pm$1.7  & 13.92 & 6.1163$\pm$0.0006            & 56.3$\pm$1.1  & 59.8$\pm$0.9  &                 &    & puls      & 20,75 \\
\hline
\end{tabular}\\
\end{minipage}
\end{center}
\end{table}
\end{landscape}

\addtocounter{table}{-1}
\begin{landscape}
\begin{table}
\begin{center}
\vspace{-0.3cm}
\begin{minipage}{230mm}%
\renewcommand{\footnoterule}{\vspace*{-15pt}}
\caption{{\it continued}}
\vspace{-0.3cm}
\begin{tabular}{lcccccccccll}
\hline
Object            & R.A.      & Dec.        & $\mu_{\alpha}\cos{\delta}\ ,\mu_{\delta}$ & $V$ & Period & $\gamma$      & $K$           & $e$ & Sec. & Variable & Ref. \\
                  & (J2000)     & (J2000)   & (mas yr$^{-1}$)                           & (mag) & (d)  & (\kmps) & (\kmps) &     & type    &          &           \\
\hline
PG1032+406       & 10 35 16.6 & $+$40 21 14 & -84.1$\pm$3.4, -38.1$\pm$3.9& 11.47 & 6.779$\pm$0.001              & 24.5$\pm$0.5  & 33.7$\pm$0.5  &                 &    &           & 20 \\
SDSSJ0952+6258   & 09 52 38.9 & $+$62 58 18 & 1.9$\pm$2.7, -13.8$\pm$3.3  & 14.69 & 6.98$\pm$0.04                & -35.4$\pm$3.6 & 62.5$\pm$3.4  &                 & WD &           & 27 \\
HE1448-0510      & 14 51 13.1 & $-$05 23 17 & 1.1$\pm$2.4, -6.1$\pm$2.7   & 14.61 & 7.1588$\pm$0.0130            & -45.5$\pm$0.8 & 53.7$\pm$1.1  &                 &    &           & 40 \\
PG1439-013       & 14 42 27.5 & $-$01 32 46 & -9.3$\pm$1.8, -1.4$\pm$2.2  & 13.87 & 7.2914$\pm$0.0005            & -53.7$\pm$1.6 & 50.7$\pm$1.5  &                 &    &           & 24 \\
SDSSJ0321+0538   & 03 21 38.8 & $+$05 38 40 & 0.5$\pm$3.7, -5.8$\pm$4.4   & 15.05 & 7.4327$\pm$0.0004            & -16.7$\pm$2.1 & 39.7$\pm$2.8  &                 &    &           & 27 \\
PHL861           & 00 51 03.9 & $-$19 59 59 & 1.5$\pm$2.6, -28.8$\pm$1.7  & 14.83 & 7.4436$\pm$0.0150            & -26.5$\pm$0.4 & 47.9$\pm$0.4  &                 &    &           & 40 \\
PG0940+068       & 09 42 55.0 & $+$06 35 37 & 11.3$\pm$2.4, -4.0$\pm$2.7  & 13.69 & 8.330$\pm$0.003              & -16.7$\pm$1.4 & 61.2$\pm$1.4  &                 &    &           & 59 \\
Feige108         & 23 16 12.4 & $-$01 50 35 & -0.4$\pm$1.0, -14.1$\pm$1.0 & 13.00 & 8.7465$\pm$0.0010            & 45.8$\pm$0.6  & 50.2$\pm$1.0  &                 &    &           & 63 \\
EC20260-4757     & 20 29 34.1 & $-$47 47 26 & -3.6$\pm$1.3, 0.0$\pm$1.3   & 13.80 & 8.952$\pm$0.002              & 56.5$\pm$1.6  & 57.1$\pm$1.9  &                 &    &           & 24 \\
FF Aqr           & 22 00 36.4 & $-$02 44 27 & 36.0$\pm$0.6, -10.7$\pm$0.8 &  9.57 & 9.20803$\pm$0.00004          & 24.5$\pm$1.7  & 116.5$\pm$2.1 &              & K0III & ecl,refl  & 76,77 \\
PG1110+294       & 11 13 04.5 & $+$29 07 46 & -7.3$\pm$1.3, -8.5$\pm$1.9  & 14.11 & 9.415$\pm$0.002              & -15.2$\pm$0.9 & 58.7$\pm$1.2  &                 &    &           & 20 \\
KIC11558725      & 19 26 34.2 & $+$49 30 30 & -28.6$\pm$5.5, -5.7$\pm$5.3 & 14.86 & 10.0545$\pm$0.0048           & -66.1$\pm$1.4 & 58.1$\pm$1.7  &                 &    & puls,dop  & 78 \\
PG1558-007       & 16 01 14.0 & $-$00 51 42 & -8.7$\pm$3.1, -15.8$\pm$4.5 & 13.54 & 10.3495$\pm$0.0006           & -71.9$\pm$0.7 & 42.8$\pm$0.8  &                 &    &           & 24 \\
LB1516           & 23 01 56.1 & $-$48 03 48 & 9.1$\pm$1.7, 1.2$\pm$1.4    & 12.86 & 10.3598$\pm$0.0005           & 14.3$\pm$1.1  & 48.6$\pm$4.4  &                 & WD & puls      & 22,79\\
CS1246           & 12 49 37.6 & $-$63 32 10 & -11.2$\pm$3.7, -1.2$\pm$3.7 & 14.37 & 14.104$\pm$0.011             & 67.3$\pm$1.7  & 16.6$\pm$0.6  &                 &    & puls      & 80\\
KIC7668647       & 19 05 06.4 & $+$43 18 31 & -11.2$\pm$5.6, -36.2$\pm$5.7& 15.22 & 14.1742$\pm$0.0042           & -27.4$\pm$1.3 & 38.9$\pm$1.9  &                 & WD & puls,dop  & 81 \\
PG1619+522       & 16 20 38.8 & $+$52 06 09 & -4.3$\pm$0.9, 10.4$\pm$0.7  & 13.24 & 15.357$\pm$0.008             & -52.5$\pm$1.1 & 35.2$\pm$1.1  &                 &    &           & 20 \\
PG0919+273       & 09 22 39.8 & $+$27 02 25 & 23.3$\pm$0.8, -27.1$\pm$1.1 & 12.66 & 15.5830$\pm$0.0005           & -68.6$\pm$0.6 & 41.5$\pm$0.8  &                 &    &           & 24 \\
EGB 5            & 08 11 12.8 & $+$10 57 17 & -17.9$\pm$1.5, 9.7$\pm$1.9  & 13.81 & 16.537$\pm$0.003             & 68.5$\pm$0.7  & 16.1$\pm$0.8  & 0.098$\pm$0.048 &    &           & 82 \\
HD 185510        & 19 39 38.8 & $-$06 03 49 & 22.8$\pm$1.1, -27.4$\pm$1.1 &  8.47 & 20.66187$\pm$0.00058         & -21.9$\pm$0.1 & 93.7$\pm$2.5  &                 & K0III/IV & ecl,refl? & 83,84 \\
PG0850+170       & 08 53 23.7 & $+$16 49 35 & 0.8$\pm$1.4, -6.8$\pm$1.6   & 14.00 & 27.81$\pm$0.05               & 32.2$\pm$2.8  & 33.5$\pm$3.1  &                 &    & puls      & 20,85 \\
59 Cyg           & 20 59 49.6 & $+$47 31 15 & 7.3$\pm$1.0, 2.5$\pm$1.0    &  4.75 & 28.1871$\pm$0.0011           & -10.4$\pm$0.8 & 121.3$\pm$1.1 & 0.141$\pm$0.008 & Be &           & 86 \\
FY CMa           & 07 26 59.5 & $-$23 05 10 & -7.8$\pm$1.0, 4.8$\pm$1.0   &  5.56 & 37.257$\pm$0.003             & 31.2$\pm$1.7  & 128.2$\pm$2.2 &                 & Be &           & 87 \\
$\phi$ Per       & 01 43 39.6 & $+$50 41 19 & 24.6$\pm$1.0, -14.0$\pm$1.0 &  4.06 & 126.6731$\pm$0.0071          & -6.1$\pm$0.5  & 81.3$\pm$0.6  &                 & Be &           & 88,89  \\
BD-11$^\circ$ 162& 00 52 15.1 & $-$10 39 46 & -29.6$\pm$1.0, -30.1$\pm$1.7& 11.17 & 421$\pm$3                    & 2.3$\pm$0.2   & 7.9$\pm$0.3   &                 & G  &           & 90 \\
PG1701+359       & 17 03 21.5 & $+$35 48 49 & -57.9$\pm$4.0, 20.4$\pm$0.9 & 13.20 & 738$\pm$4                    & -120.1$\pm$0.2 & ...          & 0.07$\pm$0.04   & G/K&           & 91 \\
PG1104+243       & 11 07 26.2 & $+$24 03 11 & -65.9$\pm$1.2, -25.1$\pm$1.2& 11.32 & 753.2$\pm$0.8                & -15.7$\pm$0.1 & 6.5$\pm$0.8   &                 & F/K&           & 92 \\
PG1018-047       & 10 21 10.6 & $-$04 56 20 & -15.1$\pm$2.1, -11.9$\pm$2.6& 13.32 & 755.9$\pm$5.1                & 38.0$\pm$0.9  & 12.6$\pm$0.8  & 0.246$\pm$0.052 & K4-K6 &        & 93 \\
PG1449+653       & 14 50 36.1 & $+$65 05 52 & -21.7$\pm$2.1, 13.6$\pm$1.0 & 13.62 & 909$\pm$2                    & -135.5$\pm$0.2& 12.8$\pm$1.1  & 0.11$\pm$0.02   &G/K &           & 91 \\
PG1338+611       & 13 40 14.7 & $+$60 52 48 & 14.5$\pm$0.9, -61.4$\pm$0.8 & 11.62 & 937.5$\pm$1.1                & 32.6$\pm$0.1  & 15.2$\pm$1.8  & 0.15$\pm$0.02   &G2-G7&          & 92,94 \\
BD+34$^\circ$ 1543& 07 10 07.7 & $+$34 24 54 & 35.2$\pm$1.0, -61.8$\pm$0.8& 10.16 & 972$\pm$2                    & 33.1$\pm$0.2  & 19.3$\pm$0.2  & 0.16$\pm$0.01   &MS  &           & 94 \\
PG1317+123       & 13 19 53.6 & $+$12 03 58 & -6.9$\pm$1.1, -1.6$\pm$1.1  & 11.41 & 1179$\pm$12                  & 40.3$\pm$0.2  & 15.5$\pm$1.7  &                 &G8V &           & 92 \\
BD-7$^\circ$ 5977& 23 17 46.8 & $-$06 28 31 & 7.3$\pm$1.7, 1.1$\pm$1.3    & 10.45 & 1195$\pm$30                  & -8.73$\pm$0.02& 6.1$\pm$0.8   &                 &K2III&          & 95 \\
BD+29$^\circ$ 3070& 17 38 21.2 & $+$29 08 47 & -6.4$\pm$0.5, 22.1$\pm$0.9 & 10.34 & 1283$\pm$63                  & -57.6$\pm$0.9 & 16.6$\pm$0.6  & 0.15$\pm$0.01   & MS &           & 94 \\
TYC3871-835-1    & 15 15 38.2 & $+$56 53 45 & -33.7$\pm$0.6, 3.2$\pm$0.5  & 11.41 & 1363$\pm$25                  & -15.03$\pm$0.03& 4.8$\pm$0.3  &                 & G0 &           & 95 \\
\hline
\end{tabular}\\
Variable: puls - sdB pulsator, refl - reflection effect, ecl - eclipsing binary, ell - ellipsoidal variations, dop - Doppler beaming.\\
References: (1) \citet{ven2012}; (2) \citet{gei2013a}; (3) \citet{sch2014a}; (4) \citet{max2002}; (5) \citet{oto2006a}; (6) \citet{koe1998}; (7) \citet{oro1999};
(8) \citet{max2000a}; (9) \citet{bil2000}; (10) \citet{gei2007}; (11) \citet{dre2001}; (12) \citet{gei2011b}; (13) \citet{kil2000}; (14) \citet{vuc2007};
(15) \citet{alm2012}; (16) \citet{ost2007}; (17) \citet{kup2014}; (18) \citet{ede2008}; (19) \citet{bar2013b}; (20) \citet{mor2003}; (21) \citet{pol2007}; (21) \citet{gei2014};
(23) \citet{ost2010b}; (24) \citet{cop2011}; (25) \citet{ost2011}; (26) \citet{tel2012b}; (27) \citet{kup2015}; (28) \citet{sch2013}; (29) \citet{sil2012}; (30) \citet{for2010}; (31) \citet{sch2014b}; (32) \citet{heb2004};
(33) This work; (34) \citet{ede2005}; (35) \citet{gei2011c}; (36) \citet{mor1999}; (37) \citet{kil2011}; (38) \citet{ost2013}; (39) \citet{mul2010}; (40) \citet{kar2006}; (41) \citet{kaw2010a}; 
(42) \citet{kaw2012a}; (43) \citet{sch2014c}; (44) \citet{for2008}; (45) \citet{gre2005}; (46) \citet{oto2004}; (47) \citet{gei2010a}; (48) \citet{ost2010c}; (49) \citet{ree2010}; (50) \citet{pab2011};
(51) \citet{blo2011}; (52) \citet{kaw2010b}; (53) \citet{tel2014a}; (54) \citet{kil2010}; (55) \citet{gei2011a}; (56) \citet{nap2004}; (57) \citet{sen2014}; (58) \citet{gei2008}; (59) \citet{max2000b}; 
(60) \citet{koe2009}; (61) \citet{jef2014}; (62) \citet{for2006}; (63) \citet{ede2004}; (64) \citet{nap2001}; (65) \citet{tha1970}; (66) \citet{mer2013}; (67) \citet{ree2004}; (68) \citet{gre2004};
(69) \citet{nas2012}; (70) \citet{gei2009}; (71) \citet{fos1991}; (72) \citet{saf1998}; (73) \citet{bar2014}; (74) \citet{ost2014}; (75) \citet{koe2010a}; (76) \citet{vac2003}; (77) \citet{vac2007};
(78) \citet{tel2012a}; (79) \citet{koe2010b}; (80) \citet{bar2011}; (81) \citet{tel2014b}; (82) \citet{gei2011d}; (83) \citet{jef1992}; (84) \citet{fek1993}; (85) \citet{gre2003}; (86) \citet{pet2013}; 
(87) \citet{pet2008}; (88) \citet{boz1995}; (89) \citet{gie1998}; (90) \citet{ost2012}; (91) \citet{bar2013a}; 
(92) \citet{bar2012}; (93) \citet{dec2012}; (94) \citet{vos2013}; (95) \citet{vos2014}.
\end{minipage}
\end{center}
\end{table}
\end{landscape}

\begin{landscape}
\begin{table}
\begin{center}
\vspace{-0.3cm}
\begin{minipage}{230mm}%
\renewcommand{\footnoterule}{\vspace*{-15pt}}
\caption{Kinematics ($U,V,W$),stellar parameters and absolute $V$ magnitude ($M_V$) of known binaries. \label{tbl_sample_kine}}
\begin{tabular}{lrrrcccclrrrcccc}
\hline
Name                &$U$  &$V$  &$W$ & \teff       &$\log{g}$ & $M_V$& Ref. &Name                &$U$  &$V$  &$W$  & \teff       &$\log{g}$& $M_V$& Ref.  \\
 &(\kmps)&(\kmps)&(\kmps) & (K) & (c.g.s.) &(mag)& & &(\kmps)&(\kmps)&(\kmps) & (K) & (c.g.s.) &(mag)& \\
\hline
CD-30 11223         &   41 &   -8 &    9 &30150 &  5.72 &  4.62 &    1 & PHL 457             &   56 &   -1 &   -9 &26500 &  5.38 &  4.04 &   35\\
SDSSJ1622+4730      &  248 & -209 &   69 &29000 &  5.65 &  4.53 &    2 & PG1528+104          &  -38 &  -63 &   -3 &27200 &  5.46 &  4.18 &   36\\
PG1017-086          &  -38 &   45 &   20 &30300 &  5.61 &  4.32 &    3 & PG1438-029          &  101 &  -96 &  -90 &27700 &  5.50 &  4.25 &   29\\
NGC6121-V46         &   38 &    1 &   15 &16197 &  5.75 &  6.58 &    4 & GALEX J2205-3141    &  -30 &   -7 &    3 &28650 &  5.68 &  4.63 &   24\\
KPD0422+5421        &   44 &  -54 &   -3 &25000 &  5.40 &  4.22 &    5 & PG1101+249          &  -43 &   19 &  -12 &29700 &  5.90 &  5.09 &   37\\
KPD1930+2752        &  -39 &   31 &   42 &35200 &  5.61 &  4.02 &    6 & PG1232-136          &  -90 &  -46 &   12 &29600 &  5.71 &  4.63 &   25\\
HS0705+6700         &   17 &  -26 &  -34 &28800 &  5.40 &  3.91 &    7 & Feige 48            &  -45 &  -71 &  -44 &29500 &  5.54 &  4.21 &   38\\
SDSSJ08205+0008     &   39 &  -40 &    1 &26700 &  5.48 &  4.27 &    8 & GD 687              &   60 &  -58 &  -15 &24350 &  5.32 &  4.05 &   26\\
PG1336-018          &   13 &  -45 &  -13 &31327 &  5.59 &  4.20 &    9 & KIC11179657         &   10 &    3 &    7 &26000 &  5.14 &  3.47 &   39\\
NSVS14256825        &   -6 &   -7 &  -10 &40000 &  5.50 &  3.55 &   10 & V 1405 Ori          &   59 &  -73 &   -6 &35100 &  5.66 &  4.14 &   16\\
HS2231+2441         &    8 &  -59 & -107 &28370 &  5.39 &  3.92 &   11 & KIC2438324          &    9 &    3 &    7 &27098 &  5.69 &  4.77 &   40\\
UVEXJ0328+5035      &  -16 &   41 &  -14 &28500 &  5.50 &  4.19 &   12 & KPD1946+4340        &   43 &  -17 &   53 &34200 &  5.43 &  3.61 &   25\\
HW Vir              &   18 &    8 &  -12 &28488 &  5.63 &  4.51 &   13 & SDSSJ0951+0347      &  -38 &  -61 &   79 &29800 &  5.48 &  4.04 &   19\\
EC10246-2707        &   12 &   -9 &  -28 &28900 &  5.64 &  4.51 &   14 & $[$CW83$]$ 1419-09  &   53 &  -58 &    3 &  ... &   ... & (4.3) &     \\
PG1043+760          &   18 &   50 &    9 &27600 &  5.39 &  3.98 &   15 & HE0929-0424         &    3 &  -68 &  -24 &29602 &  5.69 &  4.58 &   26\\
OGLE BUL-SC16335    &  -36:&  448:& -579:&31500 &  5.70 &  4.46 &   16 & KIC2991403          &    9 &    3 &    7 &27300 &  5.43 &  4.10 &   39\\
2M1938+4603         &   16 &   24 &    3 &29564 &  5.43 &  3.93 &   17 & HE0230-4323         &   32 &   19 &  -18 &31552 &  5.60 &  4.20 &   26\\
EC00404-4429        &  -67 &  -15 &  -34 &  ... &   ... & (4.3) &      & GALEXJ2349+3844     &   17 &    7 &    5 &23770 &  5.38 &  4.24 &   24\\
KIC7335517          &   85 &  -20 &    4 &  ... &   ... & (4.3) &      & KUV16256+4034       &   -5 &  -93 &  -28 &23100 &  5.38 &  4.30 &   36\\
ASAS102322-3737     &   -6 &   -7 &  -39 &25300 &  5.38 &  4.13 &   18 & BPS CS 22879-149    &  -10 &  -41 &  -51 &  ... &   ... & (4.3) &     \\
SDSSJ0830+4751      &  -37 &  -88 &   42 &28400 &  5.60 &  4.45 &   19 & HE1318-2111         &   67 &   -1 &   28 &36254 &  5.42 &  3.48 &   41\\
KIC6614501          & -150 &   68 & -163 &23700 &  5.70 &  5.80 &   20 & PG1544+488          & -136 &    2 &  148 &32800 &  5.33 &  3.65 &   42\\
2M1533+3759         &   64 &  -36 &    7 &29230 &  5.58 &  4.33 &   21 & SDSSJ1726+2744      &   57 &  -21 &  -88 &32600 &  5.84 &  4.74 &   28\\
SDSSJ1920+3722      &  -33 &   35 &   15 &27500 &  5.40 &  4.01 &   22 & PG1743+477          &  -48 &  -41 &  -24 &27600 &  5.57 &  4.43 &   43\\
HS2333+3927         &    7 &  -20 &   40 &36500 &  5.70 &  4.18 &   23 & GALEX J0507+0348    &  -74 &  -66 &    8 &23990 &  5.42 &  4.33 &   24\\
GALEX J0805-1058    &  -20 &  -49 &  -12 &22320 &  5.68 &  5.86 &   24 & PG0001+275          &   42 &  -68 &  -29 &25400 &  5.30 &  3.92 &   25\\
BPS CS 22169-0001   &    9 &   16 &    2 &39300 &  5.60 &  3.82 &   25 & PG1519+640          &  -18 &   50 &  -42 &30600 &  5.72 &  4.58 &   36\\
GALEX J0751+0925    &   -5 &  -24 &  -34 &30620 &  5.74 &  4.63 &   24 & HE1059-2735         & -143 &    8 &  -57 &40966 &  5.38 &  3.24 &   41\\
HE1415-0309         &   69 &  -15 &   89 &29520 &  5.56 &  4.26 &   26 & PG0101+039          &   15 &  -39 &  -22 &27500 &  5.53 &  4.34 &   44\\
HS1741+2133         &  -85 &  -98 &   40 &35600 &  5.30 &  3.21 &   27 & EC20182-6534        &   28 &  -29 &   35 &  ... &   ... & (4.3) &     \\
SDSSJ0823+1136      &  -90 &  -60 &   66 &31200 &  5.79 &  4.71 &   19 & PG1725+252          &  -58 &  -69 &   63 &26560 &  5.03 &  3.15 &   45\\
SDSSJ1138-0035      &   34 & -144 &  -63 &31200 &  5.54 &  4.08 &   28 & PG1247+554          &  -60 &  -42 &   22 &32366 &  6.11 &  5.42 &   46\\
PG1432+159          &   55 &  -47 &  -30 &26900 &  5.75 &  4.93 &   29 & HD 188112           &   25 &   20 &  -11 &21500 &  5.66 &  5.88 &   47\\
SDSSJ1625+3632      &  -27 &  -55 &  -60 &23570 &  6.12 &  6.86 &   30 & PG1648+536          &   68 &  -62 &  -36 &31400 &  5.62 &  4.27 &   36\\
PG2345+318          &    1 &  -17 &   -1 &27500 &  5.70 &  4.76 &   29 & SDSSJ1522-0130      &  -49 &    4 &  -49 &25200 &  5.47 &  4.36 &   19\\
SDSSJ2046-0454      &   39 &  -56 & -212 &31600 &  5.54 &  4.05 &   28 & SDSSJ2256+0656      &   24 &   -2 &   15 &28500 &  5.64 &  4.54 &   28\\
PG1329+159          &  -24 &  -48 &  -12 &29100 &  5.62 &  4.44 &   15 & EC22202-1834        &   -4 &  -59 &  -20 &  ... &   ... & (4.3) &     \\
FBS0117+396         &   72 &  -13 &    8 &29370 &  5.48 &  4.07 &   31 & PG1248+164          &   67 &    5 &  -18 &26600 &  5.68 &  4.78 &   15\\
SDSSJ1654+3037      &  100 &   32 &  -30 &24900 &  5.39 &  4.18 &   28 & JL 82               &  -35 &  -36 &    3 &26500 &  5.22 &  3.64 &   25\\
AA Dor              &  -66 &    8 &  -19 &37800 &  5.51 &  3.65 &   32 & PG0849+319          &  -77 &  -65 &  -18 &28900 &  5.37 &  3.83 &   15\\
HE0532-4503         &  182 &  -91 &   52 &25710 &  5.33 &  3.97 &   26 & SDSSJ1505+1108      &   19 & -232 &  -59 &33200 &  5.80 &  4.60 &   28\\
GALEXJ0321+4727     & -104 &  -39 &   45 &27990 &  5.34 &  3.83 &   24 & EQ Psc              &   74 &  -60 &  -27 &  ... &   ... & (4.3) &     \\
CPD-64$^\circ$481   &   48 &  -70 &  -41 &27500 &  5.60 &  4.51 &   33 & EC02200-2338        &  -20 &  -50 &    8 &  ... &   ... & (4.3) &     \\
KBS 13              &   32 &    7 &  -15 &33970 &  5.87 &  4.73 &   34 & KPD2215+5037        &  -40 &  -11 &   30 &29600 &  5.64 &  4.45 &   36\\
SDSSJ1021+3010      &   25 &   11 &  -19 &30400 &  5.67 &  4.47 &   19 & Ton S 183           &   46 &  -21 &  -38 &27600 &  5.43 &  4.08 &   25\\
HS2043+0615         &   21 &  -89 &  -42 &26157 &  5.28 &  3.81 &   26 & EC13332-1424        &  -68 &   40 &    9 &  ... &   ... & (4.3) &     \\
PG0941+280          &  -41 & -151 &   11 &29400 &  5.43 &  3.94 &   16 & PG1627+017          &  -25 &  -30 &  -29 &23500 &  5.40 &  4.32 &   25\\
\hline
\end{tabular}
\end{minipage}
\end{center}
\end{table}
\end{landscape}

\label{lastpage}

\addtocounter{table}{-1}
\begin{landscape}
\begin{table}
\begin{center}
\vspace{-0.3cm}
\begin{minipage}{230mm}%
\renewcommand{\footnoterule}{\vspace*{-15pt}}
\caption{{\it continued}}
\begin{tabular}{lrrrcccclrrrcccc}
\hline
Name                &$U$  &$V$  &$W$ & \teff       &$\log{g}$ & $M_V$& Ref. &Name                &$U$  &$V$  &$W$  & \teff       &$\log{g}$& $M_V$& Ref.  \\
 &(\kmps)&(\kmps)&(\kmps) & (K) & (c.g.s.) &(mag)& & &(\kmps)&(\kmps)&(\kmps) & (K) & (c.g.s.) &(mag)& \\
\hline
EC21556-5552        &   22 &   13 &  -25 &  ... &   ... & (4.3) &      &EC20369-1804        &    3 &  -13 &  -26 &  ... &   ... & (4.3) &     \\
PG1230+052          &    2 &  -37 &  -57 &27100 &  5.47 &  4.22 &   36 &SDSSJ1832+6309      &  -61 &  -24 &  -19 &26800 &  5.29 &  3.79 &   19\\
PG1116+301          &  -23 &  -41 &  -12 &32500 &  5.85 &  4.77 &   15 &PG0839+399          &  -22 &  -62 &   -3 &37800 &  5.53 &  3.70 &   43\\
PG0918+029          &  -57 &  -93 &   15 &31700 &  6.03 &  5.27 &   29 &PG1244+113          &   44 &   21 &   12 &36300 &  5.54 &  3.78 &   36\\
EC12408-1427        &  -64 &   15 &  -24 &  ... &   ... & (4.3) &      &CD-24$^\circ$ 731   &  -33 & -103 &    7 &35400 &  5.90 &  4.73 &   53\\
HE2135-3749         &  -21 &   -1 &  -77 &30000 &  5.84 &  4.92 &   48 &HE1115-0631         &  -40 & -130 &    1 &40443 &  5.80 &  4.30 &   41\\
PB 5333             &  -22 & -121 &   29 &37900 &  5.81 &  4.40 &   29 &PG0907+123          &  -43 &  -34 &   10 &27280 &  5.54 &  4.38 &   45\\
HS2359+1942         &  141 &  -21 &   81 &31434 &  5.56 &  4.11 &   26 &PG1032+406          &  -69 &  -57 &  -13 &31290 &  5.78 &  4.68 &   45\\
PG1452+198          &   26 &  -18 &   -8 &29400 &  5.75 &  4.74 &   36 &SDSSJ0952+6258      &   31 &  -74 &   13 &27700 &  5.59 &  4.47 &   19\\
SDSSJ1508+4940      &    5 &  -30 &  -44 &29600 &  5.73 &  4.68 &   19 &HE1448-0510         &   -1 &  -14 &  -52 &34760 &  5.53 &  3.84 &   26\\
PG1000+408          &  -20 &  -56 &   52 &36400 &  5.54 &  3.78 &   45 &PG1439-013          &  -45 &  -17 &  -23 &  ... &   ... & (4.3) &     \\
SDSSJ1132-0636      &   10 &    1 &   12 &46400 &  5.89 &  4.53 &   19 &SDSSJ0321+0538      &   37 &  -22 &    3 &30700 &  5.74 &  4.62 &   19\\
GALEX J1731+0647    &  -23 &  -73 &   73 &27780 &  5.35 &  3.87 &   24 &PHL 861             &  108 & -162 &   11 &29668 &  5.50 &  4.10 &   26\\
HE1421-1206         &  -72 &  -54 &  -67 &29600 &  5.55 &  4.23 &   25 &PG0940+068          &   56 &    8 &   15 &  ... &   ... & (4.3) &     \\
PG2331+038          &   96 &  -41 &  -13 &27200 &  5.58 &  4.48 &   36 &Feige 108           &   25 &   14 &  -38 &35880 &  6.26 &  5.61 &   45\\
HE1047-0436         &  -26 &  -20 &    8 &30200 &  5.66 &  4.46 &   49 &EC20260-4757        &   61 &   -2 &  -14 &  ... &   ... & (4.3) &     \\
GALEX J2254-5515    &  -26 &    0 &  -16 &31070 &  5.80 &  4.74 &   24 &FF Aqr              &    0 &    8 &  -29 &32000 &  6.00 &  3.47 &   54\\
PG0133+114          &   -4 &  -42 &   -6 &29600 &  5.66 &  4.50 &   43 &PG1110+294          &    4 &  -31 &  -19 &30100 &  5.72 &  4.62 &   15\\
PG1512+244          &  -50 &  -58 &   56 &29900 &  5.74 &  4.68 &   15 &KIC11558725         &  114 & -120 &  149 &27910 &  5.41 &  4.01 &   55\\
$[$CW83$]$ 1735+22  &   21 &   -9 &   56 &38000 &  5.54 &  3.71 &   25 &PG1558-007          &  -29 &  -86 &  -44 &20300 &  5.00 &  3.58 &   56\\
SDSSJ0118-0025      &    0 &  -44 &  -46 &27900 &  5.55 &  4.36 &   19 &LB 1516             &   -5 &   -1 &  -14 &25200 &  5.41 &  4.21 &   35\\
HE2150-0238         &   -5 &  -16 &   30 &30200 &  5.83 &  4.88 &   25 &CS 1246             &   -5 &  -82 &    0 &28500 &  5.46 &  4.09 &   57\\
KPD2040+3955        &  103 &  -28 &   14 &27900 &  5.54 &  4.33 &   36 &KIC7668647          &  273 &  -91 &  -29 &27700 &  5.50 &  4.25 &   58\\
SDSSJ0023-0029      & -177 &  -30 &    5 &29200 &  5.69 &  4.61 &   28 &PG1619+522          &  -19 &  -33 &  -27 &32300 &  5.98 &  5.11 &   39\\
HD 49798            &    5 &   -4 &    3 &  ... &   ... & (4.3) &      &PG0919+273          &   95 &  -20 &  -22 &32900 &  5.90 &  4.87 &   36\\
KIC7664467          &    9 &    3 &    7 &26800 &  5.17 &  3.49 &   39 &EGB 5               &  -83 &    9 &   -2 &34500 &  5.85 &  4.65 &   59\\
HD 171858           &   73 &   -3 &    3 &27200 &  5.30 &  3.78 &   25 &HD185510            &   -8 &  -30 &  -32 &31000 &  6.50 &  1.10 &   60\\
PG1403+316          &  -47 &  -36 &   24 &31200 &  5.75 &  4.61 &   36 &PG0850+170          &    1 &  -37 &   17 &27100 &  5.37 &  3.97 &   15\\
PG1716+426          &   88 &   -5 &  -33 &27400 &  5.47 &  4.19 &   15 &59 Cyg              &   -4 &   -7 &   -1 &52100 &  5.00 & -3.45 &   61\\
SDSSJ1346+2817      &  -31 &  -96 &   28 &28800 &  5.46 &  4.06 &   19 &FY CMa              &  -18 &  -11 &   -3 &45000 &  4.30 & -2.51 &   62\\
NGC188/II-91        &   10 &    5 &    6 &  ... &   ... & (4.3) &      &$\phi$ Per          &   -6 &  -22 &    0 &53000 &  4.20 & -2.80 &   63\\
PG1300+279          &    1 &  -41 &    4 &29600 &  5.65 &  4.48 &   15 &BD-11$^\circ$ 162   &   58 &   -4 &   -4 &35000 &  5.90 &  4.16 &   64\\
CPD-20$^\circ$ 1123 &   54 &  -28 &   17 &23500 &  4.90 &  3.09 &   50 &PG1701+359          & -110 & -145 &   61 &33010 &  5.91 &  4.53 &   45\\
HD 149382           &   30 &    7 &   20 &34200 &  5.89 &  4.77 &   29 &PG1104+243          &  -51 &  -54 &  -48 &33500 &  5.85 &  3.95 &   65\\
PG1538+269          &   43 &   13 &  -19 &25200 &  5.30 &  3.94 &   29 &PG1018-047          &  -22 &  -59 &  -19 &30500 &  5.50 &  3.97 &   66\\
GALEX J1632+0759    &   -3 &    3 &  -34 &38110 &  5.38 &  3.31 &   24 &PG1449+653          &  -68 & -125 &  -82 &28150 &  5.50 &  3.85 &   67\\
PG1253+284          &  -13 &   -9 &   24 &  ... &   ... & (4.3) &      &PG1338+611          &   85 &  -24 &   82 &27400 &  5.54 &  3.89 &   68\\
PG0958-073          & -116 &  -84 &  -20 &26100 &  5.58 &  4.57 &   35 &BD+34$^\circ$ 1543  &  -12 &  -66 &   28 &36700 &  5.92 &  3.55 &   68\\
KIC10553698A        & -167 &  111 & -155 &27423 &  5.44 &  4.12 &   51 &PG1317+123          &   12 &   -9 &   44 &37000 &  5.80 &  3.75 &   69\\
KPD0025+5402        &   57 &   15 &  -27 &28200 &  5.37 &  3.88 &   43 &BD-7$^\circ$ 5977   &  -28 &   -9 &    5 &29000 &  5.02 &   .21 &   65\\
PB7 352             &   11 &    8 &   14 &25000 &  5.35 &  4.07 &   25 &BD+29$^\circ$ 3070  &  -42 &  -30 &   -8 &28500 &  5.76 &  3.59 &   68\\
PG0934+186          &  -18 &  -27 &  -26 &35800 &  5.65 &  4.08 &   36 &TYC3871-835-1       &  -31 &  -49 &   28 &22500 &  5.12 &  3.24 &   65\\
Ton S 135           &   21 &  -38 &    8 &25000 &  5.60 &  4.70 &   52 &                    &      &      &      &      &       &       &     \\
\hline
\end{tabular}\\
$^a$ $M_V$ estimated for the spectral composite.\\
\end{minipage}
\end{center}
\end{table}
\end{landscape}

\addtocounter{table}{-1}
\begin{landscape}
\begin{table}
\begin{center}
\vspace{-0.3cm}
\begin{minipage}{230mm}%
\renewcommand{\footnoterule}{\vspace*{-15pt}}
\caption{{\it continued}}
\begin{tabular}{lrrrcccclrrrcccc}
\hline
\hline
\end{tabular}\\
References: (1) \citet{ven2012}; (2) \citet{sch2014a}; (3) \citet{max2002}; (4) \citet{oto2006a}; (5) \citet{koe1998}; (6) \citet{gei2007};
(7) \citet{dre2001}; (8) \citet{gei2011b}; (9) \citet{vuc2007}; (10) \citet{alm2012}; (11) \citet{ost2007}; (12) \citet{ver2012}; (13) \citet{ede2008}; (14) \citet{bar2013b};
(15) \citet{max2001}; (16) \citet{gei2014}; (17) \citet{ost2010b}; (18) \citet{sch2013}; (19) \citet{kup2015}; (20) \citet{sil2012}; (21) \citet{for2010}; (22) \citet{sch2014b};
(23) \citet{heb2004}; (24) \citet{nem2012}; (25) \citet{gei2010b}; (26) \citet{lis2005}; (27) \citet{ede2003b}; (28) \citet{gei2011c}; (29) \citet{saf1994}; (30) \citet{kil2011};
(31) \citet{ost2013}; (32) \citet{mul2010}; (33) \citet{oto2005}; (34) \citet{for2008}; (35) \citet{gei2013b}; (36) \citet{cop2011}; (37) \citet{ede1999}; (38) \citet{heb2000}; 
(39) \citet{ost2010c}; (40) \citet{lie1994}; (41) \citet{str2007}; (42) \citet{sen2014}; (43) \citet{mor2003}; (44) \citet{gei2008}; (45) \citet{bil2002}; (46) \citet{kep1995}; 
(47) \citet{heb2003}; (48) \citet{kar2006}; (49) \citet{nap2001}; (50) \citet{nas2012}; (51) \citet{ost2014}; (52) \citet{heb1986}; (53) \citet{oto2006b}; (54) \cite{vac2007}; 
(55) \citet{tel2012a}; (56) \citet{heb2002}; (57) \citet{bar2010}; (58) \citet{tel2014b}; (59) \citet{gei2011d}; (60) \citet{jef1997}; (61) \citet{pet2013}; (62) \citet{pet2008}; 
(63) \citet{gie1998}; (64) \citet{ull1998}; (65) \citet{vos2014}; (66) \citet{dec2012}; (67) \citet{azn2001}; (68) \citet{vos2013}; (69) \citet{the1995}.
\end{minipage}
\end{center}
\end{table}
\end{landscape}

\end{document}